\documentstyle[amssymb,thmsa,11pt,sw20lart]{article}

\input{tcilatex}
\begin{document}

\author{C. Bagnuls\thanks{%
Service de Physique de l'Etat Condens\'{e}} and C.\ Bervillier\thanks{%
Service de Physique Th\'{e}orique e-mail: bervil@spht.saclay.cea.fr} \\
C. E. Saclay, F91191 Gif-sur-Yvette Cedex, France}
\title{Exact Renormalization Group Equations.\\
An Introductory Review.}
\date{\today}
\maketitle

\begin{abstract}
We critically review the use of the exact renormalization group equations
(ERGE) in the framework of the scalar theory. We lay emphasis on the
existence of different versions of the ERGE and on an approximation method
to solve it: the derivative expansion. The leading order of this expansion
appears as an excellent textbook example to underline the nonperturbative
features of the Wilson renormalization group theory. We limit ourselves to
the consideration of the scalar field (this is why it is an introductory
review) but the reader will find (at the end of the review) a set of
references to existing studies on more complex systems.

\newpage
\end{abstract}

\tableofcontents

\newpage

\section{Introduction}

``{\sl The formal discussion of consequences of the renormalization group
works best if one has a differential form of the renormalization group
transformation. Also, a differential form is useful for the investigation of
properties of the }$\varepsilon ${\sl \ expansion to all orders (...) A
longer range possibility is that one will be able to develop approximate
forms of the transformation which can be integrated numerically; if so one
might be able to solve problems which cannot be solved any other way.}'' 
\cite{440}

By ``exact renormalization group equation (ERGE)'', we mean the continuous
(i.e. not discrete) realization of the Wilson renormalization group (RG)
transformation of the action in which no approximation is made and also no
expansion is involved with respect to some small parameter of the action 
\footnote{%
On the meaning of the word ``exact'', see a discussion following the talk by
Halperin in \cite{4390}.}. Its formulation --- under a differential form ---
is known since the early seventies \cite{4495,440,414}. However, due to its
complexity (an integro-differential equation), its study calls for the use
of approximation (and/or truncation) methods. For a long time it was natural
to use a perturbative approach (based on the existence of a small parameter
like the famous $\varepsilon $-expansion for example). But, the standard
perturbative field theory (e.g. see \cite{3533}) turned out to be more
efficient and, in addition the defenders of the nonperturbative approach
have turned towards the discrete formulation of the RG due to the problem of
the ''stiff'' differential equation (see a discussion following a talk given
by Wegner \cite{4389}). This is why it is only since the middle of the
eighties that substantial studies have been carried out via:

\begin{itemize}
\item  the truncation procedures in the scaling field method \cite{319,4405}
(extended studies of \cite{4404})

\item  the explicit consideration of the local potential approximation \cite
{2085} and of the derivative expansion \cite{212}

\item  an appealing use, for field theoreticians, of the ERGE \cite{354}
\end{itemize}

In the nineties there has been a rapid growth of studies in all directions,
accounting for scalar (or vector) fields, spinor, gauge fields, finite
temperature, supersymmetry, gravity, etc...

In this paper we report on progress in the handling of the ERGE. Due to the
abundance of the literature on the subject and because this is an
introductory report, we have considered in detail the various versions of
the ERGE only in the scalar (or vector) case\footnote{%
However, because of the review by Wetterich and collaborators in this
volume, we have not reported on their work as it deserves. The reader is
invited to refer to their review \cite{4700} and also to \cite{4315,4248}
from which he could realize the rich variety of calculations that may be
done within the nonperturbative framework of the ERGE.}. This critical
review must also be seen as an incitement to look at the original papers of
which we give a list as complete as possible.

Let us mention that the ERGE is almost ignored in most of the textbooks on
the renormalization group except notably in \cite{3856} (see also \cite{4397}%
).

\section{Exact Renormalization Group Equations\label{PartFirst}}

\subsection{Introduction}

There are four representations of the ERGE: the functional differential
equation, the functional integral, the infinite set of partial differential
equations for the couplings $u_{n}\left( {\bf p}_{1},\cdots ,{\bf p}%
_{n};t\right) $ (eq. 11.19 of \cite{440}) which are popularly known because
they introduce the famous ``beta'' functions $\beta _{n}\left( \left\{
u_{n}\right\} \right) $: 
\[
\frac{\text{d}u_{n}}{\text{d}t}=\beta _{n}\left( \left\{ u_{n}\right\}
\right) 
\]
and the infinite hierarchy of the ordinary differential equations for the
scaling fields $\mu _{i}(t)$ \cite{4405} (the RG-scale parameter $t$ is
defined in section \ref{Notations} and the scaling fields $\mu _{i}(t)$ in
section \ref{SecLRGT}).

In this review we shall only consider the functional differential
representation of the ERGE.

There is not a unique form of the ERGE, each form of the equation is
characterized by the way the momentum cutoff $\Lambda $ is introduced. (In
perturbative RG, this kind of dependence is known as the
regularization-scheme dependence or ``scheme dependence'' in short \cite
{4772}.) The important point is that the various forms of the ERGE embody a
unique physical content in the sense that they all preserve the same physics
at large distances and, via the recourse to a process of limit, yield the
same physics at small distances (continuum limits). The object of this part
is to present the main equations used in the literature and connections
between their formally (but not necessarily practically) equivalent forms.
We do not derive them in detail here since one may find the derivations in
several articles or reviews (that we indicate below).

Although the (Wilson) renormalization group theory owes much to the
statistical physics as recently stressed by M. E. Fisher \cite{4191} we
adopt here the notations and the language of field theory.

Before considering explicitly the various forms of the ERGE (in sections \ref
{WHERGE}-\ref{LegendreSec}), we find it essential to fix the notations and
to remind some fundamental aspects of the RG.

\subsection{Notations, reminders and useful definitions\label{Notations}}

We denote by $\Lambda $ the momentum cutoff and the RG-``time'' $t$ is
defined by $\frac{\Lambda }{\Lambda _{0}}=$e$^{-t}$ in which $\Lambda _{0}$
stands for some initial value of $\Lambda .$

We consider a scalar field $\phi ({\bf x})$ with ${\bf x}$ the coordinate
vector in an Euclidean space of dimension $d$. The Fourier transformation of 
$\phi ({\bf x})$ is defined as:

\[
\phi ({\bf x})=\int_{p}\phi _{p}\text{e}^{i{\bf p\cdot x}} 
\]
in which 
\begin{equation}
\int_{p}\equiv \int \frac{d^{d}p}{(2\pi )^{d}}  \label{nota1}
\end{equation}
and $\phi _{p}$ stands for the function $\phi ({\bf p})$ where ${\bf p} $ is
the momentum vector (wave vector).

The norms of ${\bf x}$ and ${\bf p}$ are noted respectively $x$ and $p$ ($%
x\equiv \sqrt{{\bf x}\cdot {\bf x}}$). However, when no confusion may arise
we shall denote the vectors ${\bf x}$ and ${\bf p}$ by simply $x$ and $p$ as
in (\ref{nota1}) for example. Sometimes the letters $k$ and $q$ (or ${\bf k}$
and ${\bf q}$) will refer also to momentum variables.

It is useful to define $K_{d}$ as the surface of the $d$-dimensional unit
sphere divided by $\left( 2\pi \right) ^{d}$, i.e.:

\begin{equation}
K_{d}=\frac{2\pi ^{\frac{d}{2}}}{\left( 2\pi \right) ^{d}\Gamma \left( \frac{%
d}{2}\right) }  \label{KD}
\end{equation}

We shall also consider the case where the field has $N$ components ${\bf %
\phi }=(\phi _{1},\cdots ,\phi _{N})$ that we shall also denote generically
by $\phi _{\alpha }$.

The action $S[\phi ]$ (the Hamiltonian divided by $k_{\text{B}}T$ for
statistical physics) is a general semi-local functional of $\phi $.
Semi-local means that, when it is expanded\footnote{%
We do not need to assume this expansion as the unique form of $S[\phi ]$ in
general.} in powers of $\phi $, $S[\phi ]$ involves only powers of $\phi (%
{\bf x})$ and of its derivatives with respect to $x^{\mu }$ (that we denote $%
\partial ^{\mu }\phi $ or even $\partial \phi $ instead of $\partial \phi
/\partial x^{\mu }$). This characteristics is better expressed in the
momentum-space (or wave-vector-space). So we write:

\begin{equation}
S[\phi ]=\sum_{n}^{\infty }\int_{p_{1}\cdots p_{n}}u_{n}\left( {\bf p}%
_{1},\cdots ,{\bf p}_{n}\right) \phi _{p_{1}}\cdots \phi _{p_{n}}\hat{\delta}%
\left( {\bf p}_{1}+\cdots +{\bf p}_{n}\right)  \label{expand0}
\end{equation}
in which $\hat{\delta}\left( {\bf p}\right) \equiv \left( 2\pi \right)
^{d}\delta ^{d}\left( {\bf p}\right) $ is the $d$-dimensional
delta-function: 
\[
\delta ^{d}\left( {\bf x}\right) =\int_{p}\text{e}^{i{\bf p\cdot x}} 
\]

Notice that the O(1) symmetry $\phi \rightarrow -\phi $, also called the $%
Z_{2}$ symmetry, is not assumed neither here nor in the following sections
except when it is explicitly mentioned.

The $u_{n}$'s are invariant under permutations of their arguments.

For the functional derivative with respect to $\phi $, we have the relation: 
\[
\frac{\delta }{\delta \phi _{p}}=\int \text{d}^{d}x\,\text{e}^{i{\bf p\cdot x%
}}\frac{\delta }{\delta \phi ({\bf x})}
\]
so that in performing the functional derivative with respect to $\phi _{p}$
we get rid of the $\pi $-factors involved in the definition (\ref{nota1}),
e.g..:

\begin{eqnarray*}
\frac{\delta }{\delta \phi _{p}} &&\left[ \int_{p_{1}\cdots
p_{n}}u_{n}\left( {\bf p}_{1},\cdots ,{\bf p}_{n}\right) \phi _{p_{1}}\cdots
\phi _{p_{n}}\hat{\delta}\left( {\bf p}_{1}+\cdots +{\bf p}_{n}\right)
\right] = \\
&&n\int_{p_{1}\cdots p_{n-1}}u_{n}\left( {\bf p}_{1},\cdots ,{\bf p}_{n-1},%
{\bf p}\right) \phi _{p_{1}}\cdots \phi _{p_{n-1}}\hat{\delta}\left( {\bf p}%
_{1}+\cdots +{\bf p}_{n-1}+{\bf p}\right)
\end{eqnarray*}
in order to lighten the notations we sometimes will write $\frac{\delta }{%
\delta \phi }$ instead of $\frac{\delta }{\delta \phi ({\bf x})}$ when no
confusion may arise.

Let us also introduce:

\begin{itemize}
\item  The generating functional $Z[J]$ of Green's functions:
\end{itemize}

\begin{equation}
Z[J]={\cal Z}^{-1}\int {\cal D}\phi \,\exp \left\{ -S[\phi ]+J\cdot \phi
\right\}  \label{ZJ}
\end{equation}
in which 
\[
J\cdot \phi \equiv \int d^{d}xJ({\bf x})\phi ({\bf x}) 
\]
$J({\bf x})$ is an external source, and ${\cal Z}$ is a normalization such
that $Z[0]=1$. Indeed ${\cal Z}$ is the partition function: 
\begin{equation}
{\cal Z}=\int {\cal D}\phi \exp \left\{ -S[\phi ]\right\}  \label{PartFn}
\end{equation}

\begin{itemize}
\item  The generating functional $W[J]$, of connected Green functions, is
related to $Z[J]$ as follows:
\end{itemize}

\begin{equation}
W[J]=\ln \left( Z[J]\right)  \label{WJ}
\end{equation}

Notice that if one defines ${\cal W}$

\[
\text{e}^{{\cal W}}={\cal Z} 
\]
then ${\cal W}$ is minus the free energy.

\begin{itemize}
\item  The Legendre transformation which defines the generating functional $%
\Gamma \left[ \Phi \right] $ of the one-particle-irreductible (1PI) Green
functions (or simply vertex functions):
\end{itemize}

\begin{eqnarray}
\Gamma \left[ \Phi \right] +W\left[ J\right] -J\cdot \Phi &=&0  \nonumber \\
\left. \frac{\delta \Gamma \left[ \Phi \right] }{\delta \Phi (x)}\right|
_{J} &=&J(x)  \label{Legendre}
\end{eqnarray}
in which we have introduced the notation $\Phi $ to make a distinction
between this and the (dummy) field variable $\phi $ in (\ref{ZJ}). In the
following we shall not necessarily make this distinction.

\subsubsection{Dimensions}

First let us define our conventions relative to the usual (i.e. engineering)
dimensions.

In the following, we refer to a system of units in which the dimension of a
length scale $L$ is $-1$:

\[
\left[ L\right] =-1 
\]
and a momentum scale like $\Lambda $ has the dimension $1$:

\[
\left[ \Lambda \right] =1 
\]

The usual classical dimension of the field (in momentum unit): 
\begin{equation}
d_{\phi }^{c}=\left[ \phi \right] =\frac{1}{2}\left( d-2\right)
\label{dimfcla}
\end{equation}
is obtained by imposing that the coefficient of the kinetic term $\int $d$%
^{d}x\left( \partial \phi (x)\right) ^{2}$ in $S$ is dimensionless [it is
usually set to $\frac{1}{2}$].

The dimension of the field is not always given by (\ref{dimfcla}). Indeed
one knows that the field may have an anomalous dimension:

\begin{equation}
d_{\phi }^{a}=\frac{1}{2}\left( d-2+\eta \right)  \label{dimfano}
\end{equation}
with $\eta $ a non-zero constant defined with respect to a non-trivial fixed
point.

In RG theory, the dimension of the field depends on the fixed point in the
vicinity of which the (field) theory is considered. Hence we introduce an
adjustable dimension of the field, $d_{\phi }$ , which controls the scaling
transformation of the field: 
\begin{equation}
\phi \left( s{\bf x}\right) =s^{-d_{\phi }}\phi \left( {\bf x}\right)
\label{dimfeff}
\end{equation}
For the Fourier transformation, we have:

\begin{equation}
\phi \left( s{\bf p}\right) =s^{d_{\phi }-d}\phi \left( {\bf p}\right)
\label{dimfeffFou}
\end{equation}

Since the dimension of any dimensioned (in the classical meaning for $\phi $%
) quantity is expressed in term of a momentum scale, we use $\Lambda $ to
reduce all dimensioned quantities into dimensionless quantities. In the
following we deal with dimensionless quantities and, in particular, the
notation $p$ will refer to a dimensionless momentum variable. However
sometimes, for the sake of clarity, we will need to reintroduce the explicit 
$\Lambda $-dependence, e.g. via the ratio $p/\Lambda $.

It is also useful to notice that, with a dimensionless $p$, the following
derivatives are equivalent (the derivative is taken at constant dimensioned
momentum): 
\[
\frac{\partial }{\partial t}=-\Lambda \frac{\partial }{\partial \Lambda }=p%
\frac{\partial }{\partial p}
\]

\subsubsection{Transformations of the field variable\label{SecTra}}

In order to discuss invariances in RG theory, it is useful to consider a
general transformation of the field which leaves invariant the partition
function. Following \cite{4011,2835}, we replace $\phi _{p}$ by $\phi
_{p}^{\prime }$ such that 
\begin{equation}
\phi _{p}^{\prime }=\phi _{p}+\sigma \Psi _{p}[\phi ]  \label{tra}
\end{equation}
where $\sigma $ is infinitesimally small and $\Psi _{p}$ a function which
may depend on all Fourier components of $\phi $. Then one has 
\[
S[\phi ^{\prime }]=S[\phi ]+\sigma \int_{p}\Psi _{p}[\phi ]\frac{\delta
S[\phi ]}{\delta \phi _{p}} 
\]

Moreover we have: 
\[
\int {\cal D}\phi ^{\prime }=\int {\cal D}\phi \frac{\partial \left\{ \phi
^{\prime }\right\} }{\partial \left\{ \phi \right\} }=\int {\cal D}\phi
\left( 1+\sigma \int_{p}\frac{\delta \Psi _{p}[\phi ]}{\delta \phi _{p}}%
\right) 
\]

The transformation must leave the partition function ${\cal Z}$ [eq. (\ref
{PartFn})] invariant. Therefore one obtains 
\begin{eqnarray*}
{\cal Z} &=&\int {\cal D}\phi ^{\prime }\exp \left\{ -S[\phi ^{\prime
}]\right\} \\
&=&\int {\cal D}\phi \exp \left\{ -S[\phi ]-\sigma {\cal G}_{\text{tra}%
}\left\{ \Psi \right\} S[\phi ]\right\}
\end{eqnarray*}
with

\begin{equation}
{\cal G}_{\text{tra}}\left\{ \Psi \right\} S[\phi ]=\int_{p}\left( \Psi _{p}%
\frac{\delta S}{\delta \phi _{p}}-\frac{\delta \Psi _{p}}{\delta \phi _{p}}%
\right)  \label{Gtra}
\end{equation}
which indicates how the action transforms under the infinitesimal change (%
\ref{tra}):

\[
\frac{\text{d}S}{\text{d}\sigma }={\cal G}_{\text{tra}}\left\{ \Psi \right\}
S 
\]

In the case of $N$ components, the expression (\ref{Gtra}) generalizes
obviously: 
\begin{equation}
{\cal G}_{\text{tra}}\left\{ {\bf \Psi }\right\} S[{\bf \phi }]=\sum_{\alpha
=1}^{N}\int_{p}\left( \Psi _{p}^{\alpha }\frac{\delta S}{\delta \phi
_{p}^{\alpha }}-\frac{\delta \Psi _{p}^{\alpha }}{\delta \phi _{p}^{\alpha }}%
\right)  \label{GtraN}
\end{equation}

\subsubsection{Rescaling\label{rescal}}

We consider an infinitesimal change of (momentum) scale: 
\begin{equation}
p\rightarrow p^{\prime }=s\,p=(1+\sigma )\,p  \label{dil}
\end{equation}
with $\sigma $ infinitesimally small. Introducing the rescaling operator of 
\cite{4011,2835}, the consequence on $S[\phi ]$ is written as

\begin{equation}
S\rightarrow S^{\prime }=S+\sigma {\cal G}_{\text{dil}}S  \label{ChangeDil}
\end{equation}

and thus:

\begin{equation}
\frac{\text{d}S}{\text{d}\sigma }={\cal G}_{\text{dil}}S  \label{DiffDil}
\end{equation}

Considering $S$ as given by (\ref{expand0}), then $\Delta S=\sigma {\cal G}_{%
\text{dil}}S$ may be expressed by gathering the changes induced by (\ref{dil}%
) on the various factors in the sum, namely:

\begin{description}
\item[1]  the differential volume $\prod_{i=1}^{n}$d$^{d}p_{i}=(1+\sigma
)^{-nd}\prod_{i=1}^{n}d^{d}p_{i}^{\prime }$ induces a change $\Delta S_{1}$
which may be written as
\end{description}

\[
\Delta S_{1}=-\sigma \,\left( d\,\int_{p}\phi _{p}\frac{\delta }{\delta \phi
_{p}}\right) S 
\]

\begin{description}
\item[2]  the couplings $u_{n}\left( {\bf p}_{1},\cdots ,{\bf p}_{n}\right)
=u_{n}\left( s^{-1}{\bf p}_{1}^{\prime },\cdots ,s^{-1}{\bf p}_{n}^{\prime
}\right) $ induce a change $\Delta S_{2}$ which may be written as
\end{description}

\[
\Delta S_{2}=-\sigma \,\left( \int_{p}\phi _{p}\,{\bf p}\cdot \partial
_{p}^{\prime }\frac{\delta }{\delta \phi _{p}}\right) S 
\]
where the prime on the derivative symbol ($\partial _{p}^{\prime }$)
indicates that the momentum derivative does not act on the delta-functions.

\begin{description}
\item[3]  the delta-functions{\scriptsize \ }$\hat{\delta}\left( {\bf p}%
_{1}+\cdots +{\bf p}_{n}\right) =\hat{\delta}\left( s^{-1}{\bf p}%
_{1}^{\prime }+\cdots +s^{-1}{\bf p}_{n}^{\prime }\right) $ induce a change $%
\Delta S_{3}$ which may be written as 
\[
\Delta S_{3}=\sigma \,\left( d\,S\right) 
\]
{\scriptsize \ }

\item[4]  the field itself $\phi \left( {\bf p}\right) =\phi \left( s^{-1}%
{\bf p}^{\prime }\right) $ induces a change $\Delta S_{4}$ which, according
to (\ref{dimfeffFou}), is dictated by 
\[
\phi \left( s^{-1}{\bf p}^{\prime }\right) =(1+\sigma )^{d-d_{\phi }}\phi
\left( {\bf p}\right) 
\]
\end{description}

Hence, to the first order in $\sigma $, we have 
\[
\Delta S_{4}=\sigma \,\left[ \left( d-d_{\phi }\right) \,\int_{p}\phi _{p}%
\frac{\delta }{\delta \phi _{p}}\right] S 
\]

Summing the four contributions, $\Delta S=\sum_{i=1}^{4}\Delta S_{i}$ we
obtain: 
\[
\Delta S=\sigma \,\left( d\,S-\int_{p}\phi _{p}\,{\bf p}\cdot \partial
_{p}^{\prime }\frac{\delta }{\delta \phi _{p}}-d_{\phi }\,\int_{p}\phi _{p}%
\frac{\delta }{\delta \phi _{p}}\right) S 
\]

This expression may be further simplified by allowing the momentum
derivative $\partial _{p}$ to act also on the delta-functions (this
eliminates the prime and absorbs the term $d\,S$). We thus may write 
\[
\Delta S=-\sigma \,\left( \int_{p}\phi _{p}\,{\bf p}\cdot \partial _{p}\frac{%
\delta }{\delta \phi _{p}}+d_{\phi }\,\int_{p}\phi _{p}\frac{\delta }{\delta
\phi _{p}}\right) S 
\]

hence:

\begin{equation}
{\cal G}_{\text{dil}}S=-\left( \int_{p}\phi _{p}\,{\bf p}\cdot \partial _{p}%
\frac{\delta }{\delta \phi _{p}}+d_{\phi }\int_{p}\phi _{p}\frac{\delta }{%
\delta \phi _{p}}\right) S  \label{Gdil}
\end{equation}

As in the case of ${\cal G}_{\text{tra}}$, the generalization to $N$
components is obvious [see eq. (\ref{GtraN})].

The writing of the action of ${\cal G}_{\text{dil}}$ [eq. (\ref{Gdil})] may
take on two other forms in the literature:

\begin{enumerate}
\item  due to the possible integration by parts of the first term: 
\begin{equation}
{\cal G}_{\text{dil}}S=\left( \int_{p}\left( {\bf p}\cdot \partial _{p}\phi
_{p}\right) \,\frac{\delta }{\delta \phi _{p}}+\left( d-d_{\phi }\right)
\int_{p}\phi _{p}\frac{\delta }{\delta \phi _{p}}\right) S  \label{Gdil2}
\end{equation}

\item  if one explicitly performs the derivative with respect to ${\bf p}$
acting on the $\delta $-functions: 
\begin{equation}
{\cal G}_{\text{dil}}S=dS-\left( \int_{p}\phi _{p}\,{\bf p}\cdot \partial
_{p}^{\prime }\frac{\delta }{\delta \phi _{p}}+d_{\phi }\int_{p}\phi _{p}%
\frac{\delta }{\delta \phi _{p}}\right) S  \label{Gdil3}
\end{equation}
in which now $\partial _{p}^{\prime }$ does not act on the $\delta $%
-functions.
\end{enumerate}

An other expression of the operator ${\cal G}_{\text{dil}}$ may be found in
the literature \cite{3357}, it is: 
\begin{equation}
{\cal G}_{\text{dil}}=d-\Delta _{\partial }-d_{\phi }\,\Delta _{\phi }
\label{Gdil4}
\end{equation}

where $\Delta _{\phi }=\phi .{\frac{\delta }{\delta \phi }}$ is the
`phi-ness' counting operator: it counts the number of occurrences of the
field $\phi $ in a given vertex and $\Delta _{\partial }$ may be expressed
as 
\begin{equation}
\Delta _{\partial }=d+\int_{p}\phi _{p}\,\,{\bf p}\cdot \partial _{p}\frac{%
\delta }{\delta \phi _{p}}  \label{msco}
\end{equation}
i.e. the momentum scale counting operator $+d$. Operating on a given vertex
it counts the total number of derivatives acting on the fields $\phi $ \cite
{3357}.

Notice that we have not introduced the anomalous dimension $\eta $ of the
field. This is because it naturally arises at the level of searching for a
fixed point of the ERGE. As we indicate in the following section, the
introduction of $\eta $ is related to an invariance.

\subsubsection{Linearized RG theory\label{SecLRGT}}

Following Wegner \cite{2835}, we write the ERGE under the following formal
form

\[
\frac{\text{d}S}{\text{d}t}={\cal G\,}S 
\]

Near a fixed point $S^{*}$ (such that ${\cal G\,}S^{*}=0$) we have:

\[
\frac{\text{d}\left( S^{*}+\Delta S\right) }{\text{d}t}={\cal L}\Delta S+%
{\cal Q}\Delta S 
\]

in which the RG operator has been separated into a linear ${\cal L}$ and a
quadratic ${\cal Q}$ parts.

The eigenvalue equation:

\[
{\cal L\,O}_{i}^{*}=\lambda _{i}{\cal \,O}_{i}^{*} 
\]

defines scaling exponents $\lambda _{i}$ and a set (assumed to be complete)
of eigenoperators ${\cal O}_{i}^{*}$. Hence we have for any $S(t)$:

\[
S(t)=S^{*}+\sum_{i}\mu _{i}(t){\cal \,O}_{i}^{*} 
\]

In which $\mu _{i}$ are the ``scaling fields'' \cite{413} which in the
linear approximation satisfy:

\[
\frac{\text{d}\mu _{i}(t)}{\text{d}t}=\lambda _{i}{\cal \,}\mu _{i}(t) 
\]

Which yields:

\begin{equation}
S(t)=S^{*}+\sum_{i}\mu _{i}(0){\cal \,}t^{\lambda _{i}}{\cal \,O}_{i}^{*}
\label{ScalField}
\end{equation}

\paragraph{Scaling operators}

There are three kinds of operators associated to well defined eigenvalues
and called ``scaling operators'' in \cite{413,2835} which are classified as
follows\footnote{%
In perturbative field theory, where the implicit fixed point is Gaussian,
the scaling fields are called superrenormalizable, nonrenormalizable and
strictly renormalizable respectively.}:

\begin{itemize}
\item  $\lambda _{i}>0$, the associated scaling field $\mu _{i}$ (or the
operator ${\cal O}_{i}^{*}$) is {\em relevant} because it brings the action
away from the fixed point.

\item  $\lambda _{i}<0$, the associated scaling field $\mu _{i}$ is {\em %
irrelevant} because it decays to zero when $t\rightarrow \infty $ and $S(t)$
finally approaches $S^{*}$.

\item  $\lambda _{i}=0,$ the associated scaling field $\mu _{i}$ is {\em %
marginal} and $S^{*}+\mu _{i}{\cal \,O}_{i}^{*}$ is a fixed point for any $%
\mu _{i}$. This latter property may be destroyed beyond the linear order%
\footnote{%
This is the case of the renormalized $\phi ^{4}$-coupling constant for $d=4$
with respect to the Gaussian fixed point: it is marginal in the linear
approximation and irrelevant beyond. It is marginally irrelevant.\label%
{margirr}}.
\end{itemize}

In critical phenomena, the relevant scaling fields alone are responsible for
the scaling form of the physical quantities: e.g., in its scaling form, the
free energy depends only on the scaling fields \cite{2835}. The irrelevant
scaling fields induce corrections to the scaling form \cite{2835}.

To be specific, the positive eigenvalue of a critical fixed point (once
unstable), say $\lambda _{1}$, is the inverse of the correlation length
critical exponent $\nu $ and the less negative eigenvalue is equal to
(minus) the subcritical exponent $\omega $.

In modern field theory, only the relevant (or marginally relevant) scaling
fields are of interest in the continuum limit \cite{440}: they correspond to
the renormalized couplings (or masses) of field theory the continuum limit
of which being defined ``at'' the considered fixed point (see section \ref
{SecContLimWil}).

\paragraph{Redundant operators and reparametrization invariance}

In addition to scaling operators, there are redundant operators \cite
{4011,2835}. They come out due to invariances of the RG \cite{4011,2835}
(see also \cite{4476}). Thus they can be expressed in the form ${\cal G}_{%
\text{tra}}\left\{ \Phi \right\} S^{*}$ and the associated exponents $%
\lambda _{i}$ (which, in general, have nonuniversal values) are spurious
since the free energy does not depend on the corresponding redundant fields $%
\mu _{i}$ (by construction of the transformation generator ${\cal G}_{\text{%
tra}}$ which leaves the partition function invariant, see section \ref
{SecTra}).

Although unphysical, the redundant fields cannot be neglected. For example,
a well known redundant operator is\footnote{%
A clear explanation of this may be found in \cite{4487} p.101--102.} $\frac{%
\delta S^{*}}{\delta \phi _{0}}$ which may be written under the form ${\cal G%
}_{\text{tra}}\left\{ \Phi \right\} S^{*}$ with $\Phi _{q}=\hat{\delta}(q)$.
and, most often, has the eigenvalue $\lambda =\frac{1}{2}\left( d-2+\eta
\right) $. Since $\lambda >0$ for $d=3$, this operator is relevant with
respect to the Wilson-Fisher \cite{439} (i.e. Ising-like for $d=3$) fixed
point although it is not physical. Indeed, as pointed out by Hubbard and
Schofield in \cite{3442}, the fixed point becomes unstable in presence of a $%
\int \phi ^{3}(x)$ term which, however, may be eliminated by the
substitution $\phi _{0}\rightarrow \phi _{0}+\mu $, which is controlled by
the operator $\frac{\delta S^{*}}{\delta \phi _{0}}$ since:

\[
S^{*}\left( \phi _{0}+\mu \right) =S^{*}\left( \phi _{0}\right) +\mu \frac{%
\delta S^{*}}{\delta \phi _{0}}+O\left( \mu ^{2}\right) 
\]

This redundant operator is not really annoying because it is sufficient to
consider actions that are even functional of $\phi $ ($Z_{2}$-symmetric) to
get rid of $\frac{\delta S^{*}}{\delta \phi _{0}}$.

Less obvious and more interesting for field theory is the following
redundant operator: 
\begin{equation}
{\cal O}_{1}=\int_{q}\left[ \frac{\delta ^{2}S}{\delta \phi _{q}\delta \phi
_{-q}}-\frac{\delta S}{\delta \phi _{q}}\frac{\delta S}{\delta \phi _{-q}}%
+\phi _{q}\frac{\delta S}{\delta \phi _{q}}\right]  \label{O1}
\end{equation}
which has been studied in detail by Riedel{\sl \ et al} \cite{4405}. When
the RG transforms the field variable linearly (as in the present review), $%
{\cal O}_{1}$ has, once and for all, the eigenvalue $\lambda _{1}=0$; it is
absolutely marginal \cite{4405}.

${\cal O}_{1}$ is redundant because it may be written under the form ${\cal G%
}_{\text{tra}}\left\{ \Phi \right\} S$ with:

\[
\Phi _{q}=\phi _{q}-\frac{\delta S}{\delta \phi _{-q}} 
\]

The redundant character of ${\cal O}_{1}$ is related to the invariance of
the RG transformation under a change of the overall normalization of $\phi $ 
\cite{4420,4421}. This invariance is also called the ``reparametrization
invariance'' \cite{3357}.

The most general realization of this symmetry \cite{4421} is not linear,
this explains why ${\cal O}_{1}$ is so complicated (otherwise, in the case
of a linear realization of the invariance, ${\cal O}_{1}$ would reduce to
simply $\int_{q}\phi _{q}\frac{\delta S}{\delta \phi _{q}}$).

As consequences of the reparametrization invariance \cite{4420,4421}:

\begin{itemize}
\item  a line of equivalent fixed points exists which is parametrized by the
normalization of the field,

\item  a field-rescaling parameter that enters in the ERGE must be properly
adjusted in order for the RG transformation to have a fixed point (the
exponent $\eta $ takes on a specific value).
\end{itemize}

We illustrate these two aspects with the Gaussian fixed point in section \ref
{SecLineFP} after having written down the ERGE.

\begin{itemize}
\item  Due to the complexity of ${\cal O}_{1}$, truncations of the ERGE (in
particular the derivative expansion, see section \ref{SecDeriv}) may easily
violate the reparametrization invariance \cite{4421}, in which case the line
of equivalent fixed points becomes a line of inequivalent fixed points
yielding different nonuniversal values for the exponent $\eta $. However the
search for a vestige of the invariance may be used to determine the best
approximation for $\eta $ \cite{4421,212,3836}. In the case where the
invariance is manifestly linearly realized and momentum independent (as when
a regularization with a sharp cutoff is utilized for example, see below),
then the derivative expansion may preserve the invariance and as a
consequence, $\eta $ is uniquely defined (see section \ref{SecDeriv}).
\end{itemize}

To understand why $\eta $ must take on a specific value, it is helpful to
think of a linear eigenvalue problem. ``{\sl The latter may have apparently
a solution for each arbitrary eigenvalue, but the fact that we can choose
the normalization of the eigenvector at will over-determines the system,
making that only a discrete set of eigenvalues are allowed}.'' \cite{3836}
(see section \ref{SecDeriv})

\subsection{Principles of derivation of the ERGE\label{steps}}

The Wilson RG procedure is carried out in two steps \cite{440} (see also 
\cite{301} for example):

\begin{enumerate}
\item  an integration of the fluctuations $\phi ({\bf p})$ over the range e$%
^{-t}<\left| p\right| \leq 1$ which leaves the partition function (\ref
{PartFn}) invariant,

\item  a change of the length scale by a factor e$^{-t}$ in all linear
dimensions to restore the original scale $\Lambda $ of the system, i. e. $%
{\bf p\rightarrow p}^{\prime }=$e$^{t}{\bf p}$
\end{enumerate}

For infinitesimal value of $t$, step 2 corresponds to a change in the
effective action [see eqs (\ref{dil}, \ref{ChangeDil} and \ref{DiffDil} with 
$\sigma =t$)]

\begin{equation}
S\rightarrow S^{\prime }=S+t{\cal G}_{\text{dil}}S
\end{equation}
inducing a contribution to 
\begin{equation}
\dot{S}\equiv \frac{\partial S}{\partial t}  \label{Spoint}
\end{equation}
which is equal to ${\cal G}_{\text{dil}}S$.

The step of reducing the number of degrees of freedom (step 1) is the main
part of the RG theory. It is sometimes called ``coarse grain decimation'' by
reference to a discrete realization of the RG transformation or Kadanoff's
transformation \cite{248}, it is also called sometimes ``blocking'' or
``coarsening''. It carries its own arbitrariness due to the vague notion of
``block'', i.e. in the present review the unprecised way of separating the
high from the low momentum frequencies. Step 1 may be roughly introduced as
follows.

We assume that the partition function may be symbolically written as 
\begin{equation}
{\cal Z}=\prod_{p\leq 1}\int {\cal D}\phi _{p}\exp \left\{ -S[\phi ]\right\}
\label{wea0}
\end{equation}

then after performing the integrations of step 1, we have:

\[
{\cal Z}=\prod_{p\leq e^{-t}}\int {\cal D}\phi _{p}\exp \left\{ -S^{\prime
}[\phi ]\right\} 
\]

with

\begin{equation}
\exp \left\{ -S^{\prime }[\phi ;t]\right\} =\prod_{e^{-t}<p\leq 1}\int {\cal %
D}\phi _{p}\exp \left\{ -S[\phi ]\right\}  \label{wea}
\end{equation}

$S^{\prime }[\phi ;t]$ is named the Wilson effective action. By considering
an infinitesimal value of $t$, one obtains an evolution equation for $S$
under a differential form, i.e. an explicit expression for $\dot{S}$.

As indicated by Wegner \cite{4011,2085}, the infinitesimal ``blocking''
transformation of $S$ (step 1) may sometimes\footnote{%
When the cutoff is smooth.} be expressed as a transformation of the field of
the form introduced in section \ref{SecTra}. Hence the general expression of
the ERGE may formally be written as follows \cite{4011,2085}:

\[
\dot{S}={\cal G}_{\text{dil}}S+{\cal G}_{\text{tra}}\left\{ {\bf \Psi }%
\right\} S 
\]
in which ${\bf \Psi }$ has different expressions depending on the way one
introduces the cutoff $\Lambda $. For example, in the case of the Wilson
ERGE [see eq. (\ref{WERGE}) below], ${\bf \Psi }$ has the following form 
\cite{2835}:

\[
\Psi _{p}=\left( c+2p^{2}\right) \left( \phi _{p}-\frac{\delta S}{\delta
\phi _{-p}}\right) 
\]

As it is introduced just above in (\ref{wea0}-\ref{wea}), the cutoff is said
sharp or hard\footnote{%
It corresponds to a well defined boundary between low and high momentum
frequencies, to be opposed to a smooth cutoff which corresponds to a blurred
boundary.}. It is known that a sharp boundary in momentum space introduces
non-local interactions in position space \cite{440} which one would like to
avoid. Nevertheless, a differential ERGE has been derived \cite{414} which
has been used several times with success under an approximate form. Indeed,
in the leading approximation of the derivative expansion (local potential
approximation), most of the differences between a sharp and smooth cutoff
disappear, and moreover, as stressed by Morris \cite{2520}, the difficulties
induced by the sharp cutoff may be circumvented by considering the Legendre
transformation (\ref{Legendre}) (see section \ref{LegendreSec}).

\subsection{Wegner-Houghton's sharp cutoff version of the ERGE\label{WHERGE}}

The equation has been derived in \cite{414}, one may also find an
interesting detailed presentation in \cite{3860}, it reads: 
\begin{equation}
\dot{S}=\lim_{t\rightarrow 0}\frac{1}{2t}\left[ \int_{p}^{\prime }\ln \left( 
\frac{\delta ^{2}S}{\delta \phi _{p}\delta \phi _{-p}}\right)
-\int_{p}^{\prime }\frac{\delta S}{\delta \phi _{p}}\frac{\delta S}{\delta
\phi _{-p}}\left( \frac{\delta ^{2}S}{\delta \phi _{p}\delta \phi _{-p}}%
\right) ^{-1}\right] +{\cal G}_{\text{dil}}S+\text{const}  \label{WegHou}
\end{equation}
in which we use (\ref{Spoint}) and the prime on the integral symbol
indicates that the momenta are restricted to the shell $e^{-t}<|p|\le 1$,
and ${\cal G}_{\text{dil}}S$ corresponds to any of the eqs (\ref{Gdil}--\ref
{Gdil3}) with $d_{\phi }(t)$ set to a constant in \cite{414}.

The explicit terms correspond to the step 1 (decimation or coarsening) of
section \ref{steps}, while ${\cal G}_{\text{dil}}S$ refers to step 2
(rescaling) as indicated in section \ref{rescal}. The additive constant may
be neglected in field theory [due to the normalization of (\ref{ZJ})].

\subsection{Smooth cutoff versions of the ERGE}

\subsubsection{Wilson's incomplete integration}

The first expression of the exact renormalization group equation under a
differential form has been presented as far back as 1970 \cite{4495} before
publication in the famous Wilson and Kogut review \cite{440} (see chapt.
11). The step 1 (decimation) of this version (referred to below as the
Wilson ERGE) consists in an ``incomplete'' integration in which large
momenta are more completely integrated than small momenta (see chapt. 11 of 
\cite{440} and also \cite{3856} p. 70 for the details).

The Wilson RG equation in our notations reads (with the change ${\cal H}%
\rightarrow -S$ compared to \cite{440}):

\begin{equation}
\dot{S}={\cal G}_{\text{dil}}S+\int_{p}\left( c+2p^{2}\right) \left( \frac{%
\delta ^{2}S}{\delta \phi _{p}\delta \phi _{-p}}-\frac{\delta S}{\delta \phi
_{p}}\frac{\delta S}{\delta \phi _{-p}}+\phi _{p}\frac{\delta S}{\delta \phi
_{p}}\right)  \label{WERGE}
\end{equation}

Some short comments relative to (\ref{WERGE}):

The term ${\cal G}_{\text{dil}}S$ (which comes out of the rescaling step 2)
is given by one of the eqs. (\ref{Gdil}-\ref{Gdil4}) but, in \cite{440}, the
choice $d_{\phi }=d/2$ has been made. The somewhat mysterious function $c$
(denoted $\frac{\text{d}\rho }{\text{d}t}$ in \cite{440}) must be adjusted
in such a way as to obtain a useful fixed point \cite{440,4420}. This
adjustment is related to the reparametrization invariance (see section \ref
{SecLineFP} for an example). Notice that $c$ is precisely introduced in (\ref
{WERGE}) in front of the operator ${\cal O}_{1}$ of eq. (\ref{O1}) which
controls the change of normalization of the field (see section \ref{SecLRGT}%
). Indeed, in the vicinity of the fixed point, we have \cite{212}: 
\begin{equation}
c=1-\frac{\eta }{2}  \label{cDeWilson}
\end{equation}
and most often $c$ is considered as a constant (as in \cite{4386,212} for
example). Notice that the unusual (for field theory) choice\footnote{%
See ref. \cite{4011,3856} for a further explanation of this choice.} $%
d_{\phi }=d/2$ in \cite{440}, leads to the same anomalous dimension (\ref
{dimfano}) at the fixed point.

\subsubsection{Polchinski's equation\label{PolEq}}

With a view to study field theory, Polchinski \cite{354} has derived his own
smooth cutoff version of the ERGE (see also section \ref{PolEff}). A general
ultraviolet (UV) cutoff function $K(p^{2}/\Lambda ^{2})$ is introduced (we
momentaneously restore the dimensions) with the property that it vanishes
rapidly when $p>\Lambda $. (Several kinds of explicit functions $K$ may be
chosen, the sharp cutoff would be introduced with the Heaviside step
function $K(x)=\Theta (1-x)$.) The Euclidean action reads:

\begin{equation}
{S[\phi ]\equiv }\frac{1}{2}\int_{p}\phi _{p}\phi
_{-p}p^{2}K^{-1}(p^{2}/\Lambda ^{2})+S_{\text{int}}[\phi ]  \label{PERGEint}
\end{equation}

Compared to \cite{354}, the ``mass'' term has been incorporated into $S_{%
\text{int}}[\phi ]$, this does not corrupt in any way the eventual analysis
of the massive theory because the RG naturally generates quadratic terms in $%
\phi $ and the massive or massless character of the (field) theory is not
defined at the level of eq. (\ref{PERGEint}) but a posteriori in the process
of defining the continuum limit (modern conception of the renormalization of
field theory, see sections \ref{wilcontSec} and \ref{SecTxtBrgt}).

Polchinski's ERGE is obtained from the requirement that the coarsening step
(step 1 of section \ref{steps}) leaves the generating functional $Z[J]$ [eq.
(\ref{ZJ})] invariant. The difficulty of dealing with an external source is
circumvented by imposing that $J(p)=0$ for $p>\Lambda $. As in \cite{440},
the derivation relies upon the writing of an (ad hoc) expression under the
form of a complete derivative with respect to the field in such a way as to
impose d$Z[J]/$d$\Lambda =0$. The original form of Polchinski's equation
accounts only for the step 1 and reads (for more details on this derivation,
see for example \cite{409,4442}):

\begin{equation}
\Lambda \frac{\text{d}S_{\text{int}}}{\text{d}\Lambda }=\frac{1}{2}%
\int_{p}p^{-2}\Lambda \frac{\text{d}K}{\text{d}\Lambda }\left( \frac{\delta
S_{\text{int}}}{\delta \phi _{p}}\frac{\delta S_{\text{int}}}{\delta \phi
_{-p}}-\frac{\delta ^{2}S_{\text{int}}}{\delta \phi _{p}\delta \phi _{-p}}%
\right)  \label{PERGE0}
\end{equation}
then, considering the rescaling (step 2) and the complete action, the
Polchinski ERGE is (see for example \cite{3491}):

\begin{equation}
\dot{S}={\cal G}_{\text{dil}}S-\int_{p}K^{\prime }(p^{2})\left( \frac{\delta
^{2}S}{\delta \phi _{p}\delta \phi _{-p}}-\frac{\delta S}{\delta \phi _{p}}%
\frac{\delta S}{\delta \phi _{-p}}+\frac{2p^{2}}{K(p^{2})}\phi _{p}\frac{%
\delta S}{\delta \phi _{p}}\right)  \label{PERGE}
\end{equation}
in which all quantities are dimensionless and $K^{\prime }(p^{2})$ stands
for d$K(p^{2})/$d$p^{2}$.

Let us mention that one easily arrives at eq. (\ref{PERGE0}) using the
observation that the two following functionals:

\begin{equation}
Z[J]=\int {\cal D}\phi \;\exp \left\{ -\frac{1}{2}\phi \cdot \Delta
^{-1}\cdot \phi -S[\phi ]+J\cdot \phi \right\}  \label{Trick1}
\end{equation}

and 
\begin{equation}
Z^{\prime }[J]=\int {\cal D}\phi \;\exp \left\{ -\frac{1}{2}\phi _{1}\cdot
\Delta _{1}^{-1}\cdot \phi _{1}-\frac{1}{2}\phi _{2}\cdot \Delta
_{2}^{-1}\cdot \phi _{2}-S[\phi _{1}+\phi _{2}]+J\cdot \left( \phi _{1}+\phi
_{2}\right) \right\}  \label{Trick2}
\end{equation}
are equivalent (up to a multiplicative factor) provided that $\Delta =\Delta
_{1}+\Delta _{2}$ and $\phi =\phi _{1}+\phi _{2}$ (see appendix\footnote{%
Or appendix 24 of the first edition in 1989.} 10 of \cite{3533} and also 
\cite{3359}).

\subsubsection{Redundant free ERGE}

In \cite{3359} it is proposed to make the effective dimension of the field $%
d_{\phi }$ [defined in (\ref{dimfeff})], which enters Polchinski's ERGE (\ref
{PERGE}) via the rescaling part ${\cal G}_{\text{dil}}S$ [see eqs (\ref{Gdil}%
--\ref{Gdil4})], depend on the momentum ${\bf p}$ in such a way as to keep
unchanged, along the RG flows, the initial quadratic part ${S}_{0}{[\phi
]\equiv }\frac{1}{2}\int_{p}\phi _{p}\phi _{-p}p^{2}K^{-1}(p^{2})$. This
additional condition imposed on the ERGE would completely eliminate the
ambiguities in the definition (the invariances) of the RG transformation
leaving no room for any redundant operators (see also chapt 5 of \cite{3856}%
).

If we understand correctly the procedure, it is similar to (but perhaps more
general than) that proposed in \cite{4405} to eliminate the redundant
operator (\ref{O1}). One fixes the arbitrariness associated to the
invariance (reparametrization invariance of section \ref{SecLRGT}) in order
that any RG flow remains orthogonal to the redundant direction(s).

The authors of \cite{3359} do not specify what happens in the case where the
unavoidable truncation (or approximation) used to study the ERGE breaks some
invariances\footnote{%
In particular it is known that the derivative expansion developped within
ERGE with smooth cutoffs breaks the reparametrization invariance.}. In fact,
as already mentioned at the end of section \ref{SecLRGT}, the freedom
associated with the redundant directions allows us to search for the region
of minimal break of invariance in the space of interactions \cite
{4421,212,3836}. If this freedom is suppressed one may not obtain, for
universal quantities such as the critical exponents, the optimal values
compatible with the approximation (or truncation) used.

\subsection{ERGE for the Legendre Effective Action\label{LegendreSec}}

The first derivation\footnote{%
The first mention to the Legendre transformation in the expression of the
ERGE has been formulated by E.\ Br\'{e}zin in a discussion following a talk
by Halperin \cite{4390}.} of an ERGE for the Legendre effective action $%
\Gamma \left[ \Phi \right] $ [defined in eq. (\ref{Legendre})] has been
carried out with a sharp cutoff by Nicoll and Chang \cite{4281} (see also 
\cite{4255}). Their aim was to simplify the obtention of the $\varepsilon $%
-expansion from the ERGE. More recent obtentions of this equation with a
smooth cutoff are due to Bonini, D'Attanasio and Marchesini \cite{4293},
Wetterich \cite{4374}, Morris \cite{2520} and Ellwanger \cite{4436}.

A striking fact arises with the Legendre transformation: the running cutoff $%
\Lambda $ (or intermediate-scale momentum cutoff, i.e. associated to the
``time'' $t$) acts as an IR cutoff and physical Green functions are obtained
in the limit $\Lambda \rightarrow 0$ \cite{4387,251,252,4374,2520,4293,4436}%
. The reason behind this result is simple to understand. The generating
functional $\Gamma \left[ \Phi \right] $ is obtained by integrating out {\it %
all} modes (from $\Lambda _{0}=\infty $ to $0$). If an intermediate cutoff $%
\Lambda \ll \Lambda _{0}$ is introduced and integration is performed only in
the range $\left[ \Lambda ,\Lambda _{0}\right] $, then for the integrated
modes (thus for the effective $\Gamma _{\Lambda }\left[ \Phi \right] $), $%
\Lambda $ is an IR cutoff while for the unintegrated modes (for the
effective $S_{\Lambda }\left[ \phi \right] $), $\Lambda $ is an UV\ cutoff.
There is an apparent second consequence: contrary to the ERGE for $%
S_{\Lambda }\left[ \phi \right] $, the ERGE for $\Gamma _{\Lambda }\left[
\Phi \right] $ will depend on both an IR ($\Lambda $) and an UV ($\Lambda
_{0}$) cutoffs\footnote{%
One could think a priori that as a simple differential equation, the ERGE
would be instantaneous (would depend only on $\Lambda $ and not on some
initial scale $\Lambda _{0}$), but it is an integro-differential equation.}.
However it is possible to send $\Lambda _{0}$ to infinity, see section \ref
{SecLegSCV} for a short discussion of this point.

\subsubsection{Sharp cutoff version}

In \cite{4281} an ERGE for the Legendre transformation $\Gamma \left[ \Phi
\right] $ [eq. \ref{Legendre}] is derived (See also \cite{4255}). It reads:

\begin{equation}
\dot{\Gamma}={\cal G}_{\text{dil}}\Gamma +\frac{1}{2}\dint \frac{d\Omega }{%
\left( 2\pi \right) ^{d}}\ln \left\{ \Gamma
_{q,-q}-\dint_{p}\dint_{p^{\prime }}\Gamma _{qp}\left( \Gamma ^{-1}\right)
_{pp^{^{\prime }}}\Gamma _{p^{^{\prime }},-q}\right\}  \label{ChangNic}
\end{equation}

in which: 
\[
\Gamma _{kk^{\prime }}=\frac{\delta ^{2}\Gamma }{\delta \Phi _{k}\delta \Phi
_{k^{\prime }}} 
\]
the momentum $q$ lies on the shell $q=|{\bf q}|=1$ while the integrations on 
$p$ and $p^{\prime }$ are performed inside the shell $]1,\Lambda
_{0}/\Lambda ]$ where $\Lambda _{0}$ is some initial cutoff ($\Lambda
_{0}>\Lambda $) and $\Omega $ is the surface of the $d$-dimensional unit
sphere [$\Omega =\left( 2\pi \right) ^{d}K_{d}$].

One sees that $\Lambda $ is like an IR cutoff and that (\ref{ChangNic})
depends on the initial cutoff $\Lambda _{0}$.

\subsubsection{Smooth cutoff version\label{SecLegSCV}}

We adopt notations which are close to the writing of (\ref{PERGEint}) and we
consider the Wilson effective action with an ``additive'' \cite{4374} IR
cutoff $\Lambda $ such that:

\begin{equation}
{S}_{\Lambda }{[\phi ]\equiv }\frac{1}{2}\int_{p}\phi _{p}\phi
_{-p}C^{-1}(p,\Lambda )+S_{\Lambda _{0}}[\phi ]  \label{SlambdaIR}
\end{equation}
in which $C(p,\Lambda )$ is an additive infrared cutoff function which is
small for $p<\Lambda $ (tending to zero as $p\rightarrow 0$) and $%
p^{2}C(p,\Lambda )$ should be large for $p>\Lambda $ \cite{3357}. Due to the
additive character of the cutoff function, $S_{\Lambda _{0}}[\phi ]$ is the
entire action (involving the kinetic term contrary to eq. (\ref{PERGEint})
and to \cite{2520} where the cutoff function was chosen multiplicative). In
this section, because $C$ is naturally dimensioned\footnote{%
It is not harmless that $C$ is dimensionful because the anomalous dimension $%
\eta $ may be a part of its dimension (see \cite{3357} and section \ref
{SecLTV})} [contrary to $K$ in (\ref{PERGEint})], all the dimensions are
implicitly restored in order to keep the same writing as in the original
papers. The ultra-violet regularization is provided by $\Lambda _{0}$ and
needs not to be introduced explicitly (see \cite{2520} and below). The
Legendre transformation is defined as: 
\[
\Gamma [\Phi ]+\frac{1}{2}\int_{p}\Phi p\Phi _{-p}C^{-1}(p,\Lambda
)=-W[J]+J.\Phi 
\]
in which $W[J]$ and $\Phi $ are defined as usual [see eq (\ref{Legendre})]
from (\ref{SlambdaIR}).

Then the ERGE reads:

\begin{equation}
\dot{\Gamma}={\cal G}_{\text{dil}}\Gamma +{\frac{1}{2}}\int_{p}{\frac{1}{C}}%
\Lambda {\frac{\partial C}{\partial \Lambda }}\left( 1+C{\,}\Gamma
_{p,-p}\right) ^{-1}  \label{Morris1}
\end{equation}

When the field is no longer a pure scalar but carries some supplementary
internal degrees of freedom and becomes a vector, a spinor or a gauge field
etc..., a more compact expression using the trace of operators is often
used: 
\begin{equation}
\dot{\Gamma}={\cal G}_{\text{dil}}\Gamma +{\frac{1}{2}}\text{tr}\left[ {%
\frac{1}{C}}\Lambda {\frac{\partial C}{\partial \Lambda }}\cdot \left(
1+C\cdot \frac{\delta ^{2}\Gamma }{\delta \Phi \delta \Phi }\right)
^{-1}\right]  \label{Morris2}
\end{equation}
which, for example, allows us to include the generalization to $N$
components in a unique writing ($\frac{\delta ^{2}\Gamma }{\delta \Phi
\delta \Phi }$ has then two supplementary indices $\alpha $ and $\beta $
corresponding to the derivatives with respect to the fields $\Phi _{\alpha }$
and $\Phi _{\beta }$, the trace is relative to both the momenta and the
indices).

The equations (\ref{Morris1},\ref{Morris2}) may be obtained, as in \cite
{2520}, using the trick of eqs.(\ref{Trick1} and \ref{Trick2}), but see also 
\cite{4374,4293}.

For practical computations it is actually often quite convenient (for
example, see section \ref{Consist}) to write the flow equation~(\ref{Morris1}%
) as follows \cite{4315}: 
\begin{equation}
\dot{\Gamma}={\cal G}_{\text{dil}}\Gamma +\frac{1}{2}\text{tr}\tilde{\partial%
}_{t}\ln \left( C^{-1}+\frac{\delta ^{2}\Gamma }{\delta \Phi \delta \Phi }%
\right)  \label{JungWet}
\end{equation}
with $\tilde{\partial}_{t}\equiv -\Lambda \frac{\partial }{\partial \Lambda }
$ acting only on $C$ and not on $\Gamma $, i.e. $\tilde{\partial}_{t}=\left(
\partial C^{-1}/\partial t\right) \left( \partial /\partial C^{-1}\right) $.

Wetterich's expression of the ERGE \cite{4374} is identical\footnote{%
The correspondence between Wetterich's notations and ours is as follows: $%
\Lambda \rightarrow k$; $C\rightarrow 1/R_{k}$; $t\rightarrow -t$.} to eqs. (%
\ref{Morris1}, \ref{Morris2}, \ref{JungWet}). Its originality is in the
choice of the cutoff function $C^{-1}$: to make the momentum integration in (%
\ref{Morris1}) converge, a cutoff function is introduced such that only a
small momentum interval $p^{2}\approx \Lambda ^{2}$ effectively contributes 
\cite{4374} (see also the review in \cite{4315}). This feature, which avoids
an explicit UV regularization, allows calculations in models where the UV
regularization is a delicate matter (e.g. non-Abelian gauge theories). In
fact, as noticed in \cite{2520}, the ERGE only requires momenta $p\approx
\Lambda $ and should not depend on $\Lambda _{0}\gg \Lambda $ at all.
Indeed, once a finite ERGE is obtained, the flow equation for $\Gamma
_{\Lambda }\left[ \Phi \right] $ is finite and provides us with an
``ERGE''-regularization scheme which is specified by the flow equation, the
choice of the infrared cutoff function $C$ and the ``initial condition'' $%
\Gamma _{\Lambda }$ \cite{4315}. Most often, there is no need for any UV
regularization and the limit $\Lambda _{0}\rightarrow \infty $ may be taken
safely. In this case, the cutoff function chosen by Wetterich \cite{4374}
has the following form (up to some factor):

\begin{eqnarray}
C(p,\Lambda ) &=&\frac{1-f(\frac{p^{2}}{\Lambda ^{2}})}{p^{2}f(\frac{p^{2}}{%
\Lambda ^{2}})}  \label{CutWett1} \\
f(x) &=&\text{e}^{-2ax^{b}}  \label{CutWett2}
\end{eqnarray}
in which the two parameters $a$ and $b$ may be adjusted to vary the
smoothness of the cutoff function.

Although it is almost an anticipation on the expansions (local potential
approximation and derivative expansion) considered in the parts to come, we
find it worthwhile indicating here the exact equation satisfied by the
effective potential which is often used by Wetterich and co-workers (see
their review \cite{4700} in this volume). Thus following Wetterich \cite
{4374}, we write the effective (Legendre) action $\Gamma \left[ \Phi \right] 
$ for $O(N)$-symmetric systems as follows\footnote{%
It is customary to introduce the variable $\rho $ for $O(N)$ systems because
this allows better convergences in some cases (see section \ref{SecTrunc})
but in the particular case $N=1$ the symmetry assumption is not required.}: 
\begin{eqnarray*}
\Gamma \left[ \Phi \right] &=&\int \text{d}^{d}x\left[ U(\rho )+\frac{1}{2}%
\partial ^{\mu }\Phi _{\alpha }{\cal Z}\left( \rho ,-{\square }\right)
\partial _{\mu }\Phi _{\alpha }+\frac{1}{4}\partial ^{\mu }\rho {\cal Y}%
\left( \rho ,-{\square }\right) \partial _{\mu }\rho \right] \\
\rho &=&\frac{1}{2}\Phi ^{2}
\end{eqnarray*}
where the symbol ${\square }$ stands for $\partial _{\mu }\partial ^{\mu }$
and acts only on the right (summation over repeated indices is assumed).
Then the exact evolution equation for the effective potential $U(\rho )$ 
\cite{4374} follows straightforwardly from (\ref{JungWet}): 
\begin{eqnarray}
\dot{U} &=&\frac{1}{2}\int_{p}\frac{1}{C^{2}}\Lambda \frac{\partial C}{%
\partial \Lambda }\left[ \frac{N-1}{M_{0}}+\frac{1}{M_{1}}\right]
+dU-d_{\phi }U^{\prime }  \label{PotWett1} \\
M_{0} &=&{\cal Z}\left( \rho ,p^{2}\right) p^{2}+C^{-1}+U^{\prime }
\label{PotWett2} \\
M_{1} &=&\left[ {\cal Z}\left( \rho ,p^{2}\right) +\rho {\cal Y}\left( \rho
,p^{2}\right) \right] p^{2}+C^{-1}+U^{\prime }+2\rho U^{\prime \prime }
\label{PotWett3}
\end{eqnarray}
in which $\dot{U}$ stands for $\partial U(\rho ,t)/\partial t$ and $%
U^{\prime }$ and $U^{\prime \prime }$ refer to the first and second
(respectively) derivatives with respect to $\rho $. The two last terms in (%
\ref{PotWett1}) come from ${\cal G}_{\text{dil}}\Gamma $ which was not
explicitly considered in \cite{4374}.

The interest of dealing with an additive cutoff function is that one may
easily look for the classes of $C$ that allow a {\em linear} realization of
the reparametrization invariance \cite{3357}. It is found that \cite{3357} $%
C $ must be chosen as: 
\[
C(p,\Lambda )=\Lambda ^{-2}\left( \frac{p^{2}}{\Lambda ^{2}}\right) ^{k} 
\]
with $k$ an integer such that $k>d/2-1$ to have UV convergence. With this
choice, the derivative expansion preserves the reparametrization invariance
and $\eta $ is uniquely defined \cite{3357} (see part \ref{SecReparamLinReal}%
). On the contrary, because the cutoff function corresponding to eqs. (\ref
{CutWett1}, \ref{CutWett2}) has an exponential form, the derivative
expansion does not provide us with a uniquely defined value of $\eta $ (see
section \ref{SecOtherStud}).

\paragraph{Sharp cutoff limit}

It is possible to obtain the sharp cutoff limit from eqs.(\ref{Morris1}, \ref
{Morris2}) provided one is cautious in dealing with the value at the origin
of the Heaviside function $\theta (0)$ (which is {\em not} equal to $\frac{1%
}{2}$) \cite{2520,3550}. One obtains the sharp cutoff limit of the flow
equation \cite{3550}: 
\begin{equation}
\dot{\Gamma}={\cal G}_{\text{dil}}\Gamma +\frac{1}{2}\int_{p}\frac{\delta
(p-1)}{{\gamma (}p{)}}\left[ \hat{\Gamma}\cdot \left( 1+G\cdot \hat{\Gamma}%
\right) ^{-1}\right] \left( {\bf p},-{\bf p}\right)  \label{SmoothToSharp}
\end{equation}
in which the field independent full inverse propagator $\gamma (p)$ has been
separated from the two-point function : 
\[
{{\frac{\delta ^{2}\Gamma [\Phi ]}{\delta \Phi \delta \Phi }}}\left( {\bf p},%
{\bf p}^{\prime }\right) ={\gamma (}p{)}\hat{\delta}({\bf p}+{\bf p}^{\prime
})+\hat{\Gamma}[\Phi ]\left( {\bf p},{\bf p}^{\prime }\right) 
\]
so that ${\hat{\Gamma}}[0]=0$, and $G(p)=\theta (p-1)/{\gamma (}p{)}$.

\subsection{Equivalent fixed points and reparametrization invariance.\label%
{SecLineFP}}

To illustrate the line of equivalent fixed points which arises when the
reparametrization invariance is satisfied (see the end of section \ref
{SecLRGT}), we consider here the pure Gaussian case of the Wilson ERGE (see
also the appendix of \cite{440} and \cite{3836}). No truncation is needed to
study the Gaussian case, the analysis below is thus exact.

The action is assumed to have the following pure quadratic form:

\[
S_{G}[\phi ]=\frac{1}{2}\int_{p}\phi _{p}\phi _{-p}R\left( p^{2}\right) 
\]

The effect of ${\cal G}_{\text{dil}}$ in (\ref{WERGE}) yields (the prime
denotes the derivative with respect to $p^{2}$):

\[
{\cal G}_{\text{dil}}S_{G}=\left( \frac{d}{2}-d_{\phi }\right) \int_{p}\phi
_{p}\phi _{-p}R-\int_{p}\phi _{p}\phi _{-p}p^{2}R^{\prime } 
\]
while the remaining part (coarsening) gives (up to neglected constant terms):

\[
{\cal G}_{\text{tra}}S_{G}=\int_{p}\phi _{p}\phi _{-p}\left[ \left(
c+2p^{2}\right) \left( R-R^{2}\right) -p^{2}R^{\prime }\right] 
\]

Adding the two contributions to $\dot{S}_{G}$, choosing $d_{\phi }=\frac{d}{2%
}$ as in \cite{440} and imposing that the fixed point is reached ($\dot{S}%
_{G}=0$), one obtains:

\begin{eqnarray*}
c &=&1 \\
R^{*}(p^{2}) &=&\frac{zp^{2}}{\text{e}^{-2p^{2}}+zp^{2}}
\end{eqnarray*}

This is a line of (Gaussian) fixed points parametrized by $z$. To reach this
line, the parameter $c$ must be adjusted to $1$ [i.e., following eq. (\ref
{cDeWilson}), $\eta =0$], the fixed points on the line are equivalent.

The same analysis may be done with the same kind of conclusions with the
Polchinski ERGE (see \cite{3836}). For the Wegner-Houghton version (\ref
{WegHou}), the situation is very different in nature \cite{3836}. The same
kind of considerations yields: 
\[
R^{*}(p^{2})=\,p^{2-\eta } 
\]
in which $\eta $ is undetermined. ``{\sl This phenomenon is of quite
different nature as the similar one described above. There the whole line
shares the same critical properties, here it does not; there we have
well-behaved actions throughout the fixed line, here nearly all of them are
terribly non-local (in the sense that we cannot expand the action integrand
in a power series of }$p^{2}${\sl ). What happens is that we have one
physical FP (the }$\eta =0${\sl \ case) and a line of spurious ones.}'' \cite
{3836}

\subsection{Approximations and truncations}

How to deal with integro-differential equations is not known in general. In
the case of the RG equations, one often had recourse to perturbative
expansions such as the usual perturbation in powers of the coupling but also
the famous $\varepsilon $--expansion (where $\varepsilon =4-d$), the $1/N$%
--expansion or expansions in the dimensionality $2-d$, let us mention also
an expansion exploiting the smallness of the critical exponent\footnote{%
The expansion requires also a truncation in powers of the field and some
re-expansion of $\varepsilon /4=1-d/4$ in powers of $\eta ^{1/2}$ \cite{4506}%
.} $\eta $ \cite{3348}.

When no small parameter can be identified or when one does not want to
consider a perturbative approach, one must truncate the number of degrees of
freedom involved in order to reduce the infinite system of coupled
differential equations to a finite system. The way truncations are
introduced is of utmost importance as one may learn from the development of
the scaling field method (see \cite{4405}). The ERGE is a useful starting
point to develop approximate approach. If from the beginning it has been
exclusively seen as a useful tool for the investigation of the $\varepsilon $%
-expansion \cite{4406}, two complementary approaches to nonperturbative
truncations have been proposed:

\begin{itemize}
\item  an expansion of the effective action in powers of the derivatives of
the field \cite{4468,212}: 
\begin{equation}
S[\phi ]=\int \text{d}^{d}x\left\{ V(\phi )+\frac{1}{2}Z(\phi )\left(
\partial _{\mu }\phi \right) ^{2}+O(\partial ^{4})\right\} 
\label{derivative}
\end{equation}
which is explicitly considered in this review (see sections \ref{PartLPA}
and \ref{SecDeriv})

\item  an expansion in powers of the field for the functional $\Gamma [\Phi ]
$ \cite{2836} (see also \cite{4427,4004}): 
\begin{equation}
\Gamma [\Phi ]=\sum_{n=0}^{\infty }\frac{1}{n!}\int \left( \prod_{k=0}^{n}%
\text{d}^{d}x_{k}\,\Phi (x_{k})\right) \Gamma ^{(n)}(x_{1},\ldots ,x_{n})
\label{weinberg}
\end{equation}
\end{itemize}

The flow equations for the $1$PI $n$--point functions $\Gamma ^{(n)}$ are
obtained by functional differentiation of the ERGE. The distinction between $%
S$ in (\ref{derivative}) and $\Gamma $ in (\ref{weinberg}) is not essential,
we can introduce the two approximations for both $S$ and $\Gamma $.

Because the derivative expansion corresponds to small values of $p$, it is
naturally (quantitatively) adapted to the study of the large distance or low
energy physics like critical phenomena or phase transition. We will see,
however, that at a qualitative level it is suitable to a general discussion
of many aspects of field theory (like the continuum limit, see part \ref
{PartLPA}). Obviously, when bound state formation or nonperturbative
momentum dependences are studied, the expansion (\ref{weinberg}) seems
better adapted (see, for examples, \cite{4504,4502}).

Only a few terms of such series will be calculable in practice, since the
number of invariants increases rapidly for higher orders (see section \ref
{SecDeriv}).

\subsection{Scaling field representation of the ERGE}

To date the most expanded approximate method for the solution of the ERGE
has been developed by Golner and Riedel in \cite{4404} (see also \cite{319}
and especially \cite{4405}) from the scaling-field representation. The idea
is to introduce the expansion (\ref{ScalField}) into the ERGE which is thus
transformed into an infinite hierarchy of nonlinear ordinary differential
equations for scaling fields. For evident reasons\footnote{%
The eigenoperators ${\cal O}_{i}^{*}$ and eigenvalues $\lambda _{i}$ of the
Gaussian fixed point can be determined exactly.}, the fixed point chosen for 
$S^{*}$ is the Gaussian fixed point. Approximate solutions may be obtained
by using truncations and iterations. The approximations appear to be
effective also in calculations of properties, like phase diagrams, scaling
functions, crossover phenomena and so on (see \cite{4405}).

Because they are unjustly not often mentioned in the literature, we find it
fair to extract from \cite{319} the following estimates for $N=1$ (see the
paper for estimates corresponding to other values of $N$):

\begin{eqnarray*}
\nu &=&0.626\pm 0.009 \\
\eta &=&0.040\pm 0.007 \\
\omega &=&0.855\pm 0.07 \\
\omega _{2} &=&1.67\pm 0.11 \\
\omega _{5} &=&2.4\pm 0.4
\end{eqnarray*}
in which $\omega _{2}$ is the second correction-to-scaling exponent and the
subscript ``5'' in $\omega _{5}$ refers to a $\phi ^{5}$ interaction present
in the action and which would be responsible for correction-to-scaling terms
specific to the critical behavior of fluids, as opposed to the Ising model
which satisfies the symmetry $Z_{2}$, see \cite{3124} and also section \ref
{CE-LPA}.

\subsection{Renormalizability, continuum limits and the Wilson theory\label%
{SecContLimWil}}

\subsubsection{Wilson's continuum limit\label{wilcontSec}}

In the section 12.2 of \cite{440}, a nonperturbative realization of the
renormalization of field theory is schematically presented. The illustration
is done with the example of a fixed point with one relevant direction (i.e.
a fixed point which controls the criticality of magnetic systems in zero
magnetic field). The resulting renormalized field theory is purely massive
(involving only one parameter: a mass).

We find satisfactory\footnote{%
Except the expression ``{\sl critical manifold, which consists of all bare
actions yielding a given massless continuum limit}'', see section \ref
{SecTxtBrgt}.} the presentation of this continuum limit by Morris in \cite
{3993} (relative to the discussion of his fig. 3, see fig. \ref{fig3} of the
present paper for an illustration with actual RG\ trajectories), and we
reproduce it here just as it is.

``{\sl In the infinite dimensional space of bare actions, there is the
so-called critical manifold, which consists of all bare actions yielding a
given massless continuum limit. Any point on this manifold -- i.e. any such
bare action -- flows under a given RG towards its fixed point; local to the
fixed point, the critical manifold is spanned by the infinite set of
irrelevant operators. The other directions emanating out of the critical
manifold at the fixed point, are spanned by relevant and marginally relevant
perturbations (with RG eigenvalues }$\lambda _{i}>0${\sl \ and }$\lambda
_{i}=0${\sl , respectively).} [In the example of \cite{440} and in fig. \ref
{fig3}, there is only one relevant perturbation.]{\sl \ Choosing an
appropriate parametrization of the bare action, we move a little bit away
from the critical manifold. The trajectory of the RG will to begin with,
move towards the fixed point, but then shoot away along one of the relevant
directions towards the so-called high temperature fixed point which
represents an infinitely massive quantum field theory.}

{\sl To obtain the continuum limit, and thus finite masses, one must now
tune the bare action back towards the critical manifold and at the same
time, reexpress physical quantities in renormalised terms appropriate for
the diverging correlation length. In the limit that the bare action touches
the critical manifold, the RG trajectory splits into two: a part that goes
right into the fixed point, and a second part that emanates out from the
fixed point along the relevant directions. This path is known as a
Renormalised Trajectory }\cite{440}{\sl \ (RT). The effective actions on
this path are `perfect actions' }\cite{4283}{\sl .}''

The continuum limit so defined ``at'' a critical fixed point has been used
by Wilson to show that the $\phi ^{6}$-field-theory in three dimensions has
a nontrivial continuum limit involving no coupling constant renormalization 
\cite{432} (i.e. the continuum limit involves only a mass as renormalized
parameter, but see also \cite{4498}). The fixed point utilized in the
circumstances is the Wilson-Fisher (critical) fixed point. In fact, exactly
the same limit would have been obtained starting with a $\phi ^{4}$- or a $%
\phi ^{8}$-bare-theory, since it is the symmetry of the bare action which is
important and not the specific form chosen for the initial (bare)
interaction ($\phi ^{4}$, $\phi ^{6}$, or $\phi ^{8}$ are all elements of
the same $Z_{2}$-symmetric scalar theory, see section \ref{SecTxtBrgt} for
more details).

It is noteworthy that one (mainly) exclusively presents the Wilson continuum
limit as it is illustrated in \cite{440}, i.e. relatively to a critical
point which possesses only one relevant direction. Obviously one may choose
any fixed point with several relevant directions (as suggested in the Morris
presentation reproduced above). The relevant parameters provide the
renormalized couplings of the continuum limit. For example the Gaussian
fixed point in three dimensions for the scalar theory (a tricritical fixed
point with two relevant directions) yields a continuum limit which involves
two renormalized parameters (see an interesting discussion with Br\'{e}zin
following a talk given by Wilson \cite{4196}). Indeed, that continuum limit
is nothing but the so-called $\phi ^{4}$-field theory used successfully in
the investigation, by perturbative means, of the critical properties of
statistical systems below four dimensions and which is better known as the ``%
{\em Field theoretical approach to critical phenomena}'' \cite{2706} (see 
\cite{3533} for a review). The scalar field theory below four dimensions
defined ``at'' the Gaussian fixed point involves a mass and a (renormalized) 
$\phi ^{4}$-coupling ``constant''.

``At'' the Gaussian fixed one may also define a massless renormalized
theory. To reach this massless theory, one must inhibit the direction of
instability\footnote{%
The relevant direction that points towards the most stable high-temperature
fixed point.} of the Gaussian fixed point toward the massive sector. As a
consequence, the useful space of bare interactions is limited to the
critical manifold alluded to above by Morris (there is one parameter to be
adjusted in the bare action). The discussion is as previously but with one
dimension less for the space of bare interactions: the original whole space
is replaced by the critical submanifold and this latter by the tricritical
submanifold. Notice that the massless continuum limit so defined really
involves a scale dependent parameter: the remaining relevant direction of
the Gaussian fixed point which corresponds to the $\phi ^{4}$-renormalized
coupling (see section \ref{SecTxtBrgt} for more details). It differs however
from the massless theories sketchily defined by Morris as {\em fixed point
theories }\cite{3550,3816,3828,3864,3993,3661}. Most certainly no mass can
be defined right at a fixed point \cite{3864,3993} (there is scale
invariance there and a mass would set a scale) but at the same time the
theory would also have no useful parameter at all (no scale dependent
parameter) since, by definition, right at a fixed point nothing changes,
nothing happens, there is nothing to describe.

An important aspect of the Wilson continuum limit is the resulting
self-similarity emphasized rightly in several occasions by Morris and
collaborators \cite{3816,3864,3993,3661,3817} (see section \ref{SecTxtBrgt}%
). This notion expresses the fact that in a properly defined continuum
limit, the effects of the infinite number of degrees of freedom involved in
a field theory are completely represented by a (very) small number of
flowing (scale dependent) parameters (the relevant parameters of a fixed
point): the system is self-similar in the sense that it is exactly
(completely) described by the same finite set of parameters seen at
different scales (see section \ref{SecTxtBrgt}). This is exactly what one
usually means by renormalizability in perturbative field theory. However the
question is nonperturbative in essence. For example, the $\phi ^{4}$-field
theory in four dimensions is perturbatively renormalizable, but it is not
self-similar at any scale and especially in the short distance regime due to
the UV ``renormalon'' problems \cite{789,279} (for a review see \cite{4287})
which prevent the perturbatively renormalized coupling constant to carry
exactly all the effects of the other (an infinite number) degrees of
freedom: it is not a relevant parameter for a fixed point (see section \ref
{PolEff}, \ref{SecTxtBrgt} and \cite{3554,3993}).

\subsubsection{Polchinski's effective field theories\label{PolEff}}

There has been a renewed interest of field theoreticians in the ERGE since
Polchinski's paper \cite{354} in 1984. From the properties of the RG flows
generated by an ERGE (see section \ref{PolEq}) and by using only ``{\sl very
simple bounds on momentum integrals}'', Polchinski presented an easy proof
of the perturbative renormalizability of the $\phi ^{4}$ field theory in
four dimensions (see also \cite{4316,4293}). This paper had a considerable
success. One may understand the reasons of the resulting incipient interest
of field theoreticians in the ERGE, let us cite for example:

``{\sl Proofs of renormalizability to all orders in perturbation theory were
notoriously long and complicated (involving notions of graph topologies,
skeleton expansions, overlapping divergences, the forest theorem, Weinberg's
theorem, etc.)}, ...'' \cite{4442}.

The enthusiasm of some was so great that one has sometimes referred to
Polchinski's presentation, really based on the Wilson RG theory, as the
Wilson-Polchinski theory. However the strategy relative to the construction
of the continuum limit (modern expression for ``renormalizability'') is
rather opposite to the ideas of Wilson because {\em they are perturbative in
essence}. Indeed, while the reference to a fixed point is {\em essential} in
the Wilson construction of the continuum limit, Polchinski does not need any
explicit reference to a fixed point, but in fact refers {\em implicitly} to
the Gaussian fixed point\footnote{%
As in perturbation theory.}. A classification of parameters as relevant,
irrelevant and marginal is given using a purely classical dimensional
analysis (referring to the Gaussian fixed point, see \cite{3073} for
example). In the Polchinski view, the marginal parameters are then
considered as being relevant although in some cases (as the scalar case, see
footnote \ref{margirr}) they may actually be (marginally) irrelevant.

The arguments may then lead to confusions. The notion of relevant parameter
(the natural candidate for the renormalized parameter), which, in the Wilson
theory represents an unstable direction of a fixed point (one goes away from
the fixed point which thus in the occasion displays an ultraviolet stability
or attractivity) has been replaced in the Polchinski point of view (for the $%
\phi _{4}^{4}$ field) by the least irrelevant parameter which controls the
final approach to a fixed point (one goes toward the fixed point which thus
presents an infrared stability or attractivity). Thus the renormalized
coupling resulting from the ``proof'' controls only the infrared (large
distances or low energy) regime of the scalar theory. The field theory so
constructed is actually an ``effective'' field theory \cite{3073,2731}
(valid in the infrared regime) and not a field theory well defined in the
short distance regime (e.g. even after the ``proof'' the $\phi ^{4}$ field
theory in four dimensions remains trivial due to the lack of ultraviolet
stable fixed point).

Of course, if by chance the Gaussian fixed point is ultraviolet stable
(asymptotically free field theories), then the marginal coupling is truly a
relevant parameter for the Gaussian fixed point and the perturbatively
constructed field theory exists beyond perturbation (in the Wilson sense of
section \ref{wilcontSec}). In that case, one may use the Polchinski approach
to prove the existence of a continuum limit (see some references in \cite
{3073}).

It is fair to specify that, in several occasions in \cite{354}, Polchinski
has emphasized the perturbative character of his proof which is {\em only}
equivalent to (but simpler than) the usual perturbative proof. In order to
be clear, let us precisely indicate the weak point of Polchinski's arguments
which is clearly expressed in the discussion of the fig. 2 of \cite{354}, p.
274 one may read:

``{\sl We can proceed in this way, thus defining the bare coupling }$\lambda
_{4}^{0}${\sl \ as a function of }$\lambda _{4}^{\text{R}}${\sl , }$\Lambda
_{\text{R}}${\sl , and }$\Lambda _{0}${\sl . Now take }$\Lambda
_{0}\rightarrow \infty ${\sl \ holding }$\Lambda _{\text{R}}${\sl \ and }$%
\lambda _{4}^{\text{R}}${\sl \ fixed.}''

The objection is that, in Wilson's theory (i.e. nonperturbatively) it is
impossible to make sense to the second sentence without an explicit
reference to an (eventually nontrivial) ultra-violet stable fixed point. In
perturbation theory, however, no explicit reference to a fixed point is
needed since, order by order, terms proportional to, say $(\Lambda _{\text{R}%
}/\Lambda _{0})^{2}$, give exactly zero in the limit ``$\Lambda
_{0}\rightarrow \infty ${\sl \ holding }$\Lambda _{\text{R}}${\sl \ and }$%
\lambda _{4}^{\text{R}}${\sl \ fixed}''. However this limit introduces
singularities in the perturbative expansion: the famous ultraviolet
renormalons which make it ambiguous to resum the perturbative series for the
scalar field in four dimensions (for a review on the renormalons see \cite
{4287}). See section \ref{SecTxtBrgt} for more details.

Actually, using the {\em nonperturbative} framework of the ERGE to present a
proof of the {\em perturbative} renormalizability might be seen as a
misunderstanding of the Wilson theory.

\section{Local potential approximation: A textbook example\label{PartLPA}}

\subsection{Introduction\label{SecIntroLPA}}

The local potential approximation (LPA) of the ERGE (the
momentum-independent limit of the ERGE) allows to consider all powers of $%
\phi $. The approximation still involves an infinite number of degrees of
freedom which are treated on the same footing within a nonlinear partial
differential equation for a simple function $V(\phi )$ [$V$ is the (local)
potential, $\phi $ is assumed to be a constant field and thus, except the
kinetic term, the derivatives $\partial \phi $ of $\phi (x)$ are all
neglected in the ERGE].

LPA of the ERGE is the continuous version of the Wilson approximate
recursion formula \cite{437,439} (see also \cite{440}, p. 117) which is a
discrete (and approximate) realization of the RG (the momentum scale of
reference is reduced by a factor two). As shown by Felder \cite{2625}, LPA
is also similar to a continuous version of the hierarchical model \cite{157}.

This approximation has been first considered in \cite{3480} (see also \cite
{4386}) from the sharp cutoff version of the ERGE of Wegner and Houghton 
\cite{414}, it has been rederived by Tokar \cite{4398} by using approximate
functional integrations and rediscovered by Hasenfratz and Hasenfratz \cite
{2085}.

LPA amounts to assuming that the action $S[\phi ]$ reduces to the following
form:

\begin{equation}
S[\phi ]=\int \text{d}^{d}x\frac{z}{2}\left( \partial _{\mu }\phi \right)
^{2}+V(\phi )  \label{Slpa}
\end{equation}
in which $z$ is a pure number (a constant usually set equal to 1) and , to
set the ideas, $V(\phi )$ is a simple function of $\phi _{0}$, it has the
form (symbolically) 
\begin{equation}
V(\phi )=\sum_{n}u_{n}\left( \phi _{0}\right) ^{n}\delta (0)  \label{V0}
\end{equation}

The infinity carried by the delta function would be absent in a treatment at
finite volume, it reflects the difficulties of selecting one mode out of a
continuum set, such ill-defined factors may be removed within a rescaling of 
$\phi $ \cite{3860}. In the derivation of the approximate equation, instead
of using (\ref{V0}) we find it convenient to deal with

\begin{equation}
V(\phi )=\sum_{n}^{\infty }u_{n}\int_{p_{1}\cdots p_{n}}\phi _{p_{1}}\cdots
\phi _{p_{n}}\hat{\delta}\left( {\bf p}_{1}+\cdots +{\bf p}_{n}\right)
\label{FormDeV}
\end{equation}
in which the $u_{n}$'s do not depend on the momenta [see eq. (\ref{expand0}%
)] and to project onto the zero modes $\phi _{0}$ of $\phi $ at the end of
the calculation.

Eq. (\ref{Slpa}) is identical to eq. (\ref{derivative}) in which $Z(\phi ) $
is set equal to the constant $z$. LPA may also be considered as the zeroth
order of a systematic expansion in powers of the (spatial) derivative of the
field (derivative expansion) \cite{212} (see part \ref{PartThird}).

In the following, $\varphi $ stands for $\phi _{0}$ and primes denote
derivatives with respect to $\varphi $ (at fixed $t$):

\begin{eqnarray}
\varphi &\equiv &\phi _{0}  \nonumber \\
V^{\prime }(\varphi ,t) &=&\left. \frac{\partial V}{\partial \varphi }%
\right| _{t}  \label{DerivOfV}
\end{eqnarray}

Frequently, as in the present review, $V^{\prime }(\varphi ,t)$ is replaced
by $f(\varphi ,t)$.

Let us consider the expressions of LPA for the various ERGE's introduced in
part \ref{PartFirst}.

\subsubsection{Sharp cutoff version}

The derivation of the local potential approximation for eq. (\ref{WegHou}) 
\cite{3480} (sharp cutoff version of the LPA of the ERGE) is well known, we
only give the result (for more details see \cite{2085,3860}). It reads:

\begin{equation}
\dot{V}=\frac{K_{d}}{2}\ln \left[ z+V^{\prime \prime }\right] +dV-d_{\phi
}\,\varphi V^{\prime }  \label{eqLPA-WH}
\end{equation}
in which $K_{d}$ is given by (\ref{KD}). The non logarithmic terms come from
the contribution ${\cal G}_{\text{dil}}S$ in (\ref{WegHou}). As usual in
field theory, we neglect the field independent contributions [``const'' in (%
\ref{WegHou})]) to the effective potential $V$.

The $t$-dependence is entirely carried by the coefficient $u_{n}(t)$ in (\ref
{FormDeV}) while $z$ is considered as being independent of $t$. This
condition is required for consistency of the approximation (it prevents the
ERGE from generating contribution to the kinetic term: there is no wave
function renormalization). Writing down explicitly this condition (namely $%
\dot{z}=0$) provides us with the relation:

\[
d_{\phi }=\frac{d-2}{2} 
\]
in other word $\eta =0$, i.e. the anomalous part of the dimension of the
field is zero. This is a characteristic feature of the LPA.

The dependence on $z$ in (\ref{eqLPA-WH}) may be removed (up to an additive
constant) by the simplest (or naive) change of normalization of the field $%
\varphi \rightarrow \varphi \sqrt{z}$. [The exact version (\ref{WegHou}) is
invariant under the same change.] In order to avoid the useless additive
constant terms generated in (\ref{eqLPA-WH}), it is frequent to write down
the evolution equation for the derivative $f=V^{\prime }$, it comes \cite
{2085}:

\begin{equation}
\dot{f}=\frac{K_{d}}{2}\frac{f^{\prime \prime }}{z+f^{\prime }}+\left( 1+%
\frac{d}{2}\right) f+\left( 1-\frac{d}{2}\right) \varphi f^{\prime }
\label{eqLPA'-WH}
\end{equation}

Notice that one could eliminate the factor $K_{d}$ by the change $%
f(\varphi,t)\rightarrow \lambda f(\varphi/\lambda ,t)$ with $K_{d}\cdot
\lambda ^{2}=1$.

It is interesting also to write down the ERGE in the same approximation when
the number of components $N$ of the field is variable. With a view to
eventually consider large values of $N$, it is convenient to redefine the
action and the field as follows:

\[
S\rightarrow N\,S\left[ \frac{\phi }{\sqrt{N}}\right] 
\]
then in the case of O$(N)$ symmetric potential, the LPA of (\ref{WegHou})
yields:

\begin{equation}
\dot{V}=\frac{K_{d}}{2N}\left\{ \left( N-1\right) \ln \left[ z+\frac{%
V^{\prime }}{\varphi}\right] +\ln \left[ z+V^{\prime \prime }\right]
\right\} +dV-d_{\phi }\,\varphi V^{\prime }  \label{eqLPA-WH-N}
\end{equation}

or \cite{2085}

\begin{equation}
\dot{f}=\frac{K_{d}}{2N}\left\{ \left( N-1\right) \frac{\varphi f^{\prime }-f%
}{z\varphi^{2}+\varphi f}+\frac{f^{\prime \prime }}{z+f^{\prime }}\right\}
+\left( 1+\frac{d}{2}\right) f+\left( 1-\frac{d}{2}\right) \varphi f^{\prime
}  \label{eqLPA'-WH-N}
\end{equation}

It may be also useful to express that, in the O$(N)$ symmetric case, $V$ is
a function of $\varphi ^{2}$. By setting $s=\varphi ^{2}$ and $u=2$d$V/$d$s$%
, one obtains \cite{3860}:

\begin{equation}
\dot{u}=\frac{K_{d}}{N}\left[ \frac{3u^{\prime }+2su^{\prime \prime }}{%
1+u+2su^{\prime }}+(N-1)\frac{u^{\prime }}{1+u}\right] +2u+(2-d)su^{\prime }
\label{ComTraWH}
\end{equation}

\subsubsection{The Wilson (or Polchinski) version}

Due to the originality in introducing the arbitrary scaling parameter $c$,
it is worthwhile writing down explicitly the LPA of eq. (\ref{WERGE}). This
equation has first been derived in \cite{4386}.

From the same lines as previously, it comes:

\[
\dot{V}=c\left[ V^{\prime \prime }-\left( V^{\prime }\right) ^{2}+\varphi
V^{\prime }\right] +dV-d_{\phi }\,\varphi V^{\prime } 
\]
in which $c$\ is determined by the condition implying no wave function
renormalization ($\dot{z}=0$) which reads:

\[
d-2-2d_{\phi }+2c=0 
\]
For the Wilson choice $d_{\phi }=d/2$ \cite{440}, it comes $c=1$ and from (%
\ref{cDeWilson}), $\eta =0$ (as it must). Consequently the LPA of (\ref
{WERGE}) is \cite{4386}:

\begin{equation}
\dot{V}=V^{\prime \prime }-\left( V^{\prime }\right) ^{2}+\left( 1-\frac{d}{2%
}\right) \varphi V^{\prime }+dV  \label{WilLPA}
\end{equation}
which, for the derivative $f=V^{\prime }$ yields \cite{3491}:

\[
\dot{f}=f^{\prime \prime }-2ff^{\prime }+\left( 1+\frac{d}{2}\right)
f+\left( 1-\frac{d}{2}\right) \varphi f^{\prime } 
\]

Notice that, contrary to the sharp cutoff version, this equation (as the
exact version) is not invariant under the simplest rescaling of the field $%
\varphi\rightarrow \varphi\sqrt{z}$.

In the O$(N)$-case, it comes:

\[
\dot{V}=\frac{1}{N}V^{\prime \prime }-\left( V^{\prime }\right) ^{2}+\frac{%
N-1}{N}\frac{V^{\prime }}{\varphi}+\left( 1-\frac{d}{2}\right) \varphi
V^{\prime }+dV 
\]

\subparagraph{Polchinski's version}

In the LPA, eq. (\ref{PERGE}) yields exactly the same partial differential
equation as previously \cite{3491} [eq. (\ref{WilLPA})]. For general $N$, it
has been studied by Comellas and Travesset \cite{3860} under the following
form:

\[
\dot{u}=\frac{2s}{N}u^{\prime \prime }+\left[ 1+\frac{2}{N}%
+(2-d)s-2su\right] u^{\prime }+(2-u)u 
\]
in which $s=\frac{1}{2}\varphi ^{2}$ and $u=2dV/ds$ [the definition of the
variables is different from (\ref{ComTraWH})].

\subsubsection{The Legendre transformed version\label{SecLTV}}

\paragraph{Sharp cutoff version}

LPA for the Legendre transformed ERGE has been first written down by Nicoll,
Chang and Stanley \cite{4384} (see also \cite{4387}) with a sharp cutoff.

From eq. (\ref{ChangNic}), it is easy to verify that one obtains the same
equation as in the Wegner-Houghton case (also for general $N$). This is
because at this level of approximation the effective potential coincides
with its Legendre transformation (the Helmholtz potential coincides with the
free energy). Also, the sharp cutoff limit leading to eq. (\ref
{SmoothToSharp}) yields the correct LPA (\ref{eqLPA-WH}) \cite{3550}.

\paragraph{Smooth cutoff version}

With a smooth cutoff, the LPA of the Legendre version of the ERGE [eq. (\ref
{Morris1})] keeps the integro-differential form except for particular
choices of the cutoff function and of the dimension $d$ [e.g. see eq. (\ref
{MorrisLPA}) below]. This is clearly an inconvenience.

From (\ref{PotWett1}) but without introducing the substitution $\varphi
\rightarrow \rho =\frac{1}{2}\varphi ^{2}$, we may easily write down the LPA
for general $N$, it reads: 
\begin{eqnarray}
\dot{V} &=&-\frac{1}{N}\int_{p}\frac{1}{\tilde{C}^{2}}\left( \left[ p^{2}%
\tilde{C}^{\prime }+\tilde{C}\right] \right) \left( \frac{N-1}{M_{0}^{\prime
}}+\frac{1}{M_{1}^{\prime }}\right) +dV-d_{\phi }\,\varphi V^{\prime }
\label{eqLPAWett} \\
M_{0}^{\prime } &=&p^{2}+\tilde{C}^{-1}+V^{\prime }/\varphi \\
M_{1}^{\prime } &=&p^{2}+\tilde{C}^{-1}+V^{\prime \prime }
\end{eqnarray}
in which $\tilde{C}$ is the dimensionless version of the cutoff function $C$
of section \ref{SecLegSCV} with $C=\Lambda ^{-2+\eta }\tilde{C}(p^{2})$ and $%
\tilde{C}^{\prime }=$d$\tilde{C}(p^{2})/$d$p^{2}$ ($p$ is there also
dimensionless). Notice that, following \cite{3357}, we have introduced the
anomalous dimension $\eta $ by anticipation of the anomalous scaling
behavior satisfied by the field in the close vicinity of a non trivial fixed
point. In the approximation presently considered, $\eta $ vanishes and does
not appear in (\ref{eqLPAWett}) but it would have an effect at higher orders
of the derivative expansion (see section \ref{SecDeriv}).

With the particular choice of cutoff function given by (\ref{CutWett1}, \ref
{CutWett2}), eq. (\ref{eqLPAWett}) may be written as follows: 
\begin{equation}
\dot{V}=\frac{K_{d}}{4N}\left[ \left( N-1\right) L_{0}^{d}\left( V^{\prime
}/\varphi \right) +L_{0}^{d}\left( V^{\prime \prime }\right) \right]
+dV-d_{\phi }\,\varphi V^{\prime }  \label{eqLPAWett1}
\end{equation}
in which: 
\[
L_{0}^{d}\left( w\right) =2(2a)^{\frac{2-d}{2b}}\int_{0}^{\infty }\text{d}%
y\,y^{\frac{d-2}{2b}}\,\frac{\text{e}^{-y}}{\left( 1-\text{e}^{-y}\right) }%
\frac{1}{\left[ 1+\left( \frac{2a}{y}\right) ^{\frac{1}{b}}\text{e}%
^{-y}\,w\right] } 
\]

In the sharp cutoff limit $b\rightarrow \infty $ one has: 
\[
L_{0}^{d}\left( w\right) =2\ln \left( 1+w\right) +\text{ const} 
\]
in which ``const'' is infinite, neglecting this usual infinity, one sees
that (\ref{eqLPAWett1}) gives back the expression (\ref{eqLPA-WH}) for $z=1$%
. In order to avoid the infinite ``const'', it is preferable to consider the
flow equation for the derivative $f=V^{\prime }$, in which case the function 
$L_{1}^{d}\left( w\right) =-\frac{\partial }{\partial w}L_{0}^{d}\left(
w\right) $ appears in the equation: 
\[
\dot{f}=\frac{K_{d}}{4N}\left[ \left( N-1\right) \left( f^{\prime }/\varphi
-f/\varphi ^{2}\right) L_{1}^{d}\left( f/\varphi \right) +f^{\prime \prime
}L_{1}^{d}\left( f^{\prime }\right) \right] 
\]

See \cite{3642} for more details on the function $L_{n}^{d}\left( w\right) $.

An interesting expression of the flow equation for the Legendre transformed
action $\Gamma $ is obtained from the smooth cutoff version of Morris \cite
{3357} [see eqs. (\ref{Morris1}, \ref{Morris2})] with a pure power law
(dimensionless) cutoff function of the form:

\begin{equation}
\tilde{C}(p^{2})=p^{2k}  \label{PowerLaw}
\end{equation}

For $d=3$, $k=1$ and $N=1$, the LPA reads \cite{3357}:

\begin{equation}
{\dot{V}={-}\frac{{1}}{\sqrt{2+V^{\prime \prime }}}+{3V}-}\frac{1}{2}{%
\varphi V}^{\prime }  \label{MorrisLPA}
\end{equation}
and for general $N$ \cite{3828}: 
\begin{equation}
{\dot{V}}=-\frac{1}{\sqrt{2+V^{\prime \prime }}}-\frac{N-1}{\sqrt{%
2+V^{\prime }/\varphi}}{+{3V}-}\frac{1}{2}{\varphi V}^{\prime }  \label{e:lo}
\end{equation}

The choice of the power law cutoff function (\ref{PowerLaw}) is dictated by
the will to {\em linearly} realize the reparametrization invariance \cite
{3357}. With the sharp cutoff, the power law cutoff is the only known cutoff
that satisfies the conditions required \cite{3357,3836} to preserve this
invariance.

\subsection{The quest for fixed points\label{FPLPA}}

Fixed points are essential in the RG theory. In field theory, they determine
the nature of the continuum limits; in statistical physics they control the
large distance physics of a critical system.

A fixed point is a solution of the equation 
\begin{equation}
{\dot{V}}^{*}{=0}  \label{eqFxPt}
\end{equation}

From the forms of the equations involved (see the preceding section), it is
easy to see that $V=0$ (or $V=$const) is always a solution of the fixed
point equation. This is the Gaussian fixed point. There are two other
trivial fixed points which are only accounted for with the Wilson (or
Polchinski) version. Following Zumbach \cite{3353,3346,3486}, let us write
the eq. (\ref{WilLPA}) for the quantity $\mu (\varphi,t)=\exp \left(
-V(\varphi,t)\right) $:

\begin{equation}
\dot{\mu}=\mu ^{\prime \prime }+\left( 1-\frac{d}{2}\right) \varphi \mu
^{\prime }+d\,\mu \ln \mu  \label{eqZumbach}
\end{equation}
from which the following trivial fixed point solutions are evident (the
notations differ from those used in \cite{3353,3346,3486}):

\begin{itemize}
\item  $\mu _{G}^{*}=1$, the Gaussian fixed point mentioned above

\item  $\mu _{HT}^{*}=\exp \left( -\frac{1}{2}\varphi ^{2}+\frac{1}{d}%
\right) $, the high-temperature (or infinitely massive) fixed point.

\item  $\mu _{LT}^{*}=0$, the low-temperature fixed point.
\end{itemize}

We are more interested in nontrivial fixed points. But notice that in
general, there are two generic ways fixed points can appear as $N$ or $d$ is
varied \cite{3124}:

\begin{description}
\item[(a)]  splitting off from existing fixed points (bifurcation)

\item[(b)]  appearing in pairs in any region.
\end{description}

In the case (a), the signature is the approach to marginality of some
operator representing a perturbation on an existing fixed point. The classic
example is the Wilson-Fisher fixed point \cite{439} which bifurcates from
the Gaussian as $d$ goes below four. The study of LPA (in the scalar case)
yields no other kind of fixed point, this is why we consider first the
vicinity of the Gaussian fixed point.

\subsubsection{The Gaussian fixed point\label{SecGaussFP}}

A study of the properties of the Gaussian fixed point may easily be realized
by linearization of the flow equations in the vicinity of the origin.

In this linearization, all the LPA equations mentioned in section \ref
{SecIntroLPA} reduce to a unique equation. Considering the derivative $%
f(\varphi)$ of the potential $V(\varphi)$ and a small deviation $g(\varphi)$
to a fixed point solution $f^{*}(\varphi)$: 
\[
f(\varphi)=f^{*}(\varphi)+g(\varphi) 
\]
and choosing $f^{*}(\varphi)\equiv 0$ (the Gaussian fixed point) the
equations linearized in $g$ yields\footnote{%
Up to some change of normalization for eq. (\ref{eqLPA-WH}) and eq. (\ref
{MorrisLPA}) what is authorized in the cases of the sharp cutoff and of the
power law cutoff due to (evident, see section \ref{SecIntroLPA})
reparametrization invariance.} the unique equation \cite{2085, 3836}: 
\begin{equation}
\dot{g}=g^{\prime \prime }+(1-\frac{d}{2})\varphi g^{\prime }+(1+\frac{d}{2}%
)g  \label{GaussianLPA}
\end{equation}

If one sets \cite{2085}:

\[
g(\varphi,t)=e^{\lambda t}\alpha h(\beta \varphi) 
\]
with

\[
\alpha =\frac{4}{d-2}\text{,}\qquad \beta =\left( \frac{d-2}{4}\right) ^{%
\frac{1}{2}} 
\]
then (\ref{GaussianLPA}) reads:

\begin{equation}
h^{\prime \prime }-2\varphi h^{\prime }+2\frac{2+d-2\lambda }{d-2}h=0
\label{hermite}
\end{equation}

\paragraph{Polynomial form of the potential}

If we request the effective potential to be bounded by polynomials\footnote{%
There are other possibilities, see below ``Nonpolynomial...''.} then eq. (%
\ref{hermite}) identifies \cite{2085} with the differential equation of
Hermite's polynomials of degree $n=2k-1$ for the set of discrete values of $%
\lambda $ satisfying:

\begin{equation}
\frac{2+d-2\lambda _{k}}{d-2}=2k-1\text{ \qquad }k=1,2,3,\cdots  \label{vp}
\end{equation}
since $f(\varphi,t)$ is an odd function of $\varphi$.

The same kind of considerations may be done for general $N$, in which case
the Hermite polynomials are replaced by the Laguerre polynomials \cite
{3480,4506}. Since the discussion is similar for all $N$, we limit ourselves
here to a discussion of the simple case $N=1$.

\subparagraph{Eigenvalues:}

From (\ref{vp}), the eigenvalues are defined by:

\begin{equation}
\lambda _{k}=d-k\left( d-2\right) \text{ \qquad }k=1,2,3,\cdots  \label{vp1}
\end{equation}
then it follows that

\begin{itemize}
\item  for $d=4$: $\lambda _{k}=4-2k$ \qquad $k=1,2,3,\cdots $, there are
two non-negative eigenvalues: $\lambda _{1}=2$ and $\lambda _{2}=0$

\item  for $d=3$: $\lambda _{k}=3-k$ \qquad $k=1,2,3,\cdots $, there are
three non-negative eigenvalues: $\lambda _{1}=2$, $\lambda _{2}=1$ et $%
\lambda _{3}=0$
\end{itemize}

\subparagraph{Eigenfunctions:}

If we denote by $\chi _{k}(\varphi)$ the eigenfunctions associated to the
eigenvalue $\lambda _{k}$, it comes:

\begin{itemize}
\item  $\chi _{1}^{+}=\varphi $, $\chi _{2}^{+}=\varphi ^{3}-\frac{3}{2}%
\varphi $, $\chi _{3}^{+}=\varphi ^{5}-5\varphi ^{3}+\frac{15}{4}\varphi $, $%
\cdots $, whatever the spatial dimensionality $d$. The upperscript ``$+$''
is just a reminder of the fact that the eigenfunctions are defined up to a
global factor and thus the functions $\chi _{k}^{-}(\varphi )=-\chi
_{k}^{+}(\varphi )$ are also eigenfunctions with the same eigenvalue $%
\lambda _{k}$. This seemingly harmless remark gains in importance after the
following considerations.
\end{itemize}

To decide whether the marginal operator (associated with the eigenvalue
equal to zero, i.e. $\lambda _{2}$ in four dimensions, or $\lambda _{3}$ in
three dimensions) is relevant or irrelevant, one must go beyond the linear
approximation. The analysis is presented in \cite{2085} for $d=4$. If one
considers a RG flow along $\chi _{2}^{+}$ such that $g_{2}(\varphi
,t)=c(t)\chi _{2}^{+}(\varphi )$, then one obtains, for small $c$: $%
c(t)=c(0)\left[ 1-Ac(0)t\right] $ with $A>0$. Hence the marginal parameter
decreases as $t$ grows. As is well known, in four dimensions the marginal
parameter is irrelevant. However, if one considers the direction opposite to 
$\chi _{2}^{+}$ (i.e. $\chi _{2}^{-}$) then the evolution corresponds to
changing $c\rightarrow -c$. This gives, for small values of $c$: $%
c(t)=c(0)\left[ 1+A\left| c(0)\right| t\right] $ and the parameter becomes
relevant. The parameter $c$ is the renormalized $\phi ^{4}$ coupling
constant $u_{R}$ and it is known that in four dimensions the Gaussian fixed
point is IR stable for $u_{R}>0$ but IR unstable for $u_{R}<0$ (if the
corresponding action was positive for all $\varphi $, one could say that the 
$\phi _{4}^{4}$-field theory with a negative coupling is asymptotically
free, see section \ref{SecTxtBrgt}).

For $d$ slightly smaller than four, $\lambda _{2}$ is positive and the
Gaussian fixed point becomes IR unstable in the direction $\chi _{2}^{+}$
(and remains IR unstable along $\chi _{2}^{-}$). The instability in the
direction $\chi _{2}^{+}$ is responsible for the appearance of the famous
Wilson-Fisher fixed point which remains the only known nontrivial fixed
point until $d$ becomes smaller than 3 in which case it appears a second non
trivial fixed point which bifurcates from the Gaussian fixed point. Any
dimension $d_{k}$ corresponding to $\lambda _{k}=0$, is a border dimension
below which a new fixed point appears \cite{2625} by splitting off from the
Gaussian fixed point. Eq. (\ref{vp1}) gives\footnote{%
This is a result already known from the $\epsilon $-expansion framework \cite
{4847,3480}.}: 
\[
d_{k}=\frac{2k}{k-1},\qquad k=2,3,\cdots ,\infty 
\]

\paragraph{Nonpolynomial form of the potential\label{NonPoly}}

As pointed out by Halpern and Huang \cite{3493} (see also \cite{3889}),
there exist nonpolynomial eigenfunctions for the Gaussian fixed point. In
four dimensions these nonpolynomial eigenpotentials have the asymptotic form 
$\exp \left( c\varphi ^{2}\right) $ for large $\varphi $ and provides the
Gaussian fixed point with new relevant directions (with positive
eigenvalues). From trivial, the scalar field theory in four dimensions would
become physically non trivial due to asymptotic freedom and some effort have
been made with a view to understand the physical implications of that
finding \cite{3891}.

Unfortunately, as stressed by Morris \cite{3817} (see also \cite{3816}), the
finding of Halpern and Huang implies a continuum of eigenvalues and this is
opposite to the usual formulation of the RG theory as it is applied to field
theory where the eigenvalues take on quantized values. Indeed, usually there
are a finite number (preferably small) of relevant (renormalized)
parameters, and it is precisely that property which is essential in the
renormalization of field theory: if the number of relevant parameters is
finite the theory is said renormalizable otherwise it is not. Let us
emphasize that, there is no mathematical error in the work of Halpern and
Huang (see the reply of Halpern and Huang \cite{3818} which maintain their
position except for the ``line of fixed points''\footnote{%
About the infinity of nontrivial fixed points, see also \cite{3893}.}), the
key point is that the theory of ``renormalization'' for nonpolynomial
potentials does not exist. We come back to this discussion in section \ref
{SecTxtBrgt} where we illustrate, among other notions, the notion of
self-similarity which is rightly so dear to Morris \cite{3817,3816}.

\subsubsection{Non trivial fixed points\label{SecNonTrivFP}}

As one may see from the equations presented in section \ref{SecIntroLPA},
the fixed point equation (\ref{eqFxPt}) is a second order non linear
differential equation. Hence a solution would be parametrized by two
arbitrary constants. One of these two constants may easily be determined if $%
V^{*}(\varphi)$ is expected to be an even function of $\varphi$ [O(1)
symmetry] then $V^{*\prime }(0)=0$ may be imposed\footnote{%
Or if it is an odd function of $\varphi$ then $V^{*\prime \prime }(0)=0$ may
be chosen as condition.}. It remains one free parameter: a one-parameter
family of (nontrivial) fixed points are solutions to the differential
equation. But there is not an infinity of physically acceptable fixed points.

As first\footnote{%
A discussion on the singular fixed point solutions in the case $N=\infty $
similar to that mentioned here for LPA, may be found in \cite{414}.}
indicated by Hasenfratz and Hasenfratz (private communication of H.
Leutwyler) \cite{2085}, studied in detail by Felder\footnote{%
Who demonstrates that, for $d=3$, there is only one nontrivial fixed point.} 
\cite{2625} then by Filippov and Breus \cite{176,3345,4506} and by Morris 
\cite{3358,3357}, all but a finite number of the solutions in the family are
singular at some $\varphi _{c}$. By requiring the physical fixed point to be
defined for all $\varphi $ then the acceptable fixed points (if they exist)
may be all found by adjusting one parameter in $V(\varphi )$ (see fig. \ref
{fig1}). For even fixed points, this parameter is generally chosen to be $%
V^{*\prime \prime }(0)$ ($=\sigma ^{*}$ in the following). For $N=1$ the
situation is as follows:

\begin{itemize}
\item  $d\geq 4$, no fixed point is found except for $\sigma ^{*}=0$
(Gaussian fixed point).

\item  $3\leq d<4$, one fixed point (the Wilson-Fisher fixed point \cite{439}%
) is found for a nonzero value of $\sigma ^{*}$ which depends on the
equation considered. For $d=3$ one has:
\end{itemize}

\qquad $\sigma ^{*}=-0.461533\cdots $ \cite{2085,38,3816} ($%
-0.4615413727\cdots $ in \cite{3756}) with eq. (\ref{eqLPA-WH}),

\qquad $\sigma ^{*}=-0.228601293102\cdots $ in \cite{3491} (or at $%
V^{*}(0)=0.0762099\cdots $ \cite{3345,4506}) with eq. (\ref{WilLPA}),

\qquad $\sigma ^{*}=-0.5346\cdots $ \cite{3357} with eq. (\ref{MorrisLPA}).

\begin{itemize}
\item  As indicated previously, a new nontrivial fixed point emanates from
the origin (the Gaussian fixed point) below each dimensional threshold $%
d_{k}=2k/(k-1)$, $k=2,3,\ldots ,\infty $ \cite{2625}.
\end{itemize}

We show in fig. \ref{fig1}, in the case $d=3$, how the physical fixed point
is progressively discovered by adjusting $\sigma =V^{\prime \prime }(0)$ to $%
\sigma ^{*}$ after several tries (shooting method). The knowledge of the
behavior of the solution for large $\varphi $ (obtained from the flow
equation studied) greatly facilitates the numerical determination of $\sigma
^{*}$ and of the fixed point solution $V^{*}(\varphi )$ (for example see 
\cite{3357,3358}). For an indication on the numerical methods one can use,
references \cite{3346,3828,3836} are interesting.

\subsubsection{Critical exponents in three dimensions\label{CE-LPA}}

Once the fixed point has been located, the first idea that generally occurs
to someone is to calculate the critical exponents. There is only one
exponent to calculate (e.g. $\nu $) since $\eta =0$. The other exponents are
deduced from $\nu $ by the scaling relations (e.g. $\gamma =2\nu $)\footnote{%
The right relation is $\gamma =\nu (2-\eta )$.}. The best way to calculate
the exponents is to linearize the flow equation in the vicinity of the fixed
point and to look at the eigenvalue problem. One obtains as in the case of
the Gaussian fixed point a linear second order differential equation. For
example with the Wilson (or Polchinski) version (\ref{WilLPA}), setting $%
V(\varphi ,t)=V^{*}+\;$e$^{\lambda t}v(\varphi )$, one obtains the
eigenvalue equation: 
\begin{equation}
v^{\prime \prime }+\left[ \left( 1-\frac{d}{2}\right) \varphi -2V^{*\prime
}\right] v^{\prime }+\left( d-\lambda \right) v=0  \label{EigenWil}
\end{equation}

As Morris explains in the case of eq. (\ref{MorrisLPA}) \cite{3357}, ``($%
\cdots $) {\sl again one expects solutions to} (\ref{EigenWil}) {\sl %
labelled by two parameters, however by linearity one can choose }$v(0)=1$
(arbitrary normalization of the eigenvectors) {\sl and by symmetry} $%
v^{\prime }(0)=0$ (or by asymmetry and linearity: $v(0)=0$ and $v^{\prime
}(0)=1$). {\sl Thus the solutions are unique, given }$\lambda ${\sl . Now
for large }$\varphi ${\sl , }$v(\varphi )${\sl \ is generically a
superposition\footnote{{\sl To obtain this behavior, use the large $\varphi $
behavior of the fixed point potential $V^{*}(\varphi )\simeq \frac{1}{2}%
\varphi ^{2}$ coming from eq.(\ref{WilLPA}), see \cite{3836}.}} of }$%
v_{1}\sim \varphi ^{2(d-\lambda )/(d+2)}$ and of $v_{2}\sim \exp \left( 
\frac{d+2}{4}\varphi ^{2}\right) $. {\sl Requiring zero coefficient for the
latter restricts the allowed values of }$\lambda ${\sl \ to a discret set''.}

The reason for which the exponential must be eliminated is the same as
previously mentioned in section \ref{NonPoly} to discard nonpolynomial forms
of the potential.

For $d=3$, the Wilson-Fisher fixed point possesses just one positive
eigenvalue $\lambda _{1}$ corresponding to the correlation length exponent ($%
\nu =1/\lambda _{1}$) and infinitely many negative eigenvalues. In the
symmetric case, the less negative $\lambda _{2}$ corresponds to the first
correction-to-scaling exponent $\omega =-\lambda _{2}$ while $\lambda _{3}$
provides us with the second $\omega _{2}=-\lambda _{3}$ and so on. In the
asymmetric case, which is generally not considered (see however \cite{3491}%
), one may also associate the first negative eigenvalue $\lambda _{1}^{\text{%
as}}$ to the first non-symmetric correction-to-scaling exponent $\omega
_{5}=-\lambda _{1}^{\text{as}}$ (the subscript ``5'' refers to the $\phi
^{5} $ interaction term in the action responsible for this kind of
correction, see \cite{3124}).

Another way of numerically determining the (leading) eigenvalues is the
shooting method. One chooses an initial (simple) action and tries to
approach the fixed point (one parameter of the initial action must be finely
adjusted). When the flow approaches very close to the fixed point, its rate
of approach is controlled by $\omega $ (i.e. $-\lambda _{2}$). The
adjustment cannot be perfect and the flow ends up going away from the fixed
point along the relevant direction with a rate controlled by $\lambda
_{1}=1/\nu $.

The first determination of $\nu $ and $\omega $ has been made by Parola and
Reatto \cite{2913} from eq (\ref{eqLPA-WH}) [in the course of constructing a
unified theory of fluids, for a review see \cite{3534}]. They found\footnote{%
In order to appreciate the quality of the estimates the reader may refer to
the so-called best values given, for example, in \cite{4211}.\label{GuiZin}}
(for $d=3$ and $N=1$): 
\[
\nu =0.689,\qquad \omega =0.581 
\]
estimates which are compatible with the well known results of Hasenfratz and
Hasenfratz \cite{2085} obtained from eq. (\ref{eqLPA'-WH}) using the
shooting method: 
\[
\nu =0.687(1),\qquad \omega =0.595(1) 
\]

The error is only indicative of the numerical inaccuracy of solving the
differential equation \cite{2085}. Since then, the above results have been
obtained several times. The first estimate of $\omega _{2}$ has been given
in \cite{38} from the same equation and using again the shooting method:

\[
\omega _{2}\simeq 2.8 
\]

No error was given due to the difficulty of approaching the fixed point
along the second irrelevant direction (two parameters of the initial action
must be adjusted \cite{38,3554}). More accurate estimations of this exponent
may be found in \cite{3860,4627}.

We also mention an estimate of $\omega _{5}$ from the same eq. (\ref
{eqLPA-WH}) \cite{BagBerShpo}:

\[
\omega _{5}\simeq 1.69 
\]

In themselves the estimates of critical exponents in the local potential
approximation do not present a great interest except as first order
estimates in a systematic expansion see section \ref{SecDeriv}. It is
however interesting to notice that the LPA estimates are not unique but
depend on the equation studied. Hence with the Legendre version (\ref
{MorrisLPA}) it comes \cite{3357}:

\[
\nu =0.6604,\qquad \omega =0.6285
\]
which is closer to the ``best'' values. And the closest to the ``best'' are
obtained from the Wilson (or Polchinski) version \cite{3491,3836,3860}: 
\[
\nu =0.6496,\qquad \omega =0.6557
\]

LPA estimates of exponents at various values of $N$ and $d$ have been
published (see for example \cite{3860}). It is also worth mentioning that
LPA gives the exact exponents up to $O(\varepsilon )$ \cite{4398,3553}.

\subsubsection{Other dimensions $2<d<3$\label{SecOtherD}}

If for $2<d<3$ multicritical fixed points appear at the dimensional
thresholds $d_{k}=2k/(k-1)$, $k=2,3,\ldots ,\infty $, the Wilson-Fisher
(critical) fixed point (once unstable) still exists and the same analysis as
above for $d=3$ could have been done to estimate the critical exponents.
However this kind of calculations have not been performed in the LPA despite
some considerations relative to $2<d<3$ \cite{176,3345,4506}.

The special case of $d=2$ does not yield the infinite set of nonperturbative
and multicritical fixed points expected following the conformal field
theories but only periodic solutions corresponding to critical sine-Gordon
models \cite{4424}. This is due to having discarded the nonlocal
contributions which are not small for $d=2$ ($\eta =1/4$ is not small) \cite
{176,3345,4506}.

\subsection{Truncations\label{SecTrunc}}

One may try to find solutions to the ERGE within LPA by expanding the
potential in powers of the constant field variable $\varphi $. Although it
is obviously not a convenient way of studying a non linear partial
differential equation, this programme is interesting because the study of
the ERGE necessarily requires some sort of truncation (or approximation). It
is an opportunity to study the simplest truncation scheme in the simple
configuration of the LPA in as much as there are complex systems (e.g. gauge
theories) for which the truncation in the powers of the field seems
inevitable \cite{4317}.

Margaritis, \'{O}dor and Patk\'{o}s \cite{3478} for arbitrary $N$ and
Haagensen et al \cite{3479} for $2<d<4$ have tried this kind of truncation
on eq. (\ref{eqLPA-WH-N}). The idea is as follows. One expands $V(\varphi
,t) $ in powers of $\varphi $: 
\[
V(\varphi ,t)=\sum_{m=1}^{\infty }c_{m}(t)\,\varphi ^{m} 
\]
and one reports within the flow equation to obtain, e.g. for the fixed point
equation, an infinite system of equations for the coefficients $c_{i}$. That
system may be truncated at order $M$ (i.e. $c_{i}\equiv 0$ for $i>M$) to get
a finite easily solvable set of equations (from which solutions may be
obtained analytically \cite{3479}). By considering larger and larger values
of $M$, one may expect to observe some convergence.

As one could think, the method does not generate very good results: an
apparent new (but spurious) fixed point is found \cite{3478}. Moreover, the
estimation of critical exponents (associated to the Wilson-Fisher fixed
point which, nevertheless, is identified) shows a poor oscillatory
convergence \cite{3478}. Indeed Morris \cite{3358} has shown that this poor
convergence is due to the proximity of a singularity in the complex plane of 
$\varphi $.

Surprisingly, the truncation procedure considered just above works very well
in the case of $O(N)$-symmetric systems. It appears that if one considers
the variable $\rho =\frac{1}{2}\sum_{\alpha }\varphi _{\alpha }^{2}$ and
expands $V\left( \rho ,t\right) $ about the location $\rho _{0}$ of the
minimum of the potential \cite{3642} 
\[
\left. \frac{\text{d}V(\rho ,t)}{\text{d}\rho }\right| _{\rho =\rho _{0}}=0 
\]
then one obtains an impressive apparent convergence \cite{4192} toward the
correct LPA value of the exponents (the method works also within the
derivative expansion \cite{3642, 4192}). The convergence of the method has
been studied in \cite{3553} [with eq. (\ref{eqLPA-WH})] and further in \cite
{4004} [with the Legendre effective action (\ref{MorrisLPA})] where it is
shown that the truncation scheme associated to the expansion around the
minimum of the potential (called co-moving scheme in \cite{3553,4004})
actually does not converge but finally, at a certain large order, leads also
to an oscillatory behavior.

We have seen that the estimates of the critical exponents in the LPA depend
on the ERGE chosen (see section \ref{CE-LPA}). It is thought \cite{4317}
however that for a given ERGE, the estimates should not depend on the form
of the cutoff function (the dependence would occur only at next-to-leading
order in the derivative expansion \cite{3491}). Nevertheless, the truncation
in powers of the field may violate this ``scheme independence'' and affect
the convergence of the truncation. It is the issue studied in \cite{4317}
with three smooth cutoff functions: hyperbolic tangent, exponential and
power-law. An improvement of the convergence is proposed by adjustment of
the smoothness of the cutoff function.

\subsection{A textbook example}

Despite its relative defect in precise quantitative predictions, the local
potential approximation of the ERGE is more than simply the zeroth order of
a systematic expansion in powers of derivatives (see section \ref{SecDeriv}%
). In the first place it is a pedagogical example of the way infinitely many
degrees of freedom are accounted for in RG theory. Almost all the
characteristics of the RG theory are involved in the LPA. The only lacking
features are related to phenomena highly correlated to the non local parts
neglected in the approximation. For example in two dimensions, where $\eta =%
\frac{1}{4}$ is not particularly small, LPA is unable to display the
expected fixed point structure \cite{176,3345,4506,4424}. But, when $\eta $
is small (especially for $d=4$ and $d=3$), one expects the approximation to
be qualitatively correct on all aspects of the RG theory. One may thus trust
the results presented in section \ref{FPLPA} on the search for nontrivial
fixed points in four and three dimensions.

It is a matter of fact that much of studies on RG theory are limited to the
vicinity of a fixed point. This is easily understood due to the universality
of many quantities (exponents, amplitude ratios, scaled equation of state,
etc$...$) that occurs there. However this limitation greatly curtails the
possibilities that RG theory offers. The fixed points and their local
properties (relevant and irrelevant directions) are not the only interesting
aspects of the RG theory. Let us simply quote, as an example, the crossover
phenomenon which reflects the competition between two fixed points. But what
is worse than a simple limitation in the use of the theory, is the resulting
misinterpretation of the theory. This is particularly true with respect to
the definition of the continuum limit of field theory and its relation to
the study of critical phenomena. Let us specify a bit this point (more
details may be found in \cite{38,3554})

It is often expressed that the continuum limit of field theory is defined
``at'' a fixed point and that it is sufficient to look at its relevant
directions to get the renormalized parameters, i.e. a simple linear study of
the RG theory in the vicinity of the fixed point would be sufficient to
define the continuum limit (see for example in \cite{3816,3864,3993}). This
is not wrong but incomplete and, actually void of practical meaning. Indeed
it is not enough emphasized (or understood) in the literature that, for
example, although defined ``at'' the Gaussian fixed point, the field
theoretic approach to critical phenomena \cite{2706,3533} is finally applied
``at'' the Wilson-Fisher fixed point which, in three dimensions, lies far
away from the Gaussian fixed point.\ Then if the renormalized coupling of
the $\phi _{3}^{4}$-field theory was only defined by the linear properties
of the RG theory in the vicinity of the Gaussian fixed point, one would
certainly not be able to discover the nontrivial Wilson-Fisher fixed point
by perturbative means.

Actually, the relevant directions of a fixed point provide us with
exclusively the number and the nature of the renormalized parameters
involved in the continuum limit. But the most important step of the
continuum limit is the determination of the actual scale dependence (say,
the beta functions) of those renormalized parameters. It is at this step
that the recourse to RG theory actually makes sense: the attractive RG flow
``{\sl that emanates out from the fixed point along the relevant direction} 
\cite{440}'' results from the effect of infinitely many degrees of freedom
and the problem of determining this flow is nonperturbative in essence. In
the Wilson space ${\cal S}$ of the couplings $\left\{ u_{n}\right\} $, the
flow in the continuum runs along a submanifold (of dimension one if the
fixed point has only one relevant direction) which is entirely plunged in $%
{\cal S}$. The writing of the corresponding action (the ``perfect action'' 
\cite{4283}) would require the specification of infinitely many conditions
on the action. Because it is an hopeless task to find the ``perfect action''%
\footnote{%
However a truncated expression of the perfect action could be useful in
actual studies of field theory defined on a lattice.}, one gives up all idea
of a determination of the initial absolute scale dependence\footnote{%
Called the functional form of the scale dependence in \cite{3554}.} and one
has recourse to the process of fine tuning some parameter of the (bare)
action in the vicinity of the (Gaussian) fixed point (or in the vicinity of
a critical or tricritical etc.. surface in ${\cal S}$). This allows us to
approach the ideal (or ``perfect'') flow one is looking for and which runs
along a trajectory tangent to a relevant direction. This flow is determined
under a differential form (beta function) since the initial condition is not
specified (the ``perfect action'' is not known) but it results from the
effect of all degrees of freedom involved in the theory (by virtue of the
relevant direction).

The simplicity of the ERGE in the LPA allows us to visualize the evolution
of RG trajectories in the space ${\cal S}$ of the couplings $\left\{
u_{n}\right\} $ and thus to illustrate and qualitatively discuss many
aspects directly related to the nonperturbative character of the RG theory.
In particular the approach to the continuum limit and the various domains of
attraction of a fixed point.\cite{38,3554,4627}

\subsubsection{Renormalization group trajectories\label{SecTxtBrgt}}

Following \cite{38,3554}, one considers an initial simple potential [rather
its derivative with respect to $\varphi $, see (\ref{DerivOfV})], say%
\footnote{%
The normalization of the couplings $u_{n}$ are here modified ($%
u_{n}\rightarrow n\,u_{n}$) compared to (\ref{FormDeV}).}: 
\begin{equation}
f(\varphi ,0)=u_{2}(0)\varphi +u_{4}(0)\varphi ^{3}+u_{6}(0)\varphi ^{5}
\label{fexp}
\end{equation}
corresponding to a point of coordinates $(u_{2}(0),u_{4}(0),u_{6}(0),0,0,%
\cdots )$ in ${\cal S}$, and after having numerically determined the
associated solution $f(\varphi ,t)$ of Eq. (\ref{eqLPA'-WH}) at a varying
``time'' $t$, one concretely represents the RG trajectories (entirely
plunged in ${\cal S}$) by numerically evaluating the derivatives of $f$ at
the origin ($\varphi =0$) corresponding to: $u_{2}(t)$, $u_{4}(t)$, $%
u_{6}(t) $ etc$\ldots $

We then are able to visualize the actual RG trajectories by means of
projections onto the planes $\{u_{2},u_{4}\}$ or $\{u_{4},u_{6}\}$ (for
example) of the space ${\cal S}$.

For the sake of shortness we limit ourselves to a rapid presentation of
figures. The reader is invited to read the original papers.

To approach the Wilson-Fisher fixed point in three dimensions (or $3\leq d<4$%
) starting with (\ref{fexp}), it is necessary to adjust the initial value of
one coupling (e.g. $u_{2}(0)$) to a critical value $u_{2}^{c}\left[
u_{4}(0),u_{6}(0)\right] $ (see fig. \ref{fig2}). This is because the
Wilson-Fisher fixed point has one relevant direction (which must be
thwarted). In such a case the fixed point controls the large distance
properties of a critical system and the potential corresponding to (\ref
{fexp}) with $u_{2}(0)=u_{2}^{c}\left[ u_{4}(0),u_{6}(0)\right] $ represents
some physical system at criticality. The initial $f(\varphi ,0)$ lies in the
critical submanifold ${\cal S}_{\text{c}}$ which is locally orthogonal to
the relevant eigendirection of the fixed point. As already mentioned in
section \ref{wilcontSec}, the renormalized trajectory (RT) T$_{0}$ that
emerges from the Wilson-Fisher fixed point tangentially to the relevant
direction allows to define a massive continuum limit\footnote{%
The nontrivial continuum limit proposed by Wilson in \cite{432} for the
so-called $\phi _{3}^{6}$-field-theory which involves no coupling constant
renormalization but a mass (and a wave function renormalization \cite{4498}
which cannot be evidenced in the present approximation).} (see fig. \ref
{fig3}).

In three dimensions, the Gaussian fixed point has two relevant directions.
There a field theory involving two (renormalized) parameters (a mass-like
and a $\phi ^{4}$-like coupling) may be constructed. A purely massless field
theory may also be constructed by choosing the relevant direction lying in
the critical surface (corresponding to the eigenfunction $\chi _{2}^{+}$ of
section \ref{SecGaussFP}). One obtains a one-parameter theory which
interpolates between the Gaussian and the Wilson-Fisher fixed points (the
renormalized submanifold T$_{1}$ of fig. \ref{fig4}). As already mentioned
in section \ref{wilcontSec}, this scale dependent massless theory
contradicts the Morris view of massless theories as {\em fixed point
theories }(thus scale invariant) \cite{3550,3816,3828,3864,3993,3661}. In
fact a sensible massless (renormalized) scalar theory involves only a $\phi
^{4}$-like coupling ``constant'', say $g$, as parameter. But $g$ is not at
all constant it is scale dependent and this is usually expressed via the
beta function 
\[
\beta (g)=\mu \frac{\text{d}g}{\text{d}\mu } 
\]
in which $\mu $ is some momentum scale of reference and the function $\beta
(g)$ is defined relatively to the flow running along the attractive
submanifold T$_{1}$ (the slowest flow \cite{3094,3554} in the critical
submanifold ${\cal S}_{\text{c}}$). \cite{3554}

The fact that we get a unique scale dependent parameter (renormalizability)
illustrates well the notion of ``{\em self-similarity}'' which means that
the cooperative effect of the infinite number of degrees of freedom may be
reproduced by means of a finite set of (effective or renormalized) scale
dependent parameters (the system ``looks like'' the same when probed at
different scales). In the case of the massless scalar theory, the number of
renormalized parameters is equal to one: a coupling ``constant''\footnote{%
It is, perhaps, useful to mention here that the perturbative expression of
the $\beta $-function (up to one loop order in \cite{3094} and up to two
loops in \cite{4432}) has been re-obtained from the ERGE once expanded with
respect to the number of loops (perturbative study).}.

Fig. \ref{fig4} shows other RG trajectories (different to those attracted to
T$_{1}$), see the caption and \cite{38,3554} for more details.

Less known are the RG trajectories drawn on fig. \ref{fig5} in the sector $%
u_{4}<0$ \cite{4003,4627}. They correspond to:

\begin{enumerate}
\item  The attractive submanifold T$_{1}^{\prime \prime }$ which emerges
from the Gaussian fixed point and is tangent to the eigenfunction $\chi
_{2}^{-}$ of section \ref{SecGaussFP} (symmetric to T$_{1}$ which emerges
from the Gaussian fixed point tangentially to $\chi _{2}^{+}$ ). Along this
submanifold the flows run toward larger and larger negative values of $u_{4}$%
. It is interesting to know that this submanifold still exists in four
dimensions \cite{4627}, it still corresponds to $\chi _{2}^{-}$ and the
associated eigenvalue is zero but the nonlinear analysis shows that the
associated operator ($\phi ^{4}$-like) is marginally relevant. Hence the $%
\phi _{4}^{4}$ field theory with $u_{4}<0$ would be asymptotically free if
the action had not the wrong sign for large values of $\varphi $ (the
negative sign of the $\phi _{4}^{4}$ term which is the only dominant term of
the ``perfect action'' in the vicinity of the Gaussian fixed point). It is
worthwhile mentioning that this relevant direction of the Gaussian fixed
point in four dimensions is different from those discovered by Halpern and
Huang \cite{3493} precisely because of the self-similarity displayed in the
present case and which is a consequence of the discretization of the
eigenvalues presented in section \ref{SecGaussFP} in the case of polynomial
interactions \cite{3817,3816}.

The attractive submanifold T$_{1}^{\prime \prime }$ is endless in the
infrared direction (no nontrivial fixed point lies in the sector $u_{4}<0$,
see section \ref{SecNonTrivFP}). It is customary to say in that case that T$%
_{1}^{\prime \prime }$ is associated to a first order transition. This is
because without a fixed point the correlation length $\xi $ remains finite.
However, although finite, the existence and the length of T$_{1}^{\prime
\prime }$ suggest that $\xi $ may be very large. Following Zumbach in a
study of the Stiefel nonlinear sigma model\footnote{%
The Stiefel nonlinear sigma model is a generalization of the Heisenberg
model with the field a real $N\times P$ matrix and the action is $O(N)\times
O(P)$ invariant \cite{3353,3485,3486}.} within LPA \cite{3353,3485}, one may
refer in the circumstances to ``{\em almost second order phase transition}''
or equivalently to ``{\em weakly first order phase transition}'' (see \cite
{3857,4627} for supplementary details).

\item  The ``tricritical'' attractive submanifold which approaches the
Gaussian fixed point asymptotically tangentially to $\chi _{3}^{+}$ (the
submanifold tangent to $\chi _{3}^{-}$ also exists but is not drawn)
necessitates the adjustment of two parameters in the initial action (because
in three dimensions the Gaussian fixed point is twice unstable). See the
figure caption for more comments and \cite{4627}.
\end{enumerate}

The same configuration displayed by fig. \ref{fig5} has been obtained also
by Tetradis and Litim \cite{3552} while studying analytical solutions of an
ERGE in the LPA for the $O(N)$-symmetric scalar theory in the large $N$
limit. But they were not able to determine ``{\sl the region in parameter
space which results in first order transitions}'' \cite{3552}. Fig. \ref
{fig5} shows this region for the scalar theory in three dimensions.

Fig. \ref{fig6} shows RG trajectories in the critical surface for $d=4$ and $%
u_{4}>0$. The verification of this configuration was the main aim of the
paper by Hasenfratz and Hasenfratz \cite{2085}. As emphasized by Polchinski 
\cite{354}, in the infrared regime all the trajectories approach a
submanifold of dimension one (on which runs the slowest RG flow \cite
{3094,3554}) before plunging into the Gaussian fixed point.\ This pseudo
renormalized trajectory allows to make sense to the notion of effective
field theory (a theory which only makes sense below some finite momentum
scale). Because there is no fixed point other than Gaussian, it is not
possible to adjust the initial action in such a way that the deviations
between the actual RG trajectories and the ideal one dimensional submanifold
be reduced to zero at arbitrary large momentum scales. Those irreducible
deviations are responsible for the presence of the so-called UV renormalon
singularities \cite{789,279} (for a review see \cite{4287}) in the
perturbative construction of the $\phi _{4}^{4}$ field theory. Indeed as
explained in \cite{3554}, the perturbative approach selects the ideal flow
(the slowest) and sets a priori equal to zero all possible deviation (this
is possible order by order in perturbation) without caring about the genuine
physical momentum scale dependence (that requires the explicit reference to
a relevant parameter relative to another fixed point to be well accounted
for). Especially, it is argued in \cite{3554} that the UV\ renormalon
singularities would be absent in the $\beta $-function calculated within the
minimal subtraction scheme of perturbation theory simply because in that
case the scale of reference $\mu $ is completely artificial (has no relation
with a genuine momentum scale except the dimension). See also \cite{3993},
for a discussion of the UV renormalon singularities with the help of an ERGE.

A study of the attraction of RG\ flows to infrared stable submanifolds are
also presented in \cite{4347}.

\subsubsection{Absence of infrared divergences}

It is known that the perturbation expansion of the massless scalar-field
theory with $d<4$ involves infrared divergences. However, it has been shown 
\cite{392} that theory ``develops by itself'' an infinite number of
non-perturbative terms that are adapted to make it well defined beyond
perturbation. One may show that these purely nonperturbative terms are
related to the critical parameter $u_{2}^{c}$ mentioned above \cite{32}. In
the nonperturbative framework of the ERGE there is never infrared
divergences and it is thus particularly well adapted to treat problems which
are known to develop infrared singularities in the perturbative approach
(e.g. Goldstone modes in the broken symmetry phase of $O(N)$ scalar theory
and in general super renormalizable massless theories).

\subsubsection{Other illustrations}

\paragraph{Limit $N=\infty $, large $N$}

Exact results are accessible with LPA. The limit $N=\infty $ corresponds to
the model of Berlin and Kac \cite{4513,301} which can be exactly solved.
Wegner and Houghton \cite{414} have shown that, in this limit their equation
(\ref{WegHou}) is identical to the limit $N=\infty $ of the LPA. A first
order non-linear differential equation is then obtained and studied in \cite
{414} (see also \cite{3860}). Comparison with the exact model is completely
satisfactory. More recently Breus and Filippov \cite{3345} and then Comellas
and Travesset \cite{3860} have studied the large $N$ limit of the LPA for
the Wilson (or Polchinski) ERGE (and for the Wegner-Houghton in \cite{3860})
and find also agreement with the exact results. However D'Attanasio and
Morris \cite{3815} have pointed out that the Large $N$ limit of the LPA for
the Polchinski (or Wilson) ERGE is not exact although it may give correct
results. See also \cite{3552} in which analytical solutions in the Large $N$
limit of the LPA for the Legendre ERGE are obtained and discussed..

\paragraph{The RG flow is gradient flow}

LPA allows to easily illustrate the property that the RG flow is gradient
flow. This property is important because if the RG flow is gradient flow
then only fixed points are allowed (limit cycles or more complicated
behavior are excluded) and the eigenvalues of the linearized RG in the
vicinity of a fixed point are real \cite{4511,4510}. The conditions for
gradient flow are that the beta functions $\beta _{i}\left( \left\{
u\right\} \right) $ [The infinite set of differential renormalization group
equations: $\dot{u}_{i}=\beta _{i}\left( \left\{ u\right\} \right) $] may be
written in terms of a non-singular metric $g_{ij}\left( \left\{ u\right\}
\right) $ and a scalar function $c\left( \left\{ u\right\} \right) $ \cite
{4511}: 
\[
\beta _{i}\left( \left\{ u\right\} \right) =-\sum_{j}g_{ij}\left( \left\{
u\right\} \right) \frac{\partial c\left( \left\{ u\right\} \right) }{%
\partial u_{j}} 
\]

If $g_{ij}\left( \left\{ u\right\} \right) $ is a positive-definite metric,
the function $c\left( \left\{ u\right\} \right) $ is monotonically
decreasing along the RG\ flows: 
\[
\dot{c}=\sum_{i}\beta _{i}\frac{\partial c}{\partial u_{i}}%
=-\sum_{i,\;j}g_{ij}\frac{\partial c}{\partial u_{i}}\frac{\partial c}{%
\partial u_{j}}\leq 0 
\]

Following Zumbach \cite{3346,3486} one may easily verify that the local
potential approximation of the Wilson (or Polchinski) ERGE, written in terms
of $\mu (\varphi ,t)=\exp \left( -V(\varphi ,t)\right) $ [eq. (\ref
{eqZumbach})] may be expressed as a gradient flow: 
\[
g\left( \varphi \right) \;\dot{\mu}=-\frac{\delta {\cal F}\left[ \mu \right] 
}{\delta \mu } 
\]
where 
\begin{eqnarray*}
g\left( \varphi \right) &=&\exp \left[ -\frac{1}{4}\left( d-2\right) \varphi
^{2}\right] \\
{\cal F}\left[ \mu \right] &=&\int \text{d}\varphi \,g\left( \varphi \right)
\left\{ \frac{1}{2}\mu ^{\prime \prime }+\frac{d}{4}\mu ^{2}\left( 1-2\ln
\,\mu \right) \right\}
\end{eqnarray*}

It has then been shown \cite{3830} (see also \cite{3864}) that a c-function
may be defined as ($A$ is a normalization factor) 
\[
c=\frac{1}{A}\ln \left( \frac{4{\cal F}}{d}\right) 
\]
which satisfies in any $d$ the two first properties of Zamolodchikov's
c-function \cite{4523} and has a counting property which generalizes the
third property.

Let us mention also that a c-function has been obtained in the framework of
the truncation in powers of the field of section \ref{SecTrunc} by Haagensen
et al \cite{3479,3756} and that Myers and Periwal \cite{3999} have proposed
a new form of the ERGE which is similar to a gradient flow.

\paragraph{Triviality bounds}

It has been argued \cite{2077} that an upper bound on the Higgs mass may be
estimated from the only trivial character of the scalar field theory in four
dimensions. The idea may be roughly illustrated by the following relation: 
\begin{equation}
\frac{\Lambda _{max}}{m}=\int_{g}^{\infty }\frac{dx}{\beta (x)}
\label{bound}
\end{equation}
in which $g$ is the (usual renormalized) $\phi ^{4}$-coupling of the massive
theory, $m$ is the mass parameter and $\Lambda _{max}$ the maximum value
that the momentum-scale of reference of the scalar theory can take on (it is
associated with an infinite value of the coupling $g$ since no nontrivial
fixed point exists ---this is the consequence of triviality). Hence there is
a finite relation between $m$ and $\Lambda _{max}$.

In order to determine the (triviality) upper bound for the Higgs mass $m_{H}$
(which then replaces $m$), one usually refers to the ratio $R=\frac{m_{H}}{%
m_{W}}$ in which $m_{W}$ is the mass of the vector boson $W$ of the standard
model of the electro-weak interaction \cite{210}. The ratio $R$ expresses as
a function of both $g$ (the scalar coupling) and $G$ (the gauge coupling): 
\begin{equation}
R=\frac{m_{H}}{m_{W}}=f(g,G)
\end{equation}

It appears that, at fixed $G$, $R$ is an increasing function of $g$ \cite
{2751}. For example, at tree level it comes: 
\begin{equation}
R^{2}=8\frac{g}{G^{2}}
\end{equation}

Knowing $G$ and $m_{W}$ from experiments (usually one considers $G^{2}\sim
0.4$ and $m_{W}\sim 80$~Gev, see \cite{2749} for example), the calculation
of $f(g,G)$ would thus allow us to estimate the triviality bound on $m_{H}$
from the following inequality: 
\begin{equation}
R\leq f(\infty ,G)  \label{Rmax1}
\end{equation}

However because $g$ becomes infinite at $\Lambda _{max}$, the question of
determining a bound on, say, the Higgs mass is highly a nonperturbative
issue. Hasenfratz and Nager \cite{2751} using the LPA of the Wegner-Houghton
ERGE have shown how one can proceed to estimate that bound nonperturbatively
(see also \cite{4557}).

\paragraph{Principle of naturalness.}

Some authors have invoked a ``concept of naturalness'' to argue that
fundamental scalar fields may not exist. Initiated by Wilson \cite{424},
this concept would require \cite{2268} {\sl ``the observable properties of a
theory to be stable against minute variations of the fundamental
parameters''.}

A different concept of naturalness, brought up with a view to eliminate non
asymptotically free field theories, would be that \cite{2758} {\sl ``the
effective interactions ($\cdots $) at a low energy scale $\mu _{1}$ should
follow from the properties ($\cdots $) at a much higher energy scale $\mu
_{2}$ without the requirement that various different parameters at the
energy scale $\mu _{2}$ match with an accuracy of the order of $\frac{\mu
_{1}}{\mu _{2}}$. That would be unnatural. On the other hand, if at the
energy scale $\mu _{2}$ some parameters would be very small, say $\alpha
(\mu _{2})=O(\frac{\mu _{1}}{\mu _{2}})$, then this may still be natural $%
\cdots $''}.

Anyway, the two expressions have the same consequence for the scalar field
theory which appears non natural. Let us illustrate this point with the help
of the LPA.

With a view to make the scalar field theory in four dimensions (i.e. $\phi
_{4}^{4}$) non trivial, a non Gaussian fixed point is required. Assuming
that a nontrivial fixed point exists in four dimensions, the procedure of
construction of the resulting continuum limit ``at'' this fixed point would
be similar to that described in fig. \ref{fig3} for the purely massive field
theory in three dimensions. We have seen that in order to approach the RT T$%
_{0}$, at least one parameter ($u_{2}(0)$) of the (bare) initial action as
to be finely tuned\footnote{%
It is clear that the large number of digits in the determination of $%
u_{2}^{c}$ (e.g. $-0.299586913\cdots $ as indicated in the caption of fig. 
\ref{fig2} or similarly in the determination of $\sigma
^{*}=-0.228601293102\cdots $ in \cite{3491}) is not indicative of the
accuracy in the determination of a physical parameter in the LPA but is
required to get as close as possible to the critical surface (consistently
within LPA).} to a nonzero value ($u_{2}^{c}$). Moreover an infinitely small
deviation from the actual $u_{2}^{c}$ results in a drastic change in the
scale dependence of the effective renormalized parameter. This adjustment is
unnatural: how can we justify the origin of numbers like $u_{2}^{c}$? This
unnatural adjustment will be required each time one defines a continuum
limit ``at'' a non trivial fixed point.

On the contrary, an asymptotically free field theory would appear natural
because the adjustment of the initial action is made with respect to the
Gaussian fixed point (the nonuniversal parameters like $u_{2}(0)$ are
adjusted to zero!).

\paragraph{Dynamical generation of masses.}

{\sl ``$\cdots $ one essential element of this systematic theory (a
satisfying synthesis of the theories of weak, electromagnetic and strong
interactions) has remained obscure: we must take the mass of the leptons and
quarks as input parameters, without any real idea of where they come from. $%
\cdots $ the search for a truly natural theory of the quark masses must
continue.}'' \cite{417}

In order to give masses to the intermediate vector bosons while preserving
symmetry, one has imagined the occurrence of a spontaneous symmetry breaking
mechanism. In the standard model of electro-weak interaction the
symmetry-breaking mechanism is associated with the introduction of a scalar
field ($\phi _{4}^{4}$) in the model and the vector bosons acquire masses
via the Higgs mechanism \cite{hig}. The conceptual difficulty with this
model is that one has introduced a peculiar kind of interaction (the scalar
field is self-interacting) which is not asymptotically free. The Higgs
mechanism finally appears to be convenient in the range of energy scales
over which the standard model seems to work but conceptually unnatural and
not generalizable to higher energies ($\phi _{4}^{4}$ does not make sense
above some energy). The other possibility is that the symmetry-breaking
mechanism occurs {\em dynamically}, that is to say without need for
introducing scalars but simply because the gauge fields are interacting
fields \cite{2143,2268}.

In the process of generating masses {\em dynamically}, the main interesting
feature of a non-abelian-gauge-invariant field theory is not actually
asymptotic freedom but its infrared diseases associated with the presence of
IR ``renormalons'' in perturbative series (or equivalently the existence of
a ghost in the infrared regime \cite{789}, for a review see \cite{4287}). It
is very likely that those IR ``renormalons'' convey the lack of any infrared
stable fixed point in the critical (i.e. massless) surface. This means that
a scale dependent coupling constant $G$ of a purely {\em massless} (gauge
invariant) theory is not defined below some momentum scale $\Lambda _{min}$.
It is thus expected that the appearance of massive particles below $\Lambda
_{min}$ can proceed from the existence of a symmetric (massless) theory at
momentum-scales larger than $\Lambda _{min}$ \cite{789,2268,2143}. Let us
illustrate this point with the scalar theory and LPA.

The usual scalar field theory in three dimensions is well defined but does
not present a great interest with respect to a mass generation because:

\begin{itemize}
\item  either the theory is massive and the mass is a given parameter.

\item  or the theory is purely massless but defined as such at any
(momentum) scale in the range $\left] 0,\infty \right[ $ (interpolation
between two fixed points).
\end{itemize}

On the contrary, as shown in fig. \ref{fig5}, the massless theory becomes
asymptotically free in the sector $u_{4}<0$ (the fact that the action has
the wrong sign is not important for our illustrative purposes). But more
importantly, there is no infrared stable fixed point to allow the scale
dependence to be defined at any scale along this RT. Consequently, as close
to the Gaussian fixed point as any trajectory would be initialized, the
resulting trajectory will, after a finite ``time'', end up going away from
the critical surface, i.e. within the massive sector. Finally masses would
have been generated from the momentum scale dependence of a purely massless
theory.

This mechanism illustrated here for masses may also occur for any symmetry
breaking parameter.

\section{Further developments\label{PartThird}}

\subsection{Next-to-leading order in the derivative expansion\label{SecDeriv}
}

The LPA considered in the preceding sections is the zeroth order of a
derivative expansion first proposed as a systematic expansion by Golner \cite
{212}. To be fair, the first use of the derivative expansion (or rather the
gradient expansion \cite{301}) was by Myerson \cite{4468} in conjunction
with an expansion in powers of the field. A line of fixed points with $\eta
\cong 0.045$ was obtained.

The genuine derivative expansion is a functional power series expansion of
the Wilson effective action in powers of momenta so that all powers of the
field are included at each level of the approximation. The idea is to expand
the action $S\left[ \phi ;t\right] $ in powers of momenta \cite{212}:

\[
S\left[ \phi ;t\right] =S^{(0)}\left[ \phi ;t\right] +S^{(2)}\left[ \phi
;t\right] +\sum_{i=1}^{3}S_{i}^{(4)}\left[ \phi ;t\right] +\cdots 
\]
where

\begin{eqnarray*}
S_{i}^{(2k)}\left[ \phi ;t\right]
&=&\sum_{n}a_{in}^{(2k)}(t)H_{in}^{(2k)}[\phi ]\text{,} \\
H_{in}^{(2k)}[\phi ] &=&\int_{q_{1}}\cdots \int_{q_{n}}h_{i}^{(2k)}\left( 
{\bf q}_{1},\cdots ,{\bf q}_{n}\right) \hat{\delta}\left( {\bf q}_{1}+\cdots
+{\bf q}_{n}\right) \phi _{q_{1}}\cdots \phi _{q_{n}}\text{,} \\
H_{0}^{(0)}[\phi ] &=&\delta (0)
\end{eqnarray*}
and the $h_{i}^{(2k)}\left( {\bf q}_{1},\cdots ,{\bf q}_{n}\right) $ are
homogenous monomials in $\left\{ {\bf q}_{j}\right\} $ of degree $2k$, with
the index $i$ present when needed to keep track of degeneracies. Because of
the momentum conserving $\delta $ function we have, for spatially isotropic
systems, only one linearly independent functional of degree 2: $h^{(2)}={\bf %
q}_{1}\cdot {\bf q}_{2}$, and three of degree 4: $h_{1}^{(4)}=\left( {\bf q}%
_{1}\cdot {\bf q}_{2}\right) ^{2}$, $h_{2}^{(4)}=\left( {\bf q}_{1}\cdot 
{\bf q}_{2}\right) \left( {\bf q}_{1}\cdot {\bf q}_{3}\right) $, $%
h_{3}^{(4)}=\left( {\bf q}_{1}\cdot {\bf q}_{2}\right) \left( {\bf q}%
_{3}\cdot {\bf q}_{4}\right) $, since all powers of $q_{j}^{2}$ can be
re-expressed in terms of powers of ${\bf q}_{i}\cdot {\bf q}_{j}$, $i\neq j$%
. This is better seen in the position space where the expansion up to third
order may be written as follows: 
\begin{eqnarray*}
{S[\phi ]} &=&{\int \!d^{d}x\,}\left\{ {V(\phi ,t)+}\frac{1}{2}{Z(\phi
,t)(\partial _{\mu }\phi )^{2}+H_{1}(\phi ,t)(\partial _{\mu }\phi )^{4}}%
\right. \\
&&\qquad \quad \left. {+H_{2}(\phi ,t)(\square \phi )^{2}+H_{3}(\phi
,t)(\partial _{\mu }\phi )^{2}(\square \phi )+\cdots }\right\}
\end{eqnarray*}
on which expression the integrations by parts allow to easily identify the
linearly dependent functionals (as previously the symbol ${\square }$ stands
for $\partial _{\mu }\partial ^{\mu }$).

It remains to substitute this expansion into the ERGE chosen among those
described in section \ref{PartFirst}. To our knowledge the derivative
expansion has only been explicitly written down up to the first order. This
produces two coupled nonlinear partial differential equations for $V$ and $Z$%
.

For the sake of clarity we limit ourselves to a detailed discussion of the
equations for the Polchinski version of the ERGE [of section \ref{PolEq},
eq. (\ref{PERGE})], the other forms of the ERGE\footnote{%
The discussion of the Wilson formulation given by eq. (\ref{WERGE}) is very
similar to that of Polchinski and will not be considered explicitly here
(see \cite{212} for details).} are considered in section \ref{SecOtherStud}.

The Polchinski version of the ERGE at first order of the derivative
expansion yields the following coupled equations \cite{3836} (see also \cite
{3491}):

\begin{eqnarray*}
\dot{f} &=&2K^{\prime }(0)ff^{\prime }-({\int }K^{\prime })f^{\prime \prime
}-({\int }p^{2}K^{\prime })Z^{\prime }+\frac{d+2-\eta }{2}\,f-\frac{d-2+\eta 
}{2}\,\varphi f^{\prime }, \\
\dot{Z} &=&2K^{\prime }(0)fZ^{\prime }+4K^{\prime }(0)f^{\prime
}Z+2K^{\prime \prime }(0)f^{\prime }{}^{2}-({\int }K^{\prime })Z^{\prime
\prime }-4K^{-1}(0)K^{\prime }(0)f^{\prime } \\
&&-\eta Z-\frac{d-2+\eta }{2}\,\varphi Z^{\prime },
\end{eqnarray*}
with $\varphi \equiv \phi _{0}$ and $f(\varphi )\equiv V^{\prime }(\varphi )$%
. As previously defined in section \ref{PolEq}, $K^{\prime }$ stands for d$%
K(p^{2})/$d$p^{2}$ and ${\int }K^{\prime }\equiv {\int_{p}}K^{\prime
}(p^{2}) $ etc....

It is convenient to perform the following rescalings 
\[
\varphi \longrightarrow \sqrt{-{\int }K^{\prime }}\,\varphi ,\ \ \
f\longrightarrow \frac{\sqrt{-{\int }K^{\prime }}}{-K^{\prime }(0)}\,f,\ \ \
Z\longrightarrow K^{-1}(0)Z; 
\]
so that, 
\begin{eqnarray}
\dot{f} &=&-2ff^{\prime }+f^{\prime \prime }+AZ^{\prime }+\frac{d+2-\eta }{2}%
\,f-\frac{d-2+\eta }{2}\,\varphi f^{\prime },  \label{Comellas1} \\
\dot{Z} &=&-2fZ^{\prime }-4f^{\prime }Z+2Bf^{\prime }{}^{2}+Z^{\prime \prime
}+4f^{\prime }-\eta Z-\frac{d-2+\eta }{2}\,\varphi Z^{\prime },
\label{Comellas2}
\end{eqnarray}
where 
\[
A\equiv \frac{(-K^{\prime }(0))(-{\int }p^{2}K^{\prime })}{(-{\int }%
K^{\prime })},\ \ \ B\equiv \frac{K^{\prime \prime }(0)}{(-K^{\prime
}(0))^{2}}. 
\]
Compared to \cite{3836}, we have set $K(0)=1$. These conventions coincide
also with those of \cite{3491}.

Eqs. (\ref{Comellas1}, \ref{Comellas2}) show that all cutoff (scheme)
dependence at order $p^{2}$ is reduced to a two-parameter family $(A,B)$
while at zeroth order [eq. (\ref{Comellas1}) with $Z^{\prime }=0$] there is
no explicit dependence. In general the scheme (cutoff) dependence can be
absorbed into $2k$ parameters at $k$-th order in the derivative expansion 
\cite{3491}. The set of eqs. (\ref{Comellas1}, \ref{Comellas2}) has been
considered first by Ball et al in \cite{3491} with a view to study the
scheme dependence of the estimates of critical exponents and reexamined by
Comellas \cite{3836} who emphasizes (following a remark by Morris \cite{3661}%
) the importance of the breaking of the reparametrization invariance \cite
{3550,3661} in estimating the critical exponents \cite{212} (see also \cite
{4421,4468} and section \ref{SecLRGT}). Let us report on this important
aspect of the eqs. (\ref{Comellas1}, \ref{Comellas2}).

\subsubsection{Fixed points, $\eta $ and the breaking of the
reparametrization invariance}

The distribution of the fixed points of eqs. (\ref{Comellas1}, \ref
{Comellas2}) solution of $\dot{f}^{*}=\dot{Z}^{*}=0$ is identical to that of
the leading order (LPA discussed in section \ref{FPLPA}) except for $d=2$
(see sections \ref{SecOtherD} and \ref{SecTwoD}). Let us simply present the
case of the Wilson-Fisher fixed point for $d=3$.

Following \cite{3836} and in accordance with the discussion of section \ref
{SecNonTrivFP}, to get the non trivial fixed point we impose the following
boundary conditions: 
\begin{eqnarray}
f^{*}(0) &=&0,  \label{Z2a} \\
Z^{*\prime }(0) &=&0,  \label{Z2b} \\
f^{*}(\varphi ) &\sim &\frac{2-\eta }{2}\varphi +C\varphi ^{\frac{d-2+\eta }{%
d+2-\eta }}+\cdots ,\qquad \text{as }\varphi \rightarrow \infty  \label{as1}
\\
Z^{*}(\varphi ) &\sim &D+\cdots ,\qquad \text{as }\varphi \rightarrow \infty
\label{as2}
\end{eqnarray}
where $C$ and $D$ are arbitrary constants. The first two conditions (\ref
{Z2a}, \ref{Z2b}) come from imposing $Z_{2}$-symmetry, while the last two
come directly from the fixed point eqs. (\ref{Comellas1}, \ref{Comellas2}),
once we require the solutions to exist for the whole range $0\le \varphi
<\infty $.

Hence we have three free parameters ($C$, $D$, $\eta $) which are reduced to
one after imposing eqs. (\ref{Z2a}, \ref{Z2b}). The remaining arbitrary
parameter, e.g. $z=Z(0)$, generates a line of (Wilson-Fisher) fixed points
(one fixed point for each normalization $z$). In principle these fixed
points are equivalent as a consequence of the reparameterization invariance
(see section \ref{SecLRGT}) and there is a corresponding unique value of $%
\eta $ (for any fixed point of the line). Consequently, if the
reparametrization invariance was preserved one could get rid of the
arbitrary parameter $z$ by setting it equal to 1. Unfortunately, due to the
derivative expansion, this is not the case: eqs. (\ref{Comellas1}, \ref
{Comellas2}) violate the reparametrization invariance and the estimates of $%
\eta $ (and of $\nu $) {\em depend} on $z$.

In order to get the best estimates for $\eta $, one can adjust $z$ in such a
way as to get an almost realized reparametrization invariance \cite
{4468,212,3836,4421}. The analysis is not simple \cite{3836} due to the
additional effects of the two cutoff parameters $A$ and $B$. Finally
estimates of the critical and subcritical exponents are proposed ($d=3$ and $%
N=1$) \cite{3836}:

\begin{eqnarray*}
\eta &=&0.042 \\
\nu &=&0.622 \\
\omega &=&0.754
\end{eqnarray*}

It is interesting to compare these estimates with those obtained by Golner
in \cite{212} from the Wilson version of the ERGE: 
\begin{eqnarray}
\eta &=&0.024\pm 0.007  \label{EtaGolUpTo2} \\
\nu &=&0.617\pm 0.008  \label{NuGolUpTo2}
\end{eqnarray}

The equations are essentially similar in both cases and the difference in
the estimations of $\eta $ surely originates from the way the cutoff is
introduced and used in the Polchinski case.

Although those two sets of values are close to the best values (see footnote 
\ref{GuiZin}), the procedure which involves $z$ as adjustable parameter is
less attractive than if $\eta $ was uniquely defined at each order of the
derivative expansion. It is thus interesting to look for the conditions of
preservation of the reparametrization invariance.

\subsubsection{Reparametrization invariance linearly realized and preserved%
\label{SecReparamLinReal}}

With a view to control the preservation of the reparametrization invariance,
one may impose it evidently, i.e. linearly, via a particular choice of
cutoff function and try to keep this realization through the derivative
expansion \cite{3836}. This is what has been done in \cite{3357} for the
Legendre version of the ERGE (see below). For the smooth cutoff version of
the ERGE, the only acceptable cutoff function is power-law like \cite
{3661,3836} (otherwise the cutoff should be sharp \cite{3661,3550}).
Unfortunately, for the Polchinski version, the symmetry is broken at finite
order in the derivative expansion and the regulators do not regulate, at
least not in a finite order in the derivative expansion \cite
{2520,3836,3661,3491,3550,4424}. Now considering the Legendre version of the
ERGE of section \ref{LegendreSec} is sufficient to overcome this difficulty 
\cite{2520,3661,3550,4424,3491}.

\paragraph{The smooth cutoff Legendre version and the derivative expansion}

By choosing a power-law cutoff function $\tilde{C}(q^{2})=q^{2k}$ in eq. (%
\ref{Morris1}), one is sure that the derivative expansion will preserve the
reparametrization invariance \cite{2520,3357} and that the exponent $\eta $
will be unambiguously defined.

Let us expand the Legendre (effective) action $\Gamma \left[ \Phi \right] $
as follows: 
\[
\Gamma \left[ \Phi \right] ={\int \!d^{d}x\,}\left\{ {U(\varphi ,t)+}\frac{1%
}{2}{Z(\varphi ,t)(\partial _{\mu }\Phi )^{2}}\right\} 
\]
in which $\varphi $ is independent on $x$.

For $d=3$ and $k=1$, the first order of the derivative expansion yields
(after a long but straightforward computation) the following two coupled
equations for $U$ and $Z$ \cite{3357}:

\begin{eqnarray}
\dot{U}{} &=&{-{\frac{1-\eta /4}{\sqrt{Z}\sqrt{U^{\prime \prime }+2\sqrt{Z}}}%
+}3U-{\frac{1}{2}}(1+\eta )\varphi U^{\prime }}  \nonumber \\
\dot{Z}{} &=&{-{\frac{1}{2}}(1+\eta )\varphi Z^{\prime }-\eta Z+\left( 1-{%
\frac{\eta }{4}}\right) }\left\{ {{\frac{1}{48}}{\frac{24ZZ^{\prime \prime
}-19(Z^{\prime })^{2}}{Z^{3/2}(U^{\prime \prime }+2\sqrt{Z})^{3/2}}}}\right.
\nonumber \\
{} &&{}\left. {-{\frac{1}{48}}{\frac{58U^{\prime \prime \prime }Z^{\prime }%
\sqrt{Z}+57(Z^{\prime })^{2}+(Z^{\prime \prime \prime })^{2}Z}{Z(U^{\prime
\prime }+2\sqrt{Z})^{5/2}}}+{\frac{5}{12}}{\frac{(U^{\prime \prime \prime
})^{2}Z+2U^{\prime \prime \prime }Z^{\prime }\sqrt{Z}+(Z^{\prime })^{2}}{%
\sqrt{Z}(U^{\prime \prime }+2\sqrt{Z})^{7/2}}}}\right\}
\label{LegMorrisUpTo2}
\end{eqnarray}

As expected, the search for a non trivial fixed point solution for these
equations (a solution which is nonsingular up to $\varphi \rightarrow \infty 
$) produces a unique solution with an unambiguously defined $\eta $ \cite
{3357}: 
\begin{equation}
\eta =0.05393  \label{EtaSmcUpTo2}
\end{equation}

The linearization about this fixed point yields the eigenvalues:

\begin{eqnarray}
\nu &=&0.6181  \label{NuSmcUpTo2} \\
\omega &=&0.8975  \label{OmegaSmcUpTo2}
\end{eqnarray}
and also a zero eigenvalue $\lambda =0$ \cite{3357} which corresponds to the
redundant operator ${\cal O}_{1}$ [eq. (\ref{O1})] responsible for the
moving along the line of equivalent fixed points. This is, of course, an
expected confirmation of the preservation of the reparametrization
invariance.

A generalization of the above equations (\ref{LegMorrisUpTo2}) to the $O(N)$
symmetric scalar field theory has been done by Morris and Turner in \cite
{3828}. There, estimates of $\eta $, $\nu $ and $\omega $ are provided for
various values of $N$ and it is shown that the derivative expansion
reproduces exactly known results at special values $N=\infty ,-2,-4,\ldots $
and an interesting discussion on the numerical methods used is presented in
their appendix.

\paragraph{The sharp cutoff Legendre version and the derivative expansion}

The sharp cutoff is the other kind of regularization which allows a linear
realization of the reparametrization invariance \cite{2520,3661,3491,3550}.
As in the previous case of the power-law form of the cutoff function, the
derivative expansion performed with the ERGE satisfied by the Wilson
effective action $S\left[ \phi \right] $ with a sharp cutoff induces
singularities which can be avoided by considering the Legendre transformed $%
\Gamma \left[ \Phi \right] $ \cite{2520,3550}. But the Taylor expansion in
the momenta must be replaced by an expansion in terms of homogeneous
functions of momenta of integer degree \cite{3550} (momentum-scale
expansion). A systematic series of approximations --- the $O(p^{M})$
approximations --- results \cite{3550}.

Although not absolutely necessary, an additional expansion and truncation in
powers of $\varphi $ (avoiding the truncation of the potential) have been
performed in \cite{3550} due to the complexity of the equations\footnote{%
Especially $\sim \varphi ^{8}$ terms and higher have been discarded in
non-zero momentum pieces.}. There, as in the previous case of the smooth
cutoff, the zero eigenvalue corresponding to the redundant operator ${\cal O}%
_{1}$ is found. The estimates for the exponents, however, are worse than
those obtained with smooth cutoff in \cite{3357} [see eqs.(\ref{EtaSmcUpTo2}-%
\ref{OmegaSmcUpTo2})] presumably due to the truncation of the field
dependence \cite{3550}: 
\begin{eqnarray*}
\eta &=&0.0660 \\
\nu &=&0.612 \\
\omega &=&0.91
\end{eqnarray*}

The set of equations (\ref{LegMorrisUpTo2}) together with the sharp cutoff
version of the momentum expansion [$O(p^{1}$)] have also been studied in 
\cite{3816} where, in particular, universal quantities other than the
exponents (universal coupling ratios) have been estimated.

\subsubsection{Studies in two dimensions\label{SecTwoD}}

In two dimensions it is expected that an infinite set of non-perturbative
multicritical fixed points exists corresponding to the unitary minimal
series of $(p,p+1)$ conformal field theories with $p=3,4,\ldots ,\infty $ 
\cite{4542}. As mentioned in section \ref{SecOtherD}, this infinite set
cannot be obtained at the level of LPA with which only periodic solutions
could be obtained \cite{4424}. Using the Legendre ERGE at first order of the
derivative expansion with a power law cutoff [the equations are obtained
similarly to (\ref{LegMorrisUpTo2}) but for $d=2$], Morris in \cite{4424}
(see also \cite{3661}) has found the first ten fixed points (and only these)
and computed the corresponding critical exponents (and other quantities).
The comparison with the exact results of the conformal field theory is
satisfactory (in consideration of the low --- the lowest --- order of
approximation). A similar study has been done using the Polchinski ERGE (at
first order of the derivative expansion) by Kubyshin et al \cite{4222} using
the same iteration technique as in \cite{3491}.

\subsection{A field theorist's self-consistent approach\label{Consist}}

There is an efficient short cut for obtaining the ERGE satisfied by the
(Legendre) effective action. It is based on the observation that this
(exact) equation [see (\ref{Morris2} or \ref{JungWet})] may be obtained from
the one loop (unregularized, thus formal) expression of the effective
action, which reads (up to a field independent term within the logarithm): 
\begin{equation}
\Gamma \left[ \Phi \right] =S\left[ \Phi \right] +\frac{1}{2}\text{tr}\ln
\left( \left. \frac{\delta ^{2}S}{\delta \phi \delta \phi }\right| _{\phi
=\Phi }\right) +\text{higher loop-order,}  \label{loop1}
\end{equation}
by using the following practical rules:

\begin{enumerate}
\item  add the infrared cutoff function $C(p,\Lambda )$ of (\ref{SlambdaIR})
to the action $S$, eq. (\ref{loop1}) then becomes: 
\[
\Gamma \left[ \Phi \right] =\frac{1}{2}\int_{p}\Phi _{p}\Phi
_{-p}C^{-1}(p,\Lambda )+S\left[ \Phi \right] +\frac{1}{2}\text{tr}\ln \left.
\left( C^{-1}+\frac{\delta ^{2}S}{\delta \phi \delta \phi }\right) \right|
_{\phi =\Phi }+\cdots 
\]

\item  redefine $\tilde{\Gamma}\left[ \Phi \right] =\Gamma \left[ \Phi
\right] -\frac{1}{2}\int_{p}\Phi _{p}\Phi _{-p}C^{-1}(p,\Lambda )$, then: 
\[
\tilde{\Gamma}\left[ \Phi \right] =S\left[ \Phi \right] +\frac{1}{2}\text{tr}%
\ln \left. \left( C^{-1}+\frac{\delta ^{2}S}{\delta \phi \delta \phi }%
\right) \right| _{\phi =\Phi }+\cdots 
\]

\item  perform the derivative with respect to $\Lambda $, (only the cutoff
function is concerned) and forget about the higher loop contributions: 
\[
\partial _{t}\tilde{\Gamma}={\frac{1}{2}}\text{tr}\left[ {\frac{1}{C}}%
\Lambda {\frac{\partial C}{\partial \Lambda }}\cdot \left( 1+C\cdot \frac{%
\delta ^{2}S[\Phi ]}{\delta \Phi \delta \Phi }\right) ^{-1}\right] 
\]

\item  replace\footnote{%
This step is often referred to as the ``{\em renormalization group
improvement''} of the one loop effective action.} the action $S$, in the
right hand side of the latter equation, by the effective action $\tilde{%
\Gamma}$ to get eqs. (\ref{Morris2}, \ref{JungWet}), the dilatation part $%
{\cal G}_{\text{dil}}\tilde{\Gamma}$ being obtained from usual (engineering)
dimensional considerations\footnote{%
But do not forget to introduce the anomalous dimension of the field in order
to get an eventual nontrivial fixed point.}.
\end{enumerate}

It is noteworthy that the above rules have been heuristically first used 
\cite{4429,4430} to obtain the local potential approximation of the ERGE for
the (Legendre) effective action. However the main interest of the above
considerations is that they allow introducing the (infra-red) cutoff
function independently of $S$, via the so-called ``{\em proper time}'' (or ``%
{\em heat kernel}'' or ``{\em operator}'') regularization \cite{4878}. This
kind of regularization is introduced at the level of eq. (\ref{loop1}) via
the general identity: 
\[
\text{tr}\ln \left( \frac{A}{B}\right) =-\int_{0}^{\infty }\frac{\text{d}s}{s%
}\text{tr}\left( \text{e}^{-sA}-\text{e}^{-sB}\right) 
\]

Forgetting again about the field-independent part (and, momentaneously,
about the ultra-violet regularization needed for $s\rightarrow 0$) one
introduces an infrared cutoff function $F_{\Lambda }\left( s\right) $ within
the proper time integral representation of the logarithm of $A=\left. \frac{%
\delta ^{2}S}{\delta \phi \delta \phi }\right| _{\phi =\Phi }$: 
\[
\frac{1}{2}\text{tr}\ln A\longrightarrow -\frac{1}{2}\int_{0}^{\infty }\frac{%
\text{d}s}{s}F_{\Lambda }\left( s\right) \text{tr\ e}^{-sA} 
\]

The function $F_{\Lambda }\left( s\right) $ must tend to zero sufficiently
rapidly for large values of $s$ in order to suppress the small momentum
modes and should be equal to 1 for $\Lambda =0$.

Then following the rules 3-4 above applied on $\Gamma $ (i.e., not on $%
\tilde{\Gamma}$), one obtains a new kind of ERGE\footnote{%
Notice that, because one performs a derivative with respect to $\Lambda $,
the essential contribution to $\partial _{t}\Gamma $ comes from the
integration over a small range of values of $s$ (corresponding to the rapid
decreasing of $F_{\Lambda }\left( s\right) $), hence an ultraviolet
regularization is not needed provided that the resulting RG equation be
finite.} for the effective action \cite{4858}: 
\begin{equation}
\partial _{t}\Gamma =-\frac{1}{2}\int_{0}^{\infty }\frac{\text{d}s}{s}%
\Lambda \frac{\partial F_{\Lambda }\left( s\right) }{\partial \Lambda }\exp
\left[ -s\frac{\delta ^{2}\Gamma }{\delta \Phi \delta \Phi }\right]
\label{properge}
\end{equation}

There are apparently two advantages of using this kind of ERGE:

\begin{itemize}
\item  the regularization preserves the symmetry of the action \cite{4877}

\item  the derivative expansion is slightly easier to perform than in the
conventional approach and one may preserve the reparametrization invariance 
\cite{4858}.
\end{itemize}

In \cite{4858}, Bonanno and Zappal\`{a} have considered the next-to-leading
order of the derivative expansion of (\ref{properge}) (while in \cite{4873}
only a pseudo derivative expansion, in which the wave-function
renormalization function $Z(\phi ,t)$ is field-independent\footnote{%
See below.}, was used). They have chosen $F_{\Lambda }\left( s\right) $ in
such a way that the integro-differential character of the ERGE disappears
and they have tested the preservation of the reparametrization invariance.
Moreover a scheme dependence parameter, related to the cutoff width, is at
hand in this framework which, presumably, will allow someone to look at the
best possible convergence of the derivative expansion when higher orders
will be considered.

Another kind of regularization related to this ``self-consistent'' approach
should be mentioned here. It consists in introducing the cutoff function in
eq. (\ref{loop1}) in-between the momentum integration [expressing the trace]
and the logarithm. This procedure has been considered in \cite{4857} at the
level of the local potential approximation. However the ERGE keeps its
integro-differential character and the study of \cite{4857} has then been
conducted within the constraining polynomial expansion method of section \ref
{SecTrunc}.

\subsection{Other studies up to first order of the derivative expansion\label%
{SecOtherStud}}

In this section we mention studies of the derivative expansion which,
although interesting, do not consider explicitly the reparametrization
invariance.

Filippov and Radievskii \cite{3351} have obtained a set of two coupled
equations that look like eqs. (\ref{Comellas1}, \ref{Comellas2}) but their
numerical studies were based on an approximation which consists in
neglecting the term corresponding to $AZ^{\prime }$ in (\ref{Comellas1}).
They, nevertheless, present interesting estimates of critical exponents for
several values of $d$ in the range $\left] 2,3.5\right] $.

As already mentioned, Ball et al \cite{3491} have studied the ``scheme''
dependence with the help of the Polchinski ERGE at first order in the
derivative expansion (without considering explicitly the breaking of the
reparametrization invariance). They have used an interesting simple
iteration procedure to determined the fixed point.

Bonanno et al \cite{4269} have presented a sharp version of the coupled
differential equation for $V$ and $Z$ which however yields a negative value
of $\eta $ in three dimensions. The authors claim that this failure is not
to be searched in an intrinsic weakness of the sharp cutoff. A previous
attempt had been done with the sharp cutoff version and a two loop
perturbative anomalous dimension was obtained by means of a polynomial
truncation in the field dependence \cite{4021}. Re-obtention of two loop
results from the derivative expansion of the ERGE satisfied by the effective
(Legendre) action (with smooth cutoff) are also described in \cite
{4429,3642,4432}.

\subparagraph{Pseudo derivative expansion}

Tetradis and Wetterich \cite{3642} have initiated an original strategy to
obtain systematic accurate estimates on, say, critical quantities already
from the lowest order of the derivative expansion. The idea is based on the
smallness of $\eta $ and may be roughly described as follows. At lowest
order of the derivative expansion (LPA), one assumes that ${Z(0,t)}$ already
depends on $t$. One then determines an approximate $t$-dependence (assuming $%
\eta $ is small) from the momentum dependence of the exact propagator $%
\Gamma ^{(2)}$. Hence $\eta $ is not equal to zero even at the lowest order
of the derivative expansion and this yields an ``improved'' LPA. The next
order would amount to consider an explicit $\varphi $-dependent ${Z(\varphi
,t)}$ and the following order higher derivatives of the field in the action.
This is not a genuine derivative expansion and it does not account for the
reparametrization invariance. Nevertheless the approach seems efficient
considering the estimates obtained at the leading order of that pseudo
derivative expansion. Let us first quote, for $d=3$ and $N=1$, the results
found with the supplementary help of a truncation in powers of the field
associated to an expansion around the minimum of the potential \cite{3642}: $%
\nu =0.638$, $\eta =0.045$, $\gamma =1.247$, $\beta =0.333$ and without
truncation in the field dependence \cite{4433}: $\nu =0.643$, $\eta =0.044$, 
$\gamma =1.258$, $\beta =0.336$, $\delta =4.75$. In this latter work, the
scaled equation of state has been calculated using this pseudo derivative
expansion. For more details on this approach see the review by Berges et al 
\cite{4700} in this volume.

A study for $d=2$ has also been achieved \cite{4580} following the spirit of 
\cite{3642} (i.e. with a truncation) with a view to discuss the
Kosterlitz-Thouless phase transition \cite{4581}. The aim of the authors was
to show the power of the ERGE compared to the perturbative approach (due to
IR singularities). It is amazing to notice the excellent estimation of $\eta 
$ ($\simeq 0$.$24$ instead of $\frac{1}{4}$) obtained in this work knowing
that $\eta $ was assumed to be small and that the truncation was crude. As
indicated by the authors, this result may be accidental. An extension of
this work may be found in \cite{4754}.

\subsection{Convergence of the derivative expansion?}

Comparing the estimates of the critical exponents obtained at first order of
the derivative expansion to that of, e.g., the $\varepsilon $-expansion, one
can easily see that the derivative expansion is potentially much more
effective than the perturbative (field theoretical) approach. But, to date,
it is not known whether it converges or not. Morris and Tighe \cite{4326}
have considered this question at one and two loop orders for different cases
of regularization (cutoff) functions and for either the Wilson (---
Polchinski) or Legendre effective action. It is found that the Legendre flow
equation converges at one and two loops: slowly with sharp cutoff (as a
momentum-scale expansion), and rapidly in the case of a smooth exponential
cutoff (but, in this latter case, the reparametrization invariance is not
satisfied, see above). The Wilson (--- Polchinski) version and the Legendre
flow equation with power law cutoff function do not converge.

It is possible that the derivative expansion gives rise to asymptotic series
which would be Borel summable. This is deduced from the knowledge of an
exact solution for the effective potential for QED$_{2+1}$ in a particular
inhomogeneous external magnetic field, from which it has been shown that the
derivative expansion (known at any order) is a divergent but Borel summable
asymptotic series \cite{4462}.

The annoying perspective that the derivative expansion does not converge has
prevailed on Golner to look for a method of successive approximations for
the ERGE that is not based on power series expansion \cite{3912}.

\subsection{Other models, other ERGE's, other studies...}

Up to this point, we have presented in some details various aspects
(derivation, invariances, approximations, truncations, calculations)%
\footnote{%
Let us quote, in addition, the issue of scheme dependence, already mentioned
in the text, several aspects of which are considered in \cite{SchemeDep}.}
of the ERGE for scalar systems. Since these issues are also encountered for
more complex systems (but with, potentially, a more interesting physical
content), in this section we limit ourselves to mentioning the existence of
studies based on the ERGE relative to models different from the pure scalar
theory\footnote{%
One may also refer to a recent review by Aoki \cite{4823}.}. Most often
these models involve more structure due to supplementary internal degrees of
freedom. Formally, the master equations keep essentially the same general
forms as described in section \ref{PartFirst} [owing to the trace symbol as
used in (\ref{Morris2})]. The equations involved in the studies actually
show their differences when approximations are effectively considered. The
studies are characterized by the action $S$ considered, the cutoff function
chosen and the approximation applied on the ERGE. In consideration of the
large number of publications and the variety of models studied, we choose to
classify them according to the increasing degree of complexity of the model
with respect to the field: scalar (or vector), spinor and gauge field.

\subparagraph{Other ERGE's involving pure scalars (or vectors)}

We have already mentioned the Stiefel model studied in the LPA in \cite
{3353,3485} let us quote also the Boson systems \cite{4679}, the nucleation
and spinodal decomposition \cite{4422}, the interface unbinding transitions
arising in wetting phenomena \cite{4512}, the roughening transition \cite
{4701}, transitions in magnets (non-collinear spin ordering, frustrated)
studied in \cite{4199}, disordered systems \cite{4727}, the quantum
tunnelling effect \cite{4229}, the well developed turbulence \cite{4794} and
even the one-quantum-particle system \cite{4824}. More developed are the
numerous studies of scalar theories at finite temperature \cite{Finite Temp}%
. Reviews on the finite temperature framework may be found in \cite
{4315,ReviewFT} though they cover more than scalars.

\subparagraph{Other ERGE's}

It exists some studies involving pure spinors \cite{PureSpins} and also some
mixing scalars and spinors \cite{Spin+Scal}, reviews may be found in \cite
{RevSpin}.

In addition, a rich literature on ERGE deals with systems in presence of
gauge fields: pure gauge fields \cite{4502,JaugePure}, gauge fields with
scalars \cite{Jauge with scalars} and with spins \cite{251,Jauge with spins}%
, supersymmetric gauge fields \cite{4568} and gravity \cite{gravity}.
Reviews on this theme are listed in \cite{ReviewsJaugeERGE}.\bigskip
\bigskip 

Despite the great number of studies done up to now on the ERGE, its
systematic use in nonperturbative calculations and in describing
nonuniversalities is still in its infancy. A better mastery of invariances
within the truncation procedure, the extension of series (this has required
some time in perturbation theory), the consideration of more complex and
realistic models with a view to obtain estimates of useful physical
quantities will necessitate much more investigations in the future.\newpage

\begin{center}
{\Large Figure captions}
\end{center}

\begin{enumerate}
\item  Three solutions of the sharp cutoff fixed point equation for $f(\phi
_{0})=V^{\prime }(\phi _{0})$ and $d=3$ [eq. (\ref{eqLPA'-WH}) with $\dot{f}%
=0$]. All (here two) but one ($\sigma ^{*}=-0.4615337\cdots $) of the
solutions are singular at some (not fixed) $\varphi _{c}$. The parameter $%
\sigma =V^{\prime \prime }(0)$ is adjusted to $\sigma ^{*}$ by requiring the
physical fixed point to be defined for all $\phi _{0}$ (in the text $\varphi 
$ stands for $\phi _{0}$).\label{fig1}

\item  Determination by the shooting method of the initial critical value $%
u_{2}^{c}=-0.299586913\cdots $ corresponding to the initial values $%
u_{4}(0)=3$ and $u_{n}(0)=0$ for $n>4$ [from eq. (\ref{eqLPA'-WH}) with $d=3$%
]. Open circles indicate the initial points chosen in the canonical surface
(representing simple actions) of ${\cal S}$. The illustration is made via
projections onto the plane $[u_{2},u_{4}]$. The determination of $u_{2}^{c}$
is made by iterations (shooting method) according to increasing labels.
Arrows indicate the infrared direction (decreasing of the momentum-scale of
reference). The RG trajectories follow two opposite directions according to
whether $u_{2}(0)>u_{2}^{c}$ (labels 1, 3, 5) or $u_{2}(0)<u_{2}^{c}$
(labels 2, 4, 6). The Wilson-Fisher (once infrared unstable) fixed point
(full circle) is only reached when $u_{2}(0)=u_{2}^{c}$ (dashed curve). The
corresponding RG trajectory lies in the critical surface ${\cal S}_{c}$ of
codimension 1.\label{fig2}

\item  Illustration of the simplest nonperturbative continuum limit in three
dimensions [from eq. (\ref{eqLPA'-WH}) with $d=3$].\ Approach to the purely
massive ``renormalized trajectory'' $T_{0}$ (dot-dashed curve) by RG
trajectories initialized at $u_{4}(0)=3$ and $u_{n}(0)=0$ for $n>4$ and $%
(u_{2}(0)-u_{2}^{c})\rightarrow 0^{+}$ (open circles). The trajectories
drawn correspond to $\log (u_{2}(0)-u_{2}^{c})=-1,-2,-3,-4,-5,-6$. When $%
u_{2}(0)=u_{2}^{c}$ the trajectories do not leave the critical surface and
approach the Wilson-Fisher fixed point (full circle), as in figure \ref{fig2}%
. But, ``{\sl moving a little bit away from the critical manifold, the
trajectory of the RG will to begin with, move towards the fixed point, but
then shoot away along [...] the relevant direction towards the so-called
high temperature fixed point}...{\sl ''} (see text, section \ref{wilcontSec}
and \cite{3993}){\sl .}\label{fig3}

\item  Projection onto the plane $(u_{4},u_{6})$ of some remarkable RG
trajectories for $u_{4}(0)>0$ [from eq. (\ref{eqLPA'-WH}) with $d=3$]. Full
lines represent trajectories on the critical surface ${\cal S}_{c}$. The
arrows indicate the directions of the RG flows on the trajectories. The
submanifold T$_{1}$ of one dimension to which are attracted the trajectories
with small values of $u_{4}(0)$ and which links the Gaussian fixed point to
the Wilson-Fisher fixed point corresponds to the renormalized trajectory on
which is defined the continuum limit of the massless field theory in three
dimensions. For larger values of $u_{4}(0)$ the RG trajectories approach the
Wilson-Fisher fixed point from the opposite side, they correspond to the
Ising model. The dotted line T$_{2}$ plunging into the Wilson-Fisher fixed
point does not lie on ${\cal S}_{c}$ but represents a RG trajectory
approaching the Wilson-Fisher fixed point along the second less irrelevant
direction (lying in a space of codimension 2). The corresponding critical
behavior is characterized by the absence of the first kind of correction to
scaling (that corresponding to the exponent $\omega $), it would be
representative of some Ising models with spin $s=1/2$. The two trajectories
that leave the Wilson-Fisher fixed point (dashed lines) correspond to the
unique relevant eigendirection (with two ways, due to the arbitrary
normalization, associated with the two phases of the critical point, they
also correspond to two massive RT's). The open circles represent initial
simple actions. \label{fig4}

\item  Projection onto the plane $\{u_{2},u_{4}\}$ of some remarkable RG
trajectories for $u_{4}(0)<0$ [from eq. (\ref{eqLPA'-WH}) with $d=3$]. Black
circles represent the Gaussian and Wilson-Fisher fixed points. The arrows
indicate the directions of the RG flows on the trajectories. The ideal
trajectory T$_{1}$ (dot line) which interpolates between the two fixed
points represents the RT corresponding to the so-called $\phi _{3}^{4}$
renormalized field theory in three dimensions (usual RT for $u_{4}>0$).
White circles represent the projections onto the plane of initial critical
actions. For $u_{4}(0)>0$, the effective actions (e.g. initialized at B')
run toward the Wilson-Fisher fixed point asymptotically along the usual RT.
Instead, for $u_{4}(0)<0$ and according to the initial values of the
parameters of higher order ($u_{6}$, $u_{8}$, etc.), the RG trajectories
either (A) meet an endless RT emerging from the Gaussian fixed point T$%
_{1}^{\prime \prime }$ (dashed curve) and lying entirely in the sector $%
u_{4}<0$ or (B) meet the usual RT to reach the Wilson-Fisher fixed point.
The frontier which separates these two very different cases (A and B)
corresponds to initial actions lying on the tri-critical subspace (white
square C) that are sources of RG trajectories flowing toward the Gaussian
fixed point asymptotically along the tricritical (pseudo) RT. Notice that
the coincidence of the initial point B with the RG trajectory starting at
point A is not real (it is accidentally due to the projection onto a plane
of the trajectories lying in a space of infinite dimension). See text for a
discussion and \cite{4003,4627,3552}. \label{fig5}

\item  RG trajectories on the critical surface ${\cal S}_{c}$ obtained from
integration of eq. (\ref{eqLPA'-WH}) with $d=4$ (projection onto the plane $%
(u_{4},u_{6})$). Open circles indicate the initial points chosen on the
canonical surface of ${\cal S}_{c}$ (of codimension 1)$.$The two lines which
come from the upper side of the figure are RG trajectories initialized at $%
u_{4}(0)=20$ and $u_{4}(0)=40$ respectively. The arrows indicate the
infrared direction. The trajectories are attracted to a submanifold of
dimension one before plunging into the Gaussian fixed point. This pseudo
renormalized trajectory (it has no well defined beginning) allows to make
sense to the notion of effective (here massless) field theory. Strictly
speaking, the continuum limit does not exist due to the lack of another
(nontrivial fixed point) which would allow the scale dependence (of the
renormalized parameter along the RT) to be defined in the whole range of
scale $\left] 0,\infty \right[ $. See text for a discussion (from \cite{3554}%
). \label{fig6}
\end{enumerate}

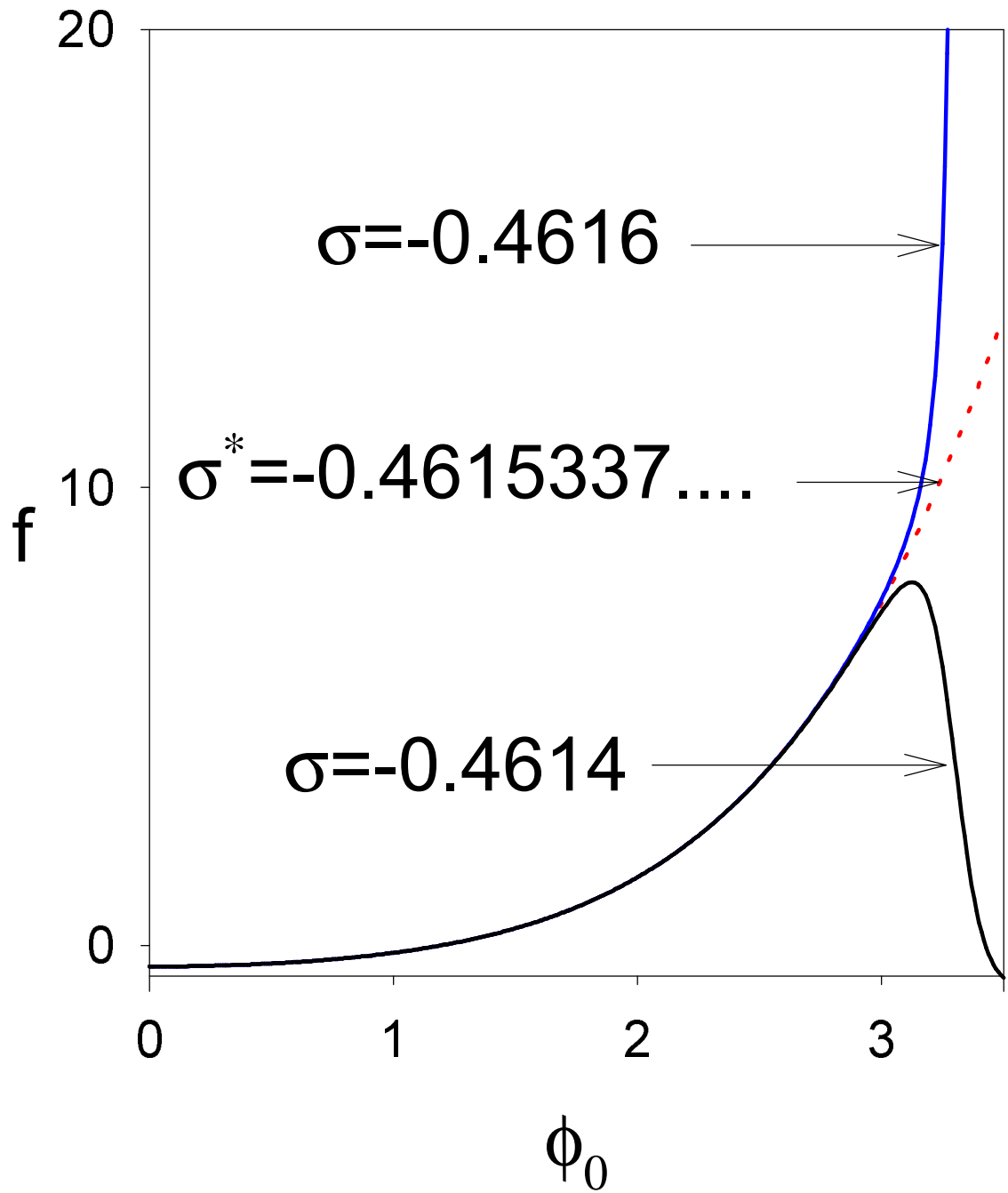
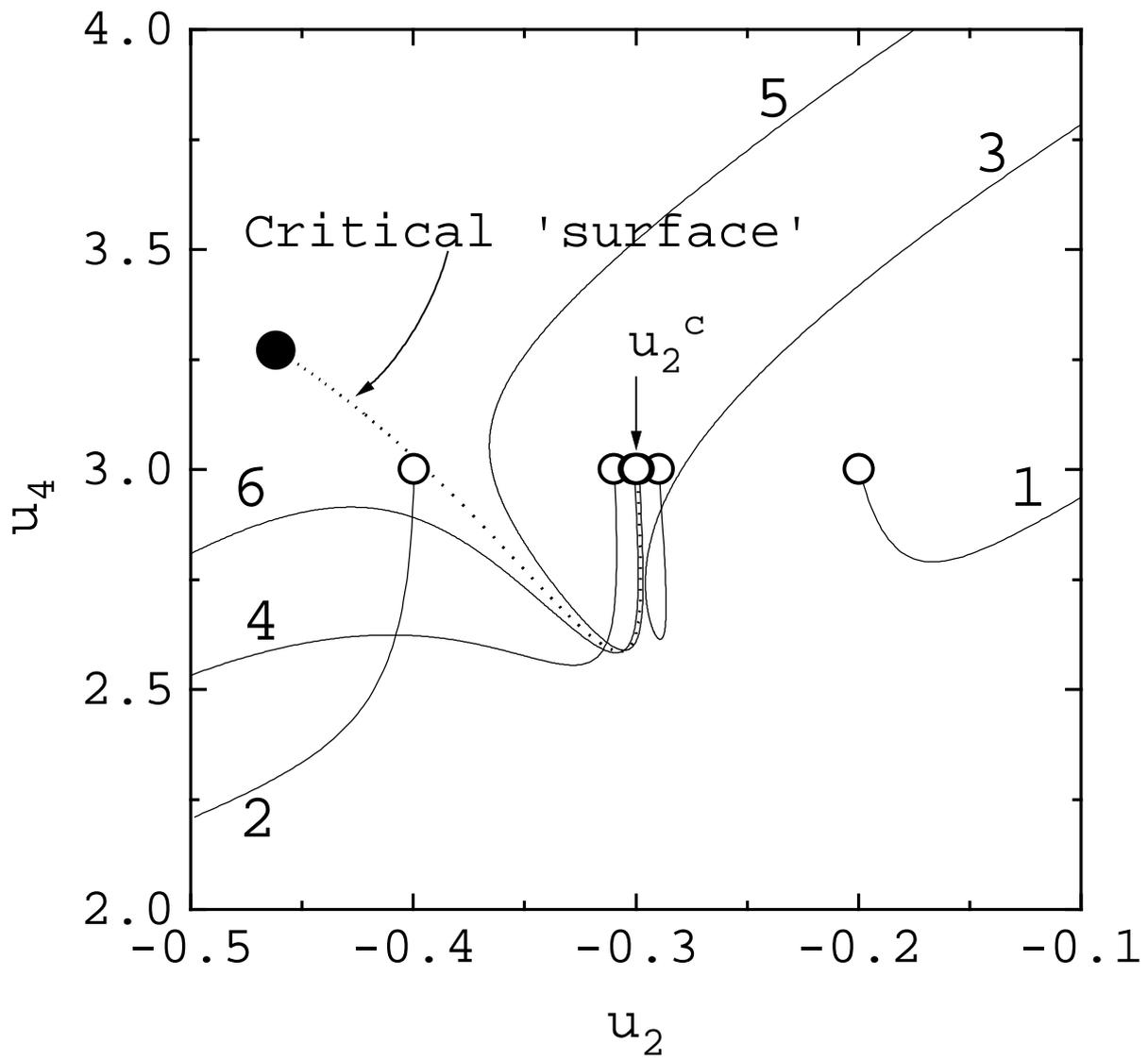
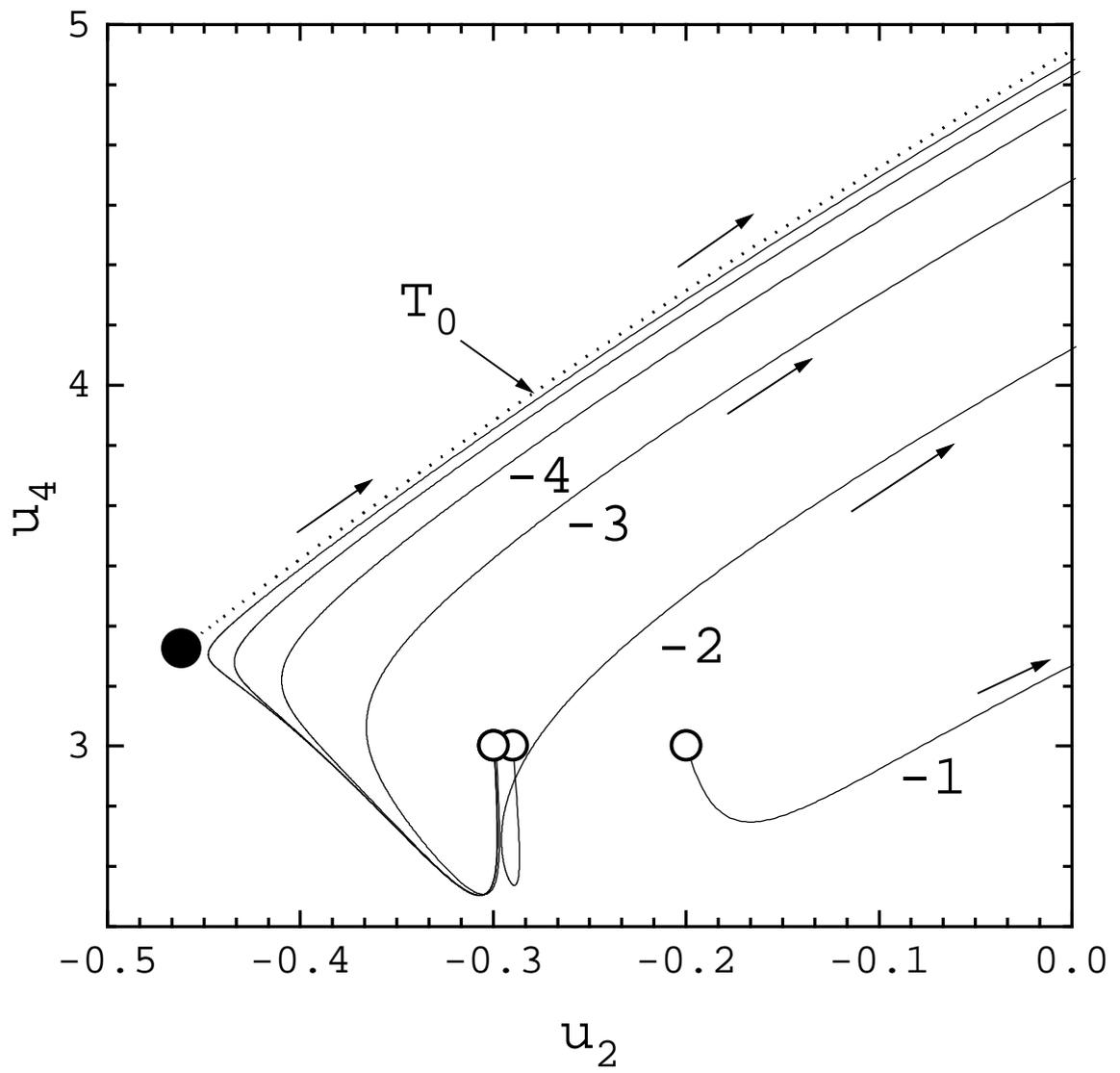
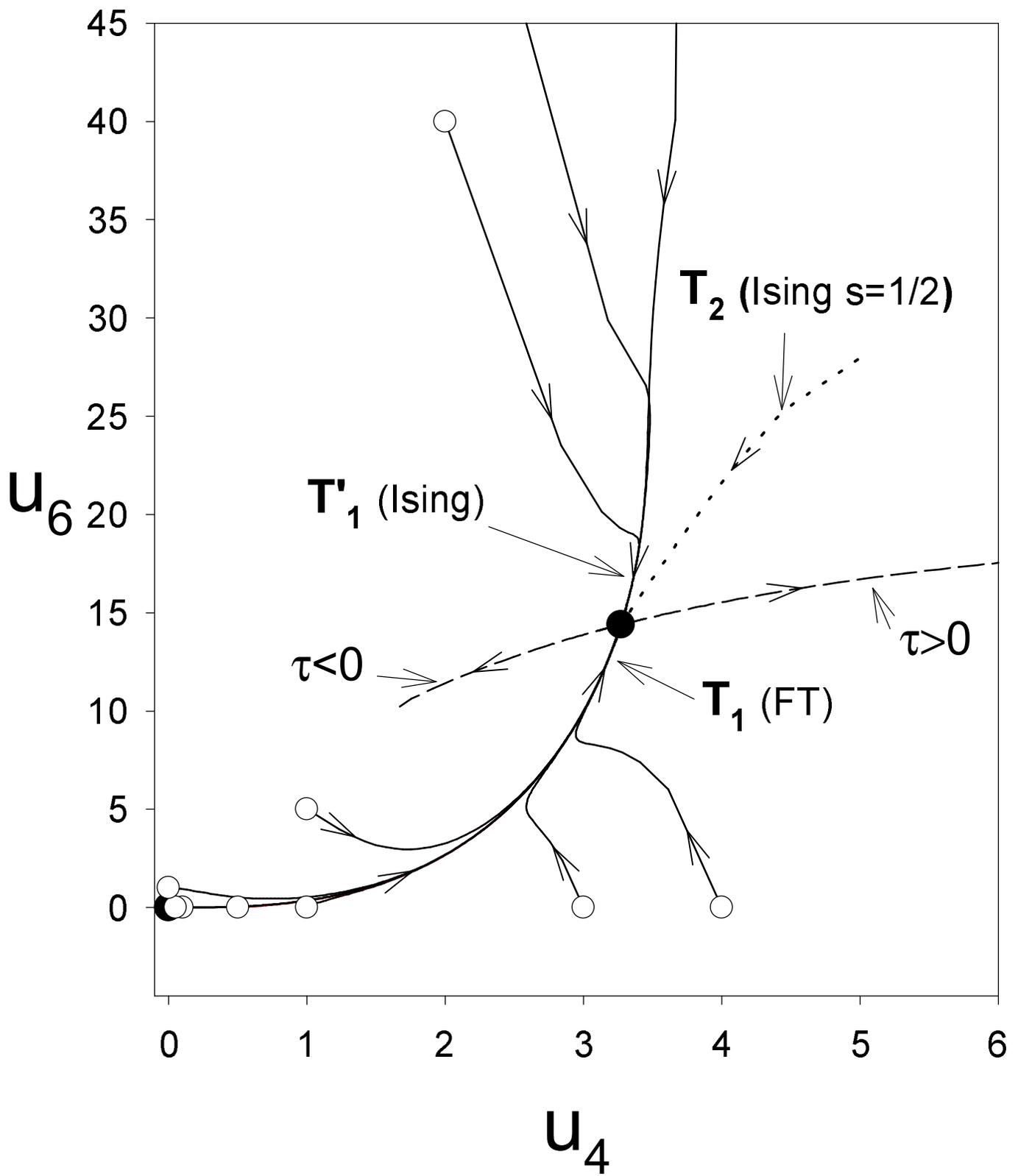
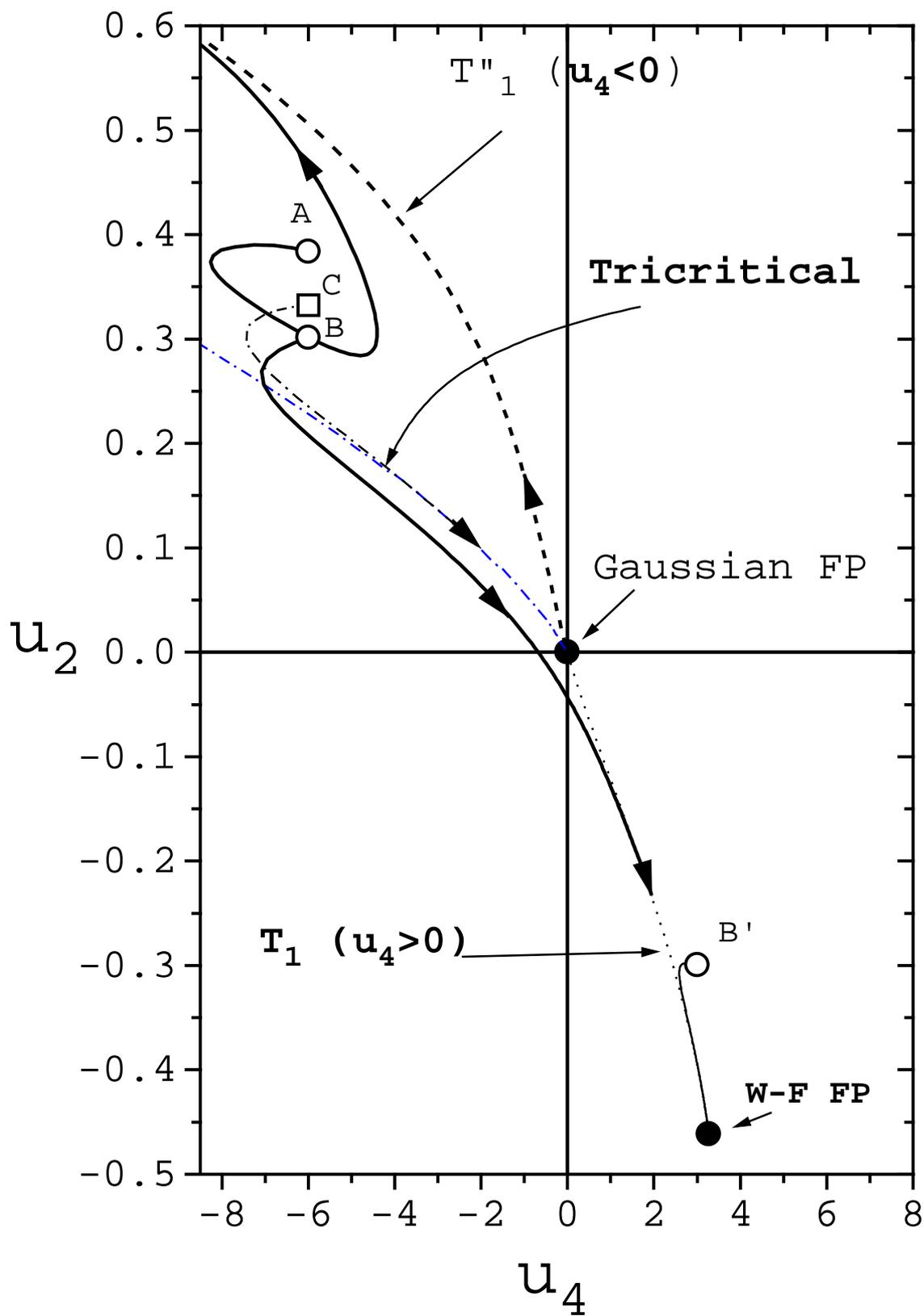
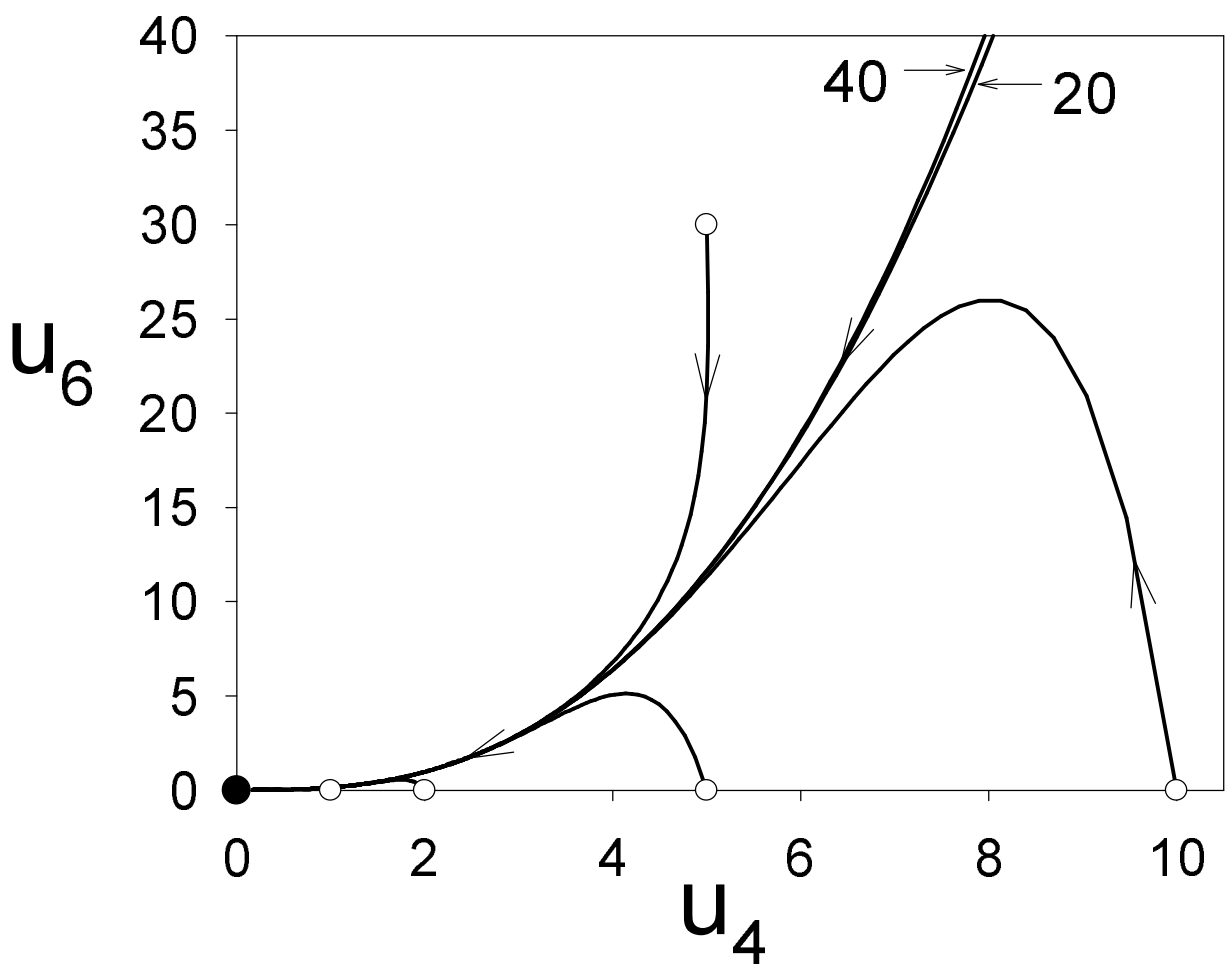
\end{document}